\documentclass[aps,twocolumn,floatfix,superscriptaddress,prb]{revtex4-2} 
\usepackage{graphicx}
\usepackage{xcolor}
\usepackage{subfigure}
\usepackage{mathtools}
\usepackage{physics,amsmath}
\usepackage{epsfig}
\usepackage{chemfig}
\usepackage[colorlinks= true,urlcolor=blue,linkcolor= blue,citecolor=blue,bookmarks=false,pdfstartview=]{hyperref}
\usepackage{siunitx}
\usepackage{bm} 
\usepackage{xr}
\usepackage{xcolor}
\usepackage{placeins}
\usepackage{braket}
\usepackage{epsfig}
\begin{document}

\title{Single-ion anisotropy driven chiral magnetic order in a spin-1 antiferromagnetic chain}

\author{S. Vaidya}
\email{s.vaidya@warwick.ac.uk}
\affiliation{Department of Physics, University of Warwick, Gibbet Hill Road, Coventry, CV4 7AL, UK}
\author{S. P. M. Curley}
\affiliation{Department of Physics, University of Warwick, Gibbet Hill Road, Coventry, CV4 7AL, UK}
\author{P. Manuel}
\author{J. Ross Stewart}
\author{M. Duc Le}
\affiliation{ISIS Pulsed Neutron Source, STFC Rutherford Appleton Laboratory, Didcot, Oxfordshire OX11 0QX, United Kingdom}
\author{A. Hern\'{a}ndez-Meli\'{a}n}
\affiliation{Department of Physics, Durham University, Durham DH1 3LE, United Kingdom}
\author{T. J. Hicken}
\affiliation{Department of Physics, Durham University, Durham DH1 3LE, United Kingdom}
\affiliation{Center for Neutron and Muon Sciences, Paul Scherrer Institut, Forschungsstrasse 111, 5232 Villigen PSI, Switzerland}
\author{C. Wang}
\author{H. Luetkens}
\author{J. Krieger}
\affiliation{Center for Neutron and Muon Sciences, Paul Scherrer Institut, Forschungsstrasse 111, 5232 Villigen PSI, Switzerland}
\author{S. J. Blundell}
\affiliation{Department of Physics, Clarendon Laboratory, University of Oxford, Parks Road, Oxford, OX1 3PU, United Kingdom}
\author{T. Lancaster}
\affiliation{Department of Physics, Durham University, Durham DH1 3LE, United Kingdom}
\author{K. A. Wheeler}
\affiliation{Department of Chemistry, Whitworth University, Spokane, Washington 99251, USA}
\author{D. Y. Villa}
\affiliation{Department of Chemistry and Biochemistry, Eastern Washington University, Cheney, Washington 99004, USA}
\affiliation{National High Magnetic Field Laboratory (NHMFL), Los Alamos National Laboratory, Los Alamos, NM, USA}
\author{Z. E. Manson}
 \author{J. A. Villa}
\author{J.~L.~Manson}\thanks{Deceased 7 June 2023.}
\affiliation{Department of Chemistry and Biochemistry, Eastern Washington University, Cheney, Washington 99004, USA}
\author{J. Singleton}
\affiliation{National High Magnetic Field Laboratory (NHMFL), Los Alamos National Laboratory, Los Alamos, NM, USA}
\author{R. D. Johnson}
\affiliation{Department of Physics and Astronomy, University College London, Gower Street, London WC1E 6BT, United Kingdom}
\affiliation{London Centre for Nanotechnology, University College London, London WC1H 0AH, United Kingdom}
\author{P. A. Goddard}
\email{p.goddard@warwick.ac.uk}
\affiliation{Department of Physics, University of Warwick, Gibbet Hill Road, Coventry, CV4 7AL, UK}

\begin{abstract}
Chirality in magnetic systems gives rise to a wide range of exotic phenomena, yet its influence in $S=1$ chains remains largely unexplored. Here, we present a comprehensive experimental study of a chiral antiferromagnetic (AFM) $S=1$ chain, [Ni(pym)(H$_{2}$O)$_{4}$]SO$_{4} \cdot$ H$_{2}$O (pym = pyrimidine), where the Ni(II) octahedral orientation exhibits a four-fold chiral periodicity. Muon spin rotation measurements indicate the onset of long-range magnetic order below $T_{\rm N} = 1.82(2)\,\mathrm{K}$. Neutron diffraction measurements reveal a chiral AFM order driven by a chiral modulation of the easy-axis anisotropy direction, rather than the typical scenario of Dzyaloshinskii-Moriya interactions, geometrical frustration or higher-order interactions. Inelastic neutron scattering (INS) measurements reveal dispersive spin-wave excitations well described by linear spin-wave theory, with Hamiltonian parameters $J_{0} = 6.81(1)\,\mathrm{K}$ (intrachain exchange), $J'_{1\rm a} = -0.091(1)\,\mathrm{K}$ (interchain exchange), and $D = -3.02(1)\,\mathrm{K}$ (easy-axis single-ion anisotropy). These parameters are further validated by Monte Carlo simulations of the magnetisation. Additionally, the INS data reveal multiple dispersionless bands, suggesting the presence of further excitations beyond the scope of our linear spin-wave theory.
\end{abstract}

\maketitle

\newpage
\section{Introduction}

Chirality arises when an object lacks inversion symmetry, preventing it from being superimposed onto its mirror image. This fundamental concept plays a pivotal role across various scientific disciplines, from the behaviour of elementary particles to the handedness of biological molecules. In condensed matter physics, chirality is at the heart of numerous exotic electronic, magnetic, and optical phenomena~\cite{bousquet_structural_2025}. Notable examples include chiral fermions in topological Weyl semimetals~\cite{Binghai_2017}, magnetic chiral dichroism in systems with chiral magnetic order~\cite{saito_magnetic_2008, sessoli_strong_2015, bordacs_chirality_2012, nakagawa_magneto-chiral_2017}, skyrmions~\cite{Muhlbauer_2009,fert_2017, Lancaster_2019} and the topological Hall effect~\cite{Mayoh_2022_THE}. These effects have drawn considerable interest in recent years, not only for their potential applications in emerging technologies but also for their impact on advancing fundamental physics~\cite{Geert_2011, Emori_2013, fert_2017, WANG_2022_skyrmions_rev, WANG_2022_THE_rev}.

In magnetic systems, chiral order~\cite{cheong_magnetic_2022} typically emerges due to the Dzyaloshinskii–Moriya (DM) interaction in non-centrosymmetric materials. In contrast, chirality in centrosymmetric systems often relies on geometrical frustration or higher-order interactions, such as biquadratic exchange~\cite{Marty_2008, Rog_2013, khanh_2020_centro}. However, these mechanisms are typically weak and challenging to control, especially in $3d$ transition metal systems where spin-orbit coupling is generally small~\cite{Huang_2004}. Moreover, the influence of chirality on low-dimensional quantum spin systems, particularly integer spin systems that can host the Haldane phase, remains largely unexplored.

To address these gaps, we investigate [Ni(pym)(H$_{2}$O)$_{4}$]SO$_{4}\cdot$H$_{2}$O (pym = pyrimidine, C$_{4}$H$_{4}$N$_{2}$), an antiferromagnetic (AFM) $S=1$ chain where the orientation of the Ni(II) octahedral environment has a four-fold chiral periodicity. In this system the single-ion anisotropy (SIA) direction is expected to follow the chiral modulation of the Ni(II) octahedra. In systems where the SIA direction alternates between neighboring ions, such as the $S=1$ staggered chain Ni(pym)(H$_{2}$O)$_{2}$(NO$_{3}$)$_{2}$~\cite{Vaidya_staggered}, it has been shown that an easy-axis SIA will result in significant spin canting even without considering the DM interaction~\cite{Feyerherm_FeCl2_2004, XCl2L_Jem_2024}. The resulting canting angle is strongly dependent on $D/J_{0}$, where $D$ is the SIA energy and $J_{0}$ is the exchange interaction strength along the chain. In contrast, in the two-fold-staggered \textit{easy-plane} case, intersection of two easy-planes is expected to result in collinear order along a \textit{pseudo-easy axis}~\cite{Povarov_2020, Facheris_2024,Vaidya_pseudo}. A chiral SIA direction may induce a chiral magnetic structure in the absence of the DM interaction.

Furthermore, a change from a two-fold staggered to a chiral structure may have a significant influence on the spin excitations. In the two-fold staggered Cu(II) $S=1/2$ chains, Cu(C$_{6}$D$_{5}$COO)$_{2}\cdot$3D$_{2}$O and Cu(pym)(H$_{2}$O)$_{2}$(NO$_{3}$)$_{2}$, the alternating orientation of the local spin environment gives rise to non-trivial staggered $g$-tensors and Dzyaloshinskii-Moriya (DM) interactions which, on application of an external magnetic field, generate a staggered field and an energy gap with a $\Delta\sim H^{2/3}$ field dependence~\cite{Dender_Cu_benzo_1997, Feyerherm_stag_2000, Zvyagin_SG_2004, Huddart_spin_transport_2021}. It was shown that mapping the Hamiltonian to the sine-Gordon quantum-field theory accounts for this behavior and the rich excitation spectra containing soliton and breather modes in the staggered $S=1/2$ chain. In the $S=1/2$ chiral chain, [Cu(pym)(H$_{2}$O)$_{4}$]SiF$_{6}\cdot$H$_{2}$O~\cite{Liu_chiral_2019}, the four-fold periodicity of the octahedral environment was found to result in a linear field dependence of the spin gap, which deviated from theoretical predictions and remains to be fully explained.~\cite{Liu_chiral_2019}.

Here, to investigate the influence of chirality on the magnetic order and spin dynamics in a $S=1$ chain, we present a comprehensive experimental study of [Ni(pym)(H$_{2}$O)$_{4}$]SO$_{4}\cdot$H$_{2}$O. We first confirm the chiral chain structure using single-crystal x-ray diffraction (XRD) and infer the magnetic Hamiltonian. Muon-spin rotation ($\mu^{+}$SR) measurements show clear signs of long-range magnetic order at low temperatures. Using elastic neutron diffraction, we indeed find an AFM chiral order consistent with an easy-axis SIA. Inelastic neutron scattering (INS) is then used to quantify the Hamiltonian parameters. While the dispersive spin-wave excitations are largely accounted for by linear spin-wave theory (LSWT), there are multiple dispersionless bands which hint at the presence of additional excitations not captured by our LSWT. Finally, we perform classical Monte Carlo (MC) simulations of the field-dependent magnetization $M(H)$ to confirm the model determined using INS data and LSWT simulations.

\section{Results and Discussion}
\subsection{X-ray diffraction}
\begin{figure}
    \centering
    \includegraphics[width= \linewidth]{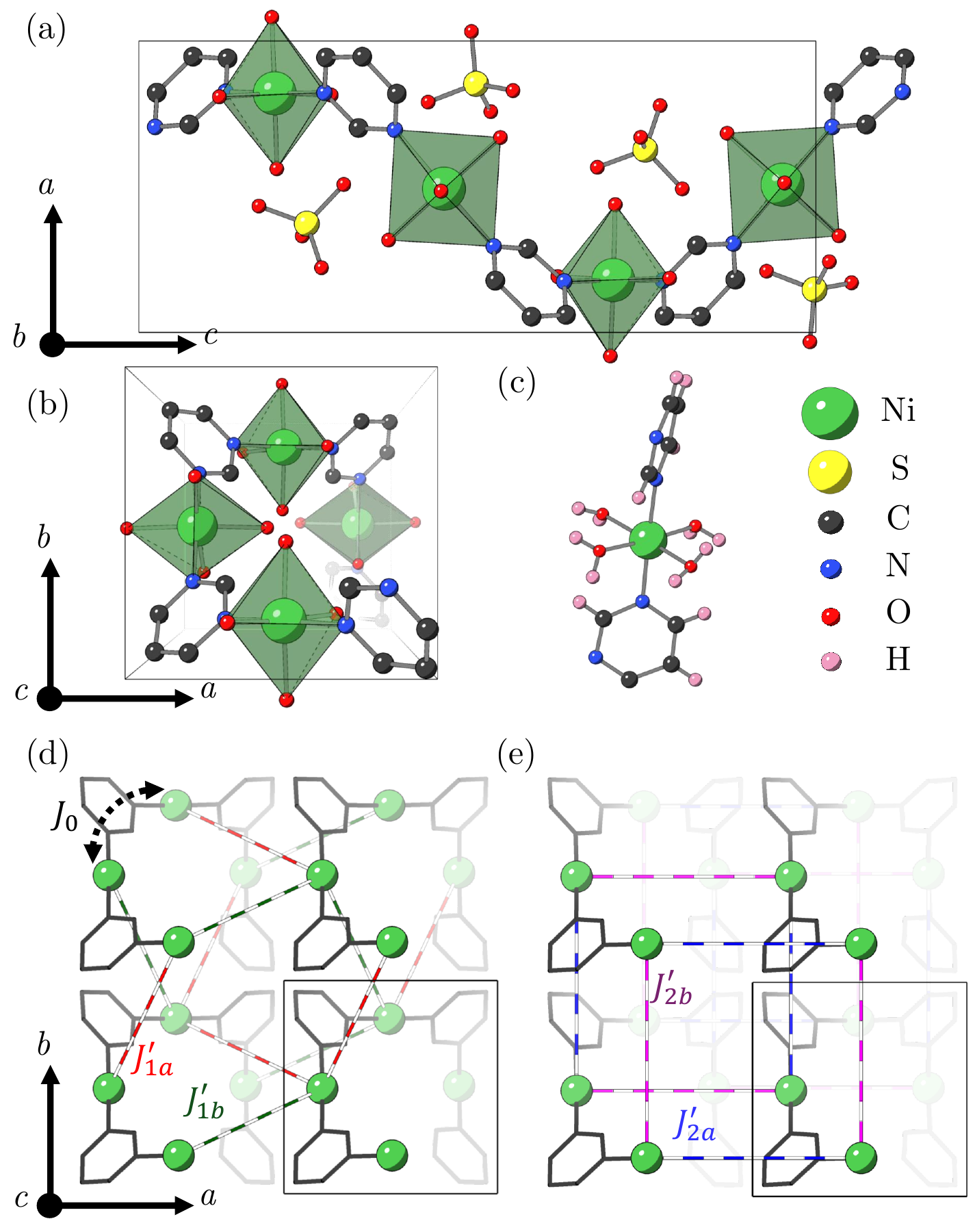}
    \caption{The structure and exchange pathways of the $S=1$ chiral chain, [Ni(pym)(H$_{2}$O)$_{4}$]SO$_{4}\cdot$H$_{2}$O. (a) View along the $b$-axis of the unit cell containing a single chiral chain connected by $J_{0}$ exchange mediating pym molecules (H atoms and H$_{2}$O molecules have been omitted for clarity). (b) View down the $c$-axis depicting the fourfold rotation of the local Ni(II) octahedral environment. (c) The local crystal environment of each Ni(II) ion. (d) and (e) View down the $c$-axis showing exchange pathways. The black dashed arrow represents the intrachain $J_{0}$ pathway. The labelled dashed lines indicate the $J'_{1a,b}$ and $J'_{2a,b}$ interchain exchange interaction mediated via H-bonds. The solid black lines indicate the unit cell boundaries.}
    \label{fig: Chiral_Struct}
\end{figure}
Single-crystal XRD measurements at $T=100$\,K reveal that [Ni(pym)(H$_{2}$O)$_{4}$]SO$_{4}\cdot$H$_{2}$O crystallises in the chiral space group $P4_{1}$, with lattice parameters $a = 7.9395(4)$\,\si{\angstrom} and $c = 18.4594(10)$\,\si{\angstrom}. As depicted in Fig.~\ref{fig: Chiral_Struct}, the Ni(II) ions are connected by pym ligands, forming Ni-pym-Ni chains that extend along the $c$-axis. Each Ni(II) ion resides in a distorted octahedral environment [see Fig.~\ref{fig: Chiral_Struct}(c)], defined by four equatorial bonds to O donors from H$_{2}$O molecules, with Ni-O bond lengths ranging from $2.060(3)$\,\si{\angstrom} to $2.078(3)$\,\si{\angstrom} and two axial bonds to nitrogen atoms of the pym ligands, with Ni–N bond lengths $2.079(3)$ \si{\angstrom} and $2.087(3)$\,\si{\angstrom}. The axial Ni-N bonds are tilted by $\alpha = 49.26(8)^{\circ}$ relative to the $c$-axis. Additionally, the $4_{1}$ screw axis of the space group induces a $90^{\circ}$ anti-clockwise rotation about $c$ in the Ni–N bond direction between neighboring Ni(II) ions connected by pym ligands. This results in a chiral chain structure, in which the orientation of the Ni(II) octahedra exhibits a four-fold chiral modulation along the chain [see Fig.~\ref{fig: Chiral_Struct}(b)].

The crystal structure of the chiral chain provides a basis for inferring the model spin Hamiltonian. The dominant AFM exchange interaction is expected to occur through the pym ligands with an interaction strength $J_{0}$~\cite{XCl2L_Jem_2024, Vaidya_pseudo, Vaidya_staggered}. Since the $J_{0}$ exchange exchange interaction pathways are not located at inversion centers, DM interaction terms are also permitted, with the DM vector having a uniform component along the $c$-axis and a four-fold staggered component in the $ab$-plane~\cite{Liu_chiral_2019}. However, in molecule-based Ni(II) systems, the DM interaction is expected to be small relative to the leading terms in the Hamiltonian~\cite{Huang_2004, XCl2L_Jem_2024, Vaidya_pseudo}. Indeed, in similar pym-containing Ni(II) systems, we have found that the DM interaction is not necessary to account for the ground-state magnetic properties~\cite{Vaidya_pseudo,Vaidya_staggered}. 

Additionally, each Ni(II) ion is coupled to eight neighboring ions in adjacent chains via the four H-bond mediated exchange pathways illustrated in Fig.~\ref{fig: Chiral_Struct}(d) and (e). The $J'_{1 \rm a}$ and $J'_{1 \rm b}$ interactions couple the $i^{\rm th}$ spin in one chain to the $i \pm 1^{\rm th}$ spins in the adjacent chain, with Ni-Ni distances of $7.5985(6)$\,\si{\angstrom} and $7.4852(6)$\,\si{\angstrom}, respectively. In contrast, the $J'_{2 \rm a}$ and $J'_{2 \rm b}$ interactions couple two spins located at the $i^{\rm th}$ position in adjacent chains, with Ni-Ni distances of $7.9395(5)$\,\si{\angstrom}. If either $J'_{1 \rm a}$ or $J'_{1 \rm b}$ share the same sign as $J'_{2 \rm a}$ or $J'_{2 \rm b}$, these interchain interactions will compete with each other, resulting in reduced net interchain coupling. Whatever their sign, these interchain interactions are expected to weak compared to $J_{0}$.

The crystal field of the Ni(II) octahedral environment introduces an SIA term in the Hamiltonian. In the local frame of the Ni(II) octahedra, defined such that the $z$-axis aligns with Ni-N bond, the SIA tensor is $K^{loc}_{i} = \text{diag}[0,0,D]$. Here $D$ represents the uniaxial anisotropy parameter. While the point group symmetry permits rhombohedral anisotropy within the Ni-O plane, variations in the equatorial Ni-O bond lengths are minimal $(<1\%)$, implying that the in-plane anisotropy ($E$) is negligible compared to $D$ and therefore the anisotropy is assumed to be predominantly uniaxial. Euler rotations are applied to transform $K^{loc}_{i}$ to the $xyz$ laboratory frame SIA tensor $K_{i}$ as shown in the Supplementary Material~\cite{supplementary}.

The spin Hamiltonian of the chiral chain can therefore be written as
\begin{equation}\label{eq: chiral_Hamitonian}  
\begin{split}
   \mathcal{H}=&J_{0}\sum_{i}\hat{\mathbf{S}}_{i}\cdot\hat{\mathbf{S}}_{i+1}
     +\sum_{\left \langle i,j \right \rangle _{\perp}}J'_{i,j}\hat{\mathbf{S}}_{i}\cdot\hat{\mathbf{S}}_{j}\\
     &+\sum_{i}\hat{\mathbf{S}}_{i}\cdot K_{i}\cdot\hat{\mathbf{S}}_{i}
     +\sum_{i}g\mu_{\text{B}}\mu_{0}\mathbf{H}\cdot\hat{\mathbf{S}}_{i},
\end{split} 
\end{equation}
where $\hat{\mathbf{S}}_{i}$ is the spin operator at site $i$ and $\left \langle i,j \right \rangle$ denotes a sum over unique interchain exchange bonds with strength $J'_{i,j}$. The final term in Eq.~\ref{eq: chiral_Hamitonian} describes the Zeeman energy in an applied magnetic field $\mathbf{H}$ and assuming an isotropic $g$ factor. For Ni(II) ions in a slightly distorted octahedral environment, a small $g$-anisotropy may arise, with magnitude $\Delta g = g_{z} - g_{xy} = 2D/\lambda$, where $\lambda \sim500$\,K is a typical value of the spin-orbit coupling parameter~\cite{Boca_2004}. In Ni(II) systems with similar local environments, $D \sim 10$\,K~\cite{Manson_2020}, yielding a tiny $g$-anisotropy $\Delta g \sim 0.04$.

\subsection{Magnetometry}

\begin{figure}
    \centering
    \includegraphics[width= \linewidth]{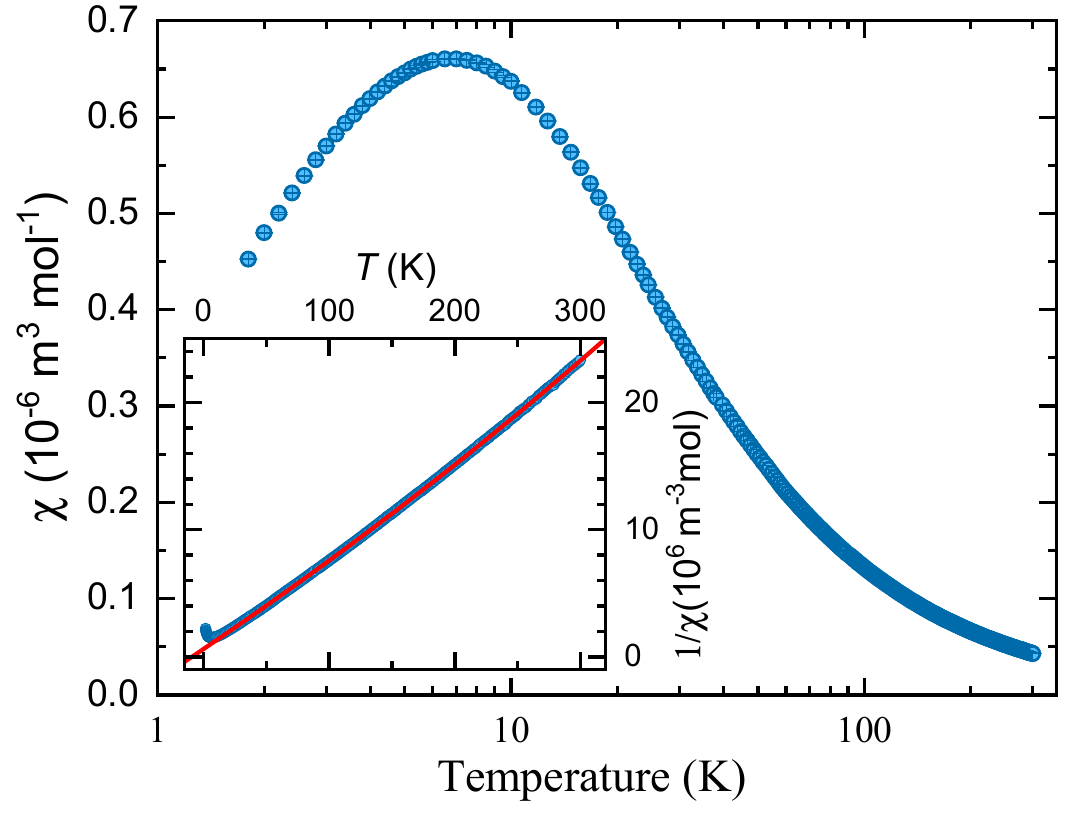}
    \caption{Zero-field-cooled temperature dependent magnetic susceptibility, $\chi(T)$, (blue circles) measured on powder samples of the $S=1$ chiral chain, [Ni(pym)(H$_{2}$O)$_{4}$]SO$_{4}\cdot$D$_{2}$O in an applied field of $0.1$\,T. Inset: The inverse susceptibility $\chi^{-1}$ with a Curie-Weiss fit (red line).}
    \label{fig: Chiral_Chi}
\end{figure}

Figure~\ref{fig: Chiral_Chi} shows the temperature-dependent zero-field-cooled magnetic susceptibility, $\chi(T)$, which was measured in an applied field of $0.1$\,T using a partially deuterated powder sample of [Ni(pym)(D$_{2}$O)$_{4}$]SO$_{4}\cdot$D$_{2}$O, where H in the H$_{2}$O molecules were replaced with D atoms. Deuteration of the system was found to have minimal effects on the magnetic properties, as evidenced by the $\chi(T)$ data shown in the Supplementary Material~\cite{supplementary}. The field-cooled susceptibility data are also shown in the Supplementary Material~\cite{supplementary} are the same within error as the ZFC data.

Above $50$\,K, the ZFC $\chi(T)$ is well described by the Curie-Weiss law, $\chi(T) = C/(T-\theta_{\text{CW}}) + \chi_{0}$, where $\chi_{0}$ is a temperature independent term, the Curie constant $C = N_{\text{A}}\mu_{0}g^{2}\mu_{\text{B}}^{2}S(S+1)/3k_{\text{B}}$ and $\theta_{\text{CW}}$ is the Curie-Weiss temperature. A fit to the inverse susceptibility $\chi^{-1}$ [see inset to  Fig.~\ref{fig: Chiral_Chi}] yields $g = 2.18(1)$, $\chi_{0}=-5.73(4) \times 10^{-9}$\,m$^{3}$mol$^{-1}$ and $\theta_{\text{CW}} = -9.19(6)$\,K. The $g$ factor is typical of Ni(II) spin-1 systems, and the negative $\theta_{\text{CW}}$ is indicative of AFM coupling between spins bridged by pym. Below $50$\,K, a broad maximum in $\chi(T)$ is observed at around $T_{\chi \rm max}\sim6.5$\,K due to the the development of AFM short-range correlations along the chain, a common feature in quasi-low-dimensional systems.
\begin{figure}
    \centering
    \includegraphics[width= \linewidth]{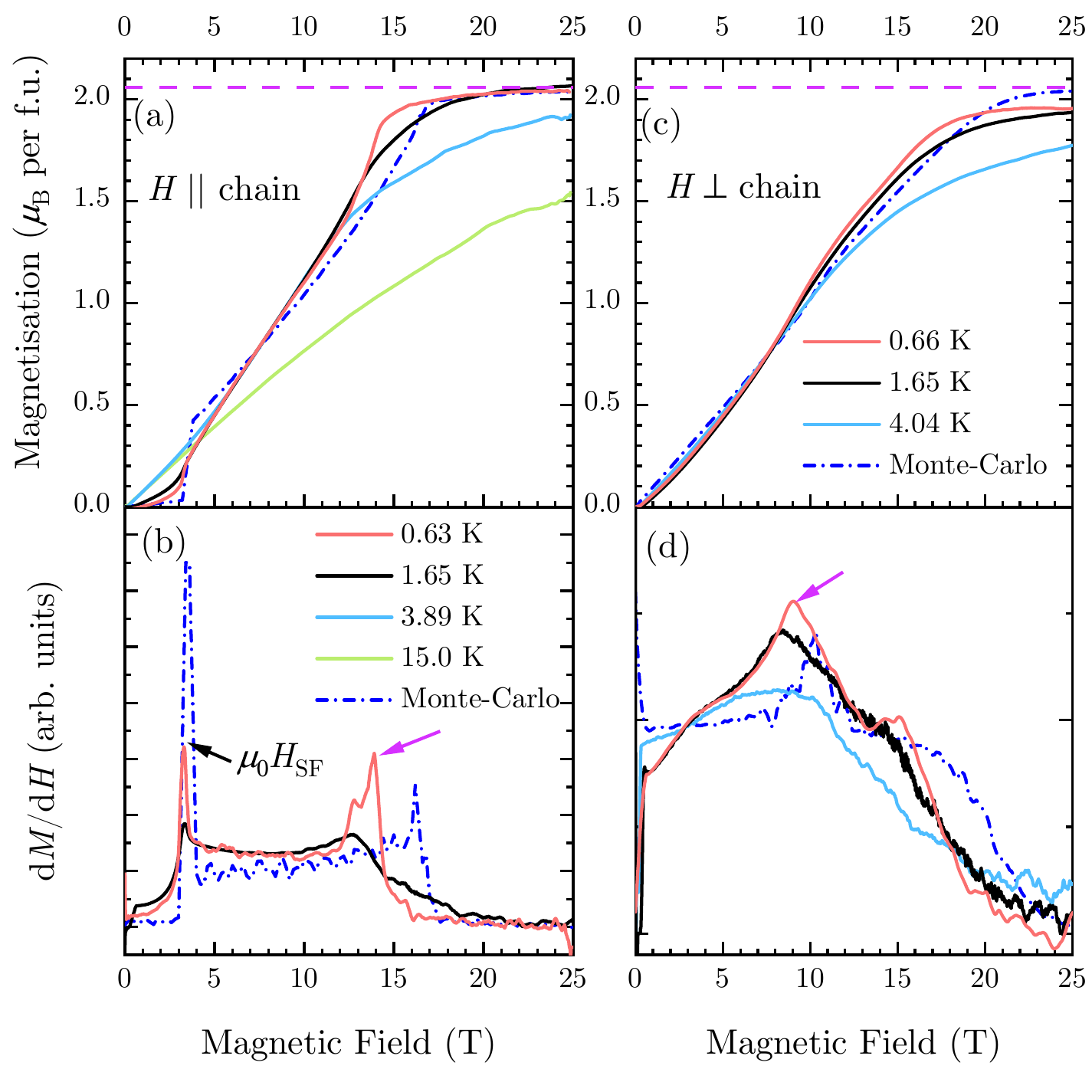}
    \caption{Field-dependent magnetization, $M(H)$, of chiral $S=1$ chain [Ni(pym)(H$_{2}$O)$_{4}$]SO$_{4}\cdot$H$_{2}$O with the applied field (a) parallel and (c) perpendicular to the chain. Differential susceptibility, d$M$/d$H$, with applied field (b) parallel and (d) perpendicular to the chain. The purple dashed line shows the $0.63$\,K projected saturation value with $H||c$. The black arrow marks the spin-flop transition and the purple arrows indicate additional meta-magnetic transitions. The blue dash-dot line in all panels shows the results of the classical Monte Carlo simulation described in Section~\ref{sec: Chiral_MC}.}
    \label{fig: Chiral_MH}
\end{figure}

Pulsed-field magnetization, $M(H)$, measurements were performed on multiple crystallites coaligned along the chain axis to obtain data with the field applied parallel and
perpendicular to the chain, as shown in Fig.~\ref{fig: Chiral_MH}(a) and (c) respectively. For the $H \perp c$ measurments, the crystallites were orientated such that the applied field was close to the crystallographic $[1\,0\,0]$ direction.

chain. Figure~\ref{fig: Chiral_MH}(b) and (d) display the corresponding differential susceptibilities, d$M$/d$H$. When the field is applied parallel to the chain, a sharp peak in d$M$/d$H$ is observed at $\mu_{0}H_{\rm SF} = 3.30(5)$,T and a temperature of $0.63$\,K, indicative of a spin-flop transition. This transition is no longer present in the $M(H)$ data for $T \geq 3.89$\,K. At $13.91(5)$\,T, another peak in d$M$/d$H$ suggests magnetic saturation. However, $M(H)$ continues to increase gradually up to the projected value of $2.06(2)$\,$\mu_{\rm B}$ per Ni(II) ion, as indicated by the purple line in Fig.~\ref{fig: Chiral_MH}(a). When the field is applied perpendicular to the chain, a peak in d$M$/d$H$ is observed at $8.9(5)$\,T, suggesting a metamagnetic transition. As the field increases, the magnetization rises gradually to a projected value of $1.95(2)$\,$\mu_{\rm B}$ per Ni(II) ion, with no evidence of a magnetic saturation transition. These observations and the Monte-Carlo simulations of $M(H)$ are discussed further in Sec.~\ref{sec: Chiral_MC}.

\subsection{Muon-spin rotation}
\begin{figure}
    \centering
    \includegraphics[width= \linewidth]{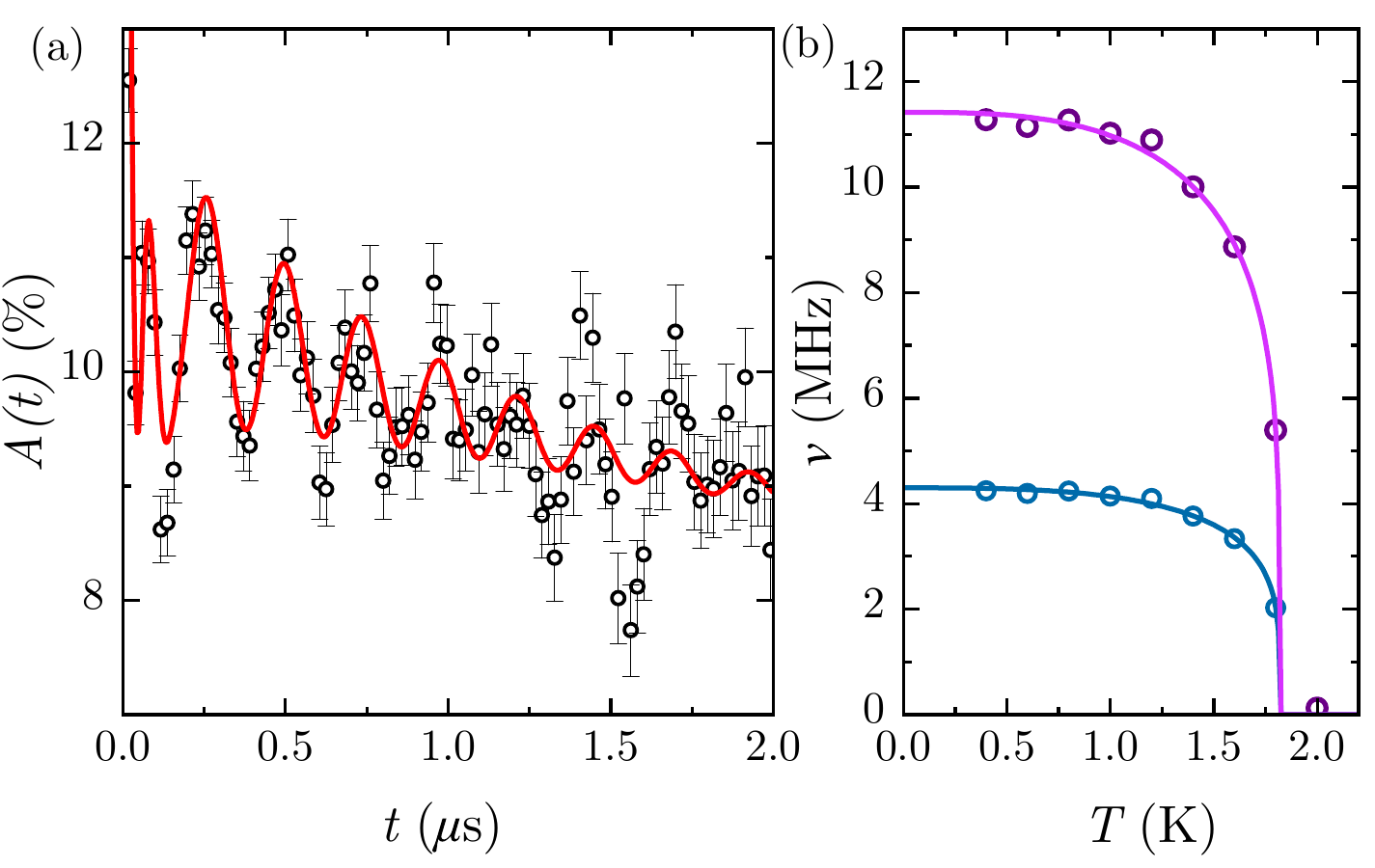}
    \sloppypar
    \caption{(a) Representative $\mu^{+}$SR spectra (black circles) measured at $T=0.2$\,K. (b) Temperature dependence of the muon spin precession frequencies (circles). Solid lines in both correspond to the fits to the data described in the text.}
    \label{fig: Chiral_muon}
\end{figure}

Zero-field (ZF) $\mu^{+}$SR measurements were performed on a powder samples of [Ni(pym)(H$_{2}$O)$_{4}$]SO$_{4}\cdot$H$_{2}$O using the FLAME instrument at the Swiss Muon Source, Paul Scherrer Institut, Switzerland. A representative ZF $\mu^{+}$SR spectrum at $T=0.2$\,K is shown in Fig.~\ref{fig: Chiral_muon}(a). The data show clear oscillations in the time dependent asymmetry, $A(t)$, indicative of coherent muon-spin precession in a quasi-static local field arising from long-range magnetic order. We find that the spectra in the magnetically ordered regime are well described by two relaxing oscillation components arising from two magnetically distinct $\mu^{+}$ stopping sites within the sample. Consequently, $A(t)$ can be fitted using a function of the form,
\begin{equation}
A(t) = \sum^{2}_{i=1} A_{i}\cos(2\pi\nu_{i} t + \phi_{i})e^{-\lambda_{i}t} + A_{3}e^{(-\lambda_{3}t)} + A_{\rm b}.
\label{eq: chiral_muon_fitting}
\end{equation}
Here, $A_{\rm b}$ is a contribution from muons stopping in environments where the muon spin is not relaxed, such as nonmagnetic sites and regions outside the sample. The $A_{i}$ are the magnitudes of each component with corresponding oscillation frequencies $\nu_{i}$, relaxation rates $\lambda_{i}$ and phase factors $\phi_{i}$. The second term, with amplitude $A_{3}$, describes a decaying signal from muons whose polarization is initially parallel to the local magnetic field. 

During fitting, the values of $A_{\rm b} = 8\,\%$ and $\lambda_{3} = 0.55$\,$\mu$s$^{-1}$ were found to be temperature-independent, and therefore fixed to their average values. The parameters $A_{3} = 2.23(7)\,\%$, $A_{1} = 11.1(1)\,\%$, $A_{2} = 2.92(2)\,\%$, $\phi_{1} = -31(2)^{\circ}$, $\phi_{2} = -20(1)^{\circ}$ and the frequency ratio $\nu_{2}/\nu_{1} = 2.66(4)$ were fitted globally across all temperatures. Non-zero values of $\phi_{i}$ are typically associated with incommensurate magnetic order. However, in molecular systems such as this one, non-zero $\phi_{i}$ values are observed even in the presence of commensurate order, and the underlying mechanism remains unclear~\cite{Schueter_2011}. 

The temperature dependence of the $\nu_{i}$ were determined by fitting the data at each temperature, and the results are shown in Fig.~\ref{fig: Chiral_muon}(b). The frequencies $\nu_{i}$ are proportional to the local magnetic field at the $\mu^{+}$ stopping sites and act as an order parameter for the system. The temperature dependence of $\nu_{i}$ are described by a phenomenological function~\cite{Blundell2022_muon}
\begin{equation}
    \nu_{i}(T)=\nu_{i}(0)\left [1 - \left ( \frac{T}{T_{N}} \right )^{\alpha}  \right ]^{\beta}.
\label{eq: freq(T)}
\end{equation}
Fitting the data in Fig.~\ref{fig: Chiral_muon}(b) yields an ordering temperature of $T_{\rm N} = 1.82(2)$\,K and $\beta = 0.22(7)$ for a fixed value of $\alpha = 3$~\footnote{The exponent $\alpha$ describes the low-temperature behavior of the order parameter, while $\beta$ characterizes its behavior near $T_{\rm N}$. For AFMs, $\alpha$ typically lies between 2 and 3. However, in our fitting procedure, $\alpha$ tended to have large non-physical values. We therefore constrained $\alpha$ to a fixed value of 3.}. Due to the sparse data near $T_{\rm N}$, the fitted value of $\beta$ should not be directly compared to theoretical models.

\subsection{Elastic neutron diffraction}
\begin{figure}
    \centering
    \includegraphics[width= \linewidth]{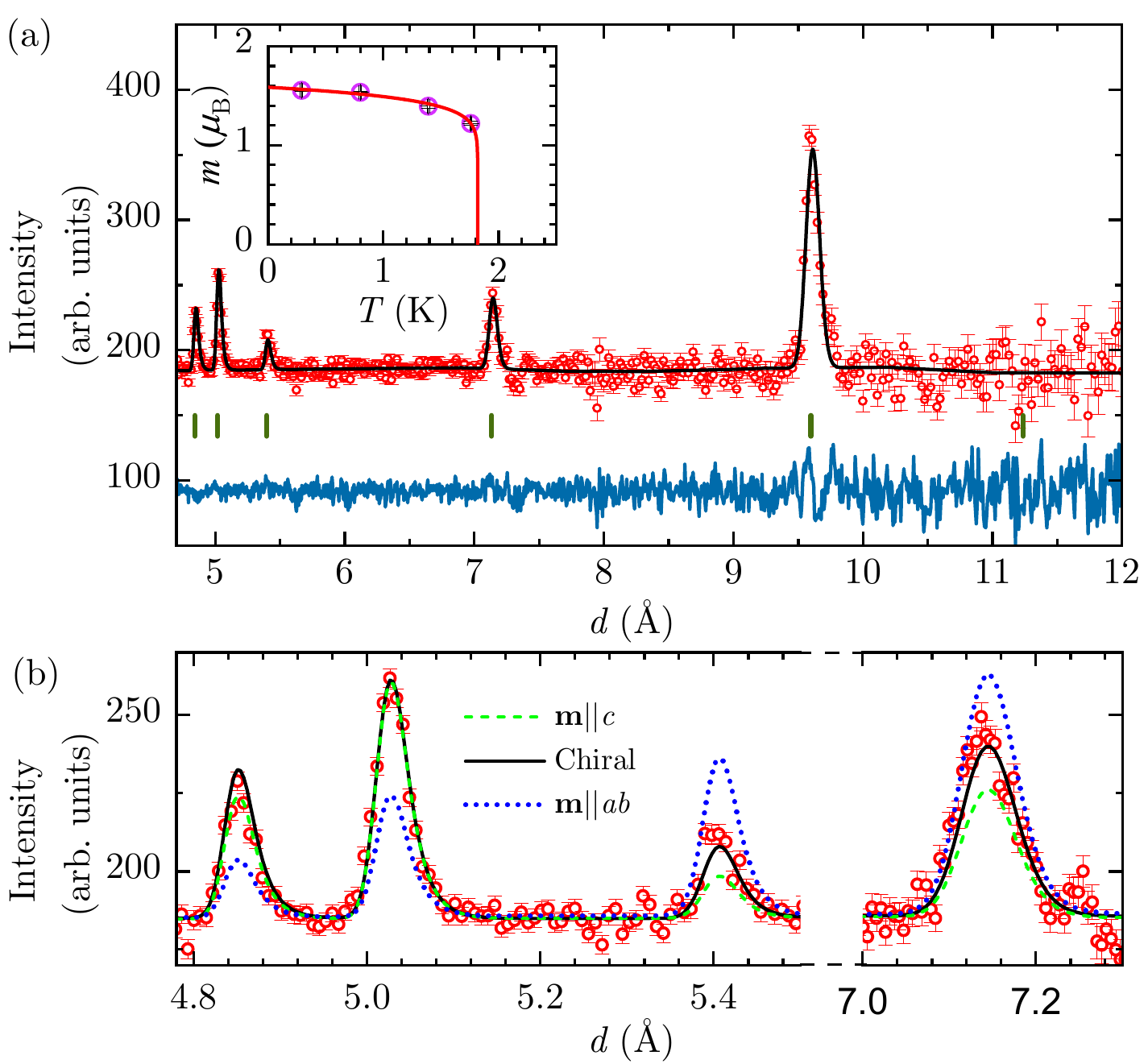}
    \caption{(a) The magnetic powder diffraction pattern of [Ni(pym)(D$_{2}$O)$_{4}$]SO$_{4}\cdot$D$_{2}$O, obtained by subtracting the nuclear structure contribution at $5$\,K from the $0.28$\,K data. The red circles represent the data, the black line shows the Rietveld fit of the chiral magnetic structure, the blue line is the difference and the green ticks mark the magnetic Bragg positions. Inset: $T$ dependence of the ordered Ni(II) magnetic moment. (b) Comparison of the diffraction patterns produced by a collinear $m||c$, chiral and $m||ab$ magnetic structures.}
    \label{fig: Chiral_Wish}
\end{figure}

To determine the magnetic structure of the Ni chiral chain below $T_{\rm N}$, elastic neutron diffraction experiments were carried out on the WISH instrument at ISIS, the UK Neutron and Muon Source~\cite{WISH, WISH_data}. To reduce incoherent background scattering, measurements were carried out on partially deuterated powder samples of [Ni(pym)(D$_{2}$O)$_{4}$]SO$_{4}\cdot$D$_{2}$O. Rietveld refinement of the nuclear structure was performed using the FULLPROF software package~\cite{Fullprof} and reveals that [Ni(pym)(D$_{2}$O)$_{4}$]SO$_{4}\cdot$D$_{2}$O retains the $P4_{1}$ structure down to $T=0.28$\,K, with the lattice parameters $a = 7.93017 (16)$\,$\si{\angstrom}$ and $c=18.45401(66)$\,$\si{\angstrom}$, which are close to the values determined using XRD. Further details on the nuclear structure refinement using powder diffraction data can be found in the Supplementary Material~\cite{supplementary}.

Data were collected above and below $T_{\rm N}$, at $T=5$\,K and $T=0.28$\,K. Due to the low magnetic scattering cross-section of $S=1$ moments relative to the nuclear scattering cross-section, the magnetic structure was determined from the difference between the 0.28\,K and 5~K data, as shown in Fig.~\ref{fig: Chiral_Wish}. The subtracted data reveals several magnetic Bragg reflections which are indexed by the commensurate propagation vector $\textbf{k}=(1/2,1/2,0)$. Symmetry analysis using \mbox{\textsc{isodistort}}~\cite{Isodistort, Isodistort_2006} reveals three candidate symmetries for the magnetic structure: irreducible representation (irreps) $mM_{1}$, $mM_{2}$ and $mM_3\oplus mM_4$. The irrep $mM_{2}$ corresponds to FM order along the chain and can be ruled out from the negative $\theta_{\rm CW}$. 

\begin{figure}
    \centering
    \includegraphics[width= \linewidth]{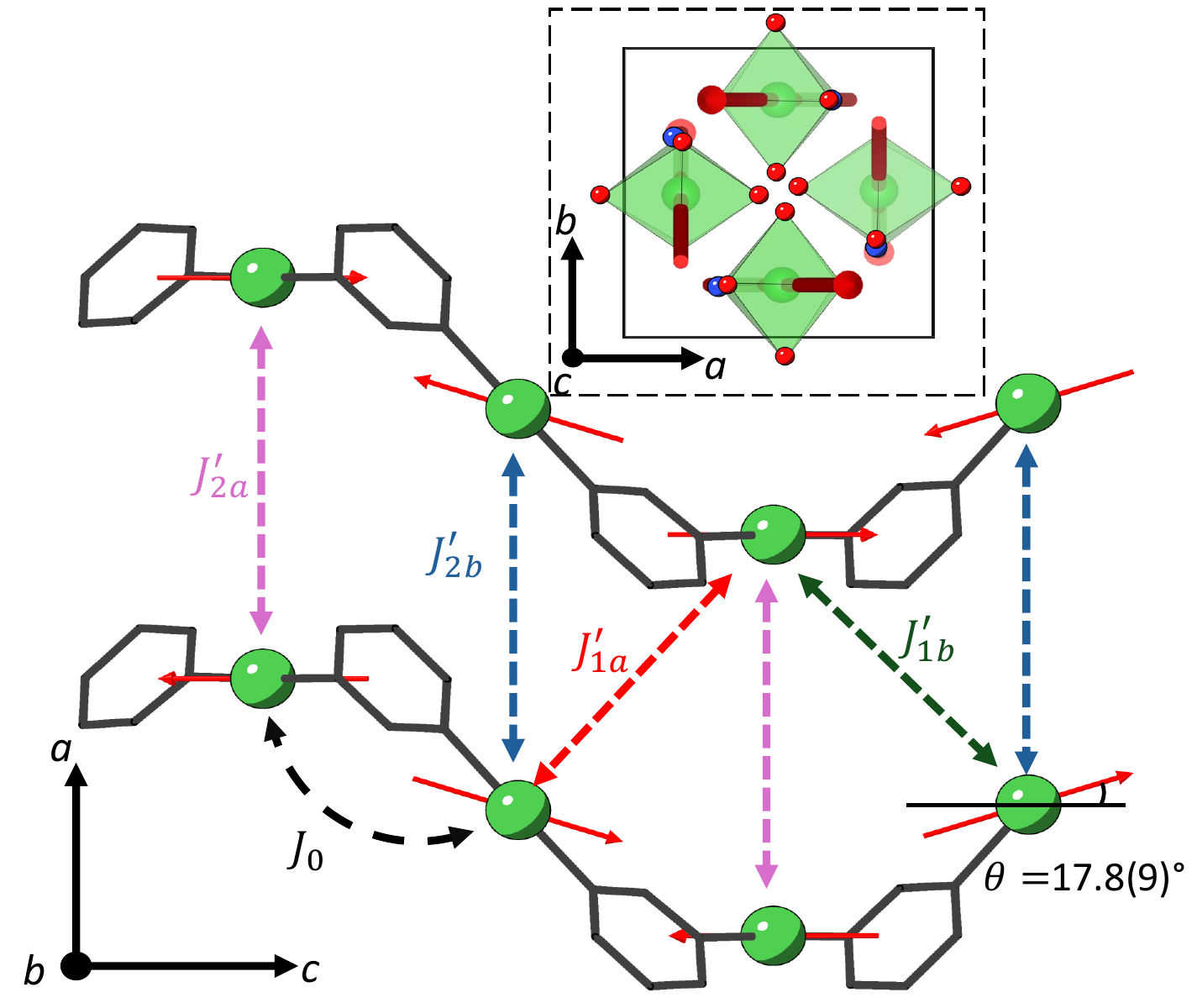}
    \caption{Zero-field magnetic structure of the $S=1$ chiral chain [Ni(pym)(D$_{2}$O)$_{4}$]SO$_{4}\cdot$D$_{2}$O determined using elastic neutron diffraction and viewed along the $b$-axis. The red arrows indicate the Ni(II) moments which are canted by $\theta=17.8(9)^{\circ}$ from the $c$-axis and whose component in the $ab$ plane rotates by $90^{\circ}$ from site to site along the chain yielding a chiral magnetic structure. Dashed lines indicate the magnetic exchange interactions. Inset: View of a single chain down the $c$-axis showing chiral arrangement of the magnetic moments.}
    \label{fig: Chiral_mag_struct}
\end{figure}

The structure associated with the $mM_{1}$ irrep can be described by a linear combination of AFM and chiral modes. The AFM mode couples the $c$-axis components of neighboring moments antiferromagnetically, while the chiral mode induces a $90^{\circ}$ clockwise rotation of the $ab$-plane components along the chain. The mixing of these two modes is parameterised by an angle $\theta$, which measures the canting of the spins away from a collinear arrangement along the $c$ direction. The collinear structure was first tested by fixing $\theta = 0$ and allowing the Ni(II) moment size to freely refine against the data. Subsequently, the chiral structure was evaluated by allowing both $\theta$ and moment size to refine against the data. Although the in-plane moment direction is in principle refinable within the $mM_{1}$ irrep, we found that constraining the moments to align along either the $a$ or $b$ axis yielded a lower magnetic R-factor ($R_{\rm mag}$).

In the $mM_3\oplus mM_4$ irrep, the magnetic moments of nearest-neighbor spins along the chain are independent, whereas the next-nearest-neighbor moments are constrained to linear combinations of an AFM component along the $c$-axis and an FM component in the $ab$-plane. In this irrep, a nearest neighbour AFM configuration is only possible when moments lie in the $ab$-plane. The moment directions within the $ab$-plane of the two independent Ni sites were freely refined while the moment sizes were constrained to be equal. 

The simulated diffraction patterns corresponding to the different ground states are shown in Fig.~\ref{fig: Chiral_Wish}(b). The intensities of the magnetic reflections are best reproduced by the $mM_{1}$ chiral structure with a $R_{\rm mag} = 8.398\%$ compared to $R_{\rm mag} = 11.93\%$ of the collinear ($\theta=0^{\circ}$) structure and $R_{\rm mag} = 36.06\%$ of the $ab$-plane structure within the $mM_3\oplus mM_4$ irrep. The resulting fit yields $\theta=17.8(9)^{\circ}$ and an ordered-moment size of $1.60(1)$\,$\mu_{\rm B}$ at $T = 0.2$\,K. The refined magnetic structure of [Ni(pym)(D$_{2}$O)$_{4}$]SO$_{4}\cdot$D$_{2}$O is depicted in Fig.~\ref{fig: Chiral_mag_struct}. Spins cant toward the local Ni--N axes, indicating an easy-axis SIA, consistent with our previous report on the non-chiral Ni staggered chain, Ni(pym)(H$_{2}$O)$_{2}$(NO$_{3}$)$_{2}$, which hosts a similar Ni(II) octahedral environment~\cite{Vaidya_staggered}. The suppression of the ordered moment size, from the classical expectation value of $gS \approx 1.99$\,$\mu_{\text{B}}$ observed at high-field, indicates the presence of quantum fluctuations typical of a quasi-one-dimensional system.

The temperature dependence of the ordered magnetic moment $m(T)$ is shown in the inset to Fig.~\ref{fig: Chiral_Wish}(a) and is modeled using a power-law function of the form:
\begin{equation}
    \label{eq: mag_moment}
    m(T) = m(0)[1-(T/T_{\rm N})]^{\beta}.
\end{equation}
Fitting the data yields $m(0) = 1.60(3)$\,$\mu_{\rm B}$, $\beta = 0.12(5)$, and $T_{\rm N} = 2.0(2)$\,K, in good agreement with the ordering temperature determined from ZF $\mu^{+}$SR measurements. However, due to the limited number of data points near $T_{\rm N}$, the extracted value of $\beta$ should be interpreted with caution and may not be directly comparable to theoretical critical exponents.

To provide an estimate of the $D/J$ ratio expected to produce a canting angle $\theta=17.8(9)^{\circ}$, we can consider a mean-field model containing only the $D$ and $J_{0}$ terms in Eq.~\ref{eq: chiral_Hamitonian}. As derived in the Supplementary Material~\cite{supplementary}, this model yields the expression
\begin{equation}
    \frac{D}{J_{0}} = \frac{\sin{2\theta}}{2\sin(\theta-\alpha)\cos(\theta-\alpha)}.
    \label{eq: chiral_DJ_ratio}
\end{equation}
Substituting $\alpha = 49.26(8)^{\circ}$ and $\theta=17.8(9)^{\circ}$ into Eq.~\ref{eq: chiral_DJ_ratio} yields an estimated ratio of $|D|/J_{0} \approx 0.65(4)$, which is reasonably similar to the D/J ratio in other Ni(II) systems~\cite{Huang_2004, XCl2L_Jem_2024, Vaidya_pseudo}.

\subsection{Inelastic neutron scattering}
\begin{figure*}
    \centering
    \includegraphics[width= \linewidth]{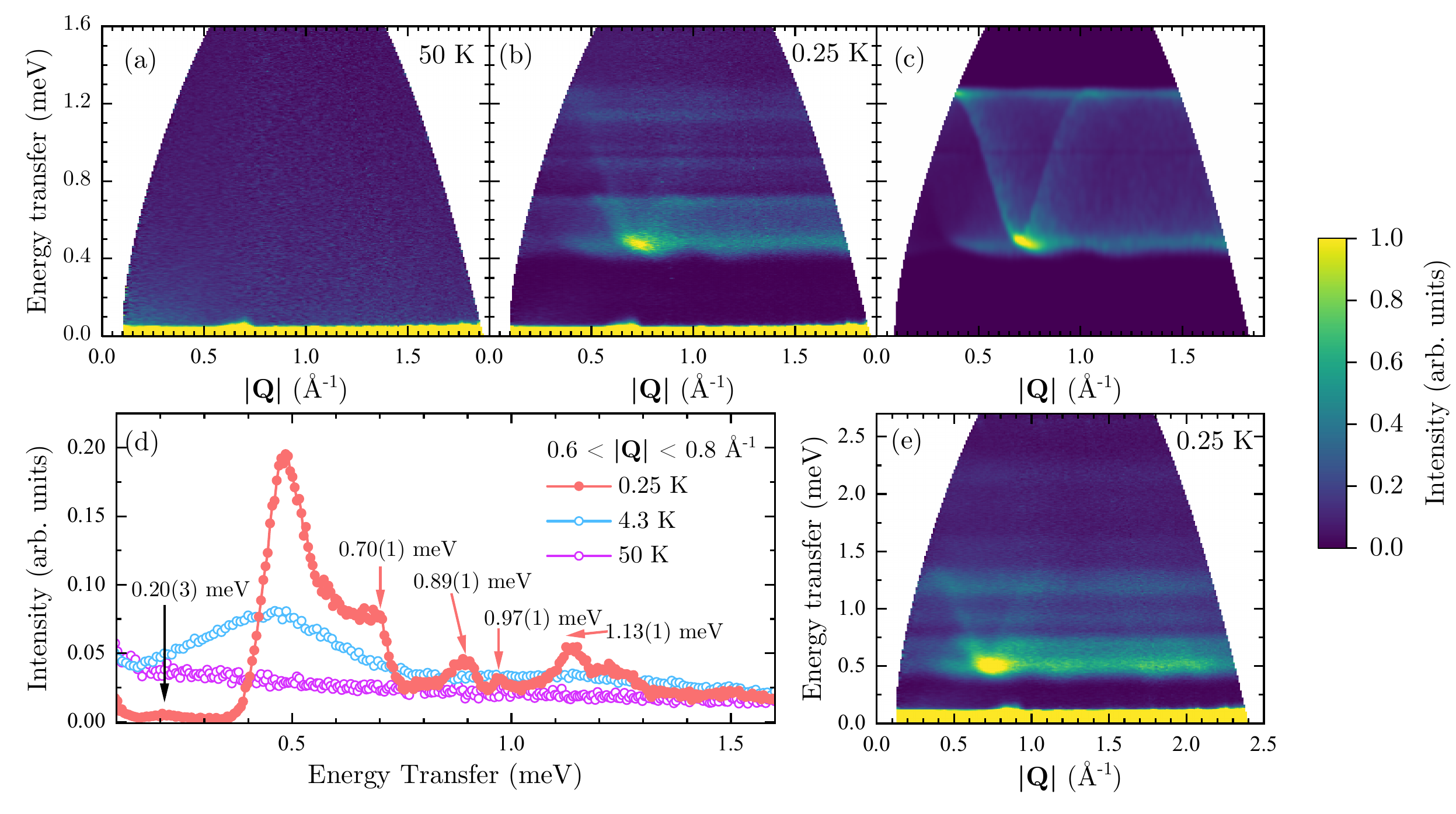}
    \caption{Time of flight powder inelastic neutron scattering (INS) spectra of [Ni(pym-D$_{4}$)(H$_{2}$O)$_{4}$]SO$_{4}\cdot$H$_{2}$O taken at (a) $50$\,K and (b) $0.25$\,K with $\epsilon_{i} = 2.04$\,meV. (c) Powder averaged linear spin wave theory simulations of the spin-wave spectra using the Hamiltonian in Eq.~\ref{eq: chiral_Hamitonian} and the parameters  $J_{0}=6.81(1)$\,K, $J'_{1 \rm a}= -0.091(1)$\,K and $D=-3.02(1)$\,K. (d) Line cut of the spectra observed at various temperatures, integrated over $0.6<|\mathbf{Q}|<0.8$\,$\si{\angstrom}^{-1}$. (e) INS spectra collected at $0.25$\,K with $\epsilon_{i} = 3.36$\,meV. }
    \label{fig: Chiral_let_maps}
\end{figure*}

To quantify the Hamiltonian in Eq.~\ref{eq: chiral_DJ_ratio}, INS experiments were carried out at $T=0.25$\,K, $4.3$\,K and $50$\,K on the LET instrument at ISIS with incident energies $\epsilon_{i} = 2.04$ and $3.36$\,meV~\cite{LET, Let_data}. Fully deuterated powder samples of [Ni(pym-D$_{4}$)(D$_{2}$O)$_{4}$]SO$_{4}\cdot$D$_{2}$O were used, where all H atoms in the pym and H$_{2}$O ligands were replaced with D.

INS spectra collected with $\epsilon_{i} = 2.04$ at $T=50$\,K and $0.25$\,K are shown in Fig.~\ref{fig: Chiral_let_maps}(a) and (b), respectively. Figure~\ref{fig: Chiral_let_maps}(d) shows the energy cuts of spectra, integrated over $0.6<|\textbf{Q}|<0.8$\,$\si{\angstrom}^{-1}$ at various temperatures. The $50$\,K data is featureless with the exception of some quasi-elastic scattering which decays with increasing $\epsilon$, consistent with a paramagnetic state. The $4.31$\,K spectra, shown in the Supplementary Material~\cite{supplementary}, exhibit a broad $|\textbf{Q}|$ and $\epsilon$ dependent scattering due to short-range correlations along the chain. The observation of short-range correlations at this $T$ is consistent with the broad maximum observed in the $\chi(T)$ data.

In the ordered regime, at $T=0.25$\,K, the INS data exhibit gapped, dispersive spin-wave excitations [Fig.~\ref{fig: Chiral_let_maps}(b)], along with several bright dispersionless bands. These appear as prominent peaks in the $|\textbf{Q}|$ integrated cuts at $0.70(1)$\,meV, $0.89(1)$\,meV, $0.97(1)$\,meV and $1.13(1)$\,meV [Fig.~\ref{fig: Chiral_let_maps}(d)]. A suppression of intensity is also evident as a dark band between the $0.89(1)$\,meV, $0.97(1)$\,meV peaks at $0.95(1)$\,meV [Fig.~\ref{fig: Chiral_let_maps}(b)]. Additionally, a low-intensity in-gap peak is observed in the integrated cuts at $0.20(3)$\,meV. This is reminiscent of the in-gap feature observed in the Ref.~\cite{Vaidya_staggered} and may originate from the transition between the SIA split $\ket{m_{s} = 0}$ and $\ket{m_{s} = \pm 1}$ states of impurity Ni(II) ions which are orphaned from the exchange network. This would suggest that $|D|\approx2.3(3)$\,K. In the $0.25$\,K spectra collected with $\epsilon_{i}=3.36$\,meV, low-intensity double magnon scattering extends up to $2.2$\,meV [Fig.~\ref{fig: Chiral_let_maps}(e)].

The main dispersive feature extends from the energy gap of $0.4$\,meV up to the top of the single-magnon band at $1.25$\,meV. LSWT simulations, as implemented in the SpinW~\cite{SpinW} program, and using Eq.~\ref{eq: chiral_Hamitonian} reveal that the $0.4$\,meV energy gap and bandwidth of the dispersion is determined by a unique combination of $J_{0}$ and $D$ values. 

A periodic modulation of the energy gap as a function of $|\textbf{Q}|$ is observed, indicating that low-energy dispersive spin-wave branches exist in this region. LSWT simulations show that this modulation arises from interchain spin-wave dispersions in the plane perpendicular to the chain. As such, the bandwidth of the modulation reflects the overall strength of the net interchain interactions. However, because four distinct interchain exchanges contribute to this gap modulation, it is not possible to uniquely determine the magnitude and sign of each individual exchange. A distinct suppression of intensity at $0.95(1)$\,meV is only reproduced in the simulations when either $J'_{1 \rm a}$ or $J'_{1 \rm b}$ interactions are the dominant interchain interactions. The shorter hydrogen-bond exchange pathway suggests that $|J'_{1 \rm a}|>|J'_{1 \rm b}|$, hence $J'_{1 \rm a}$ was refined against the data while $J'_{1 \rm b}=J'_{2 \rm a}=J'_{2 \rm b}= 0$ were fixed.

\begin{figure}
    \centering
    \includegraphics[width= \linewidth]{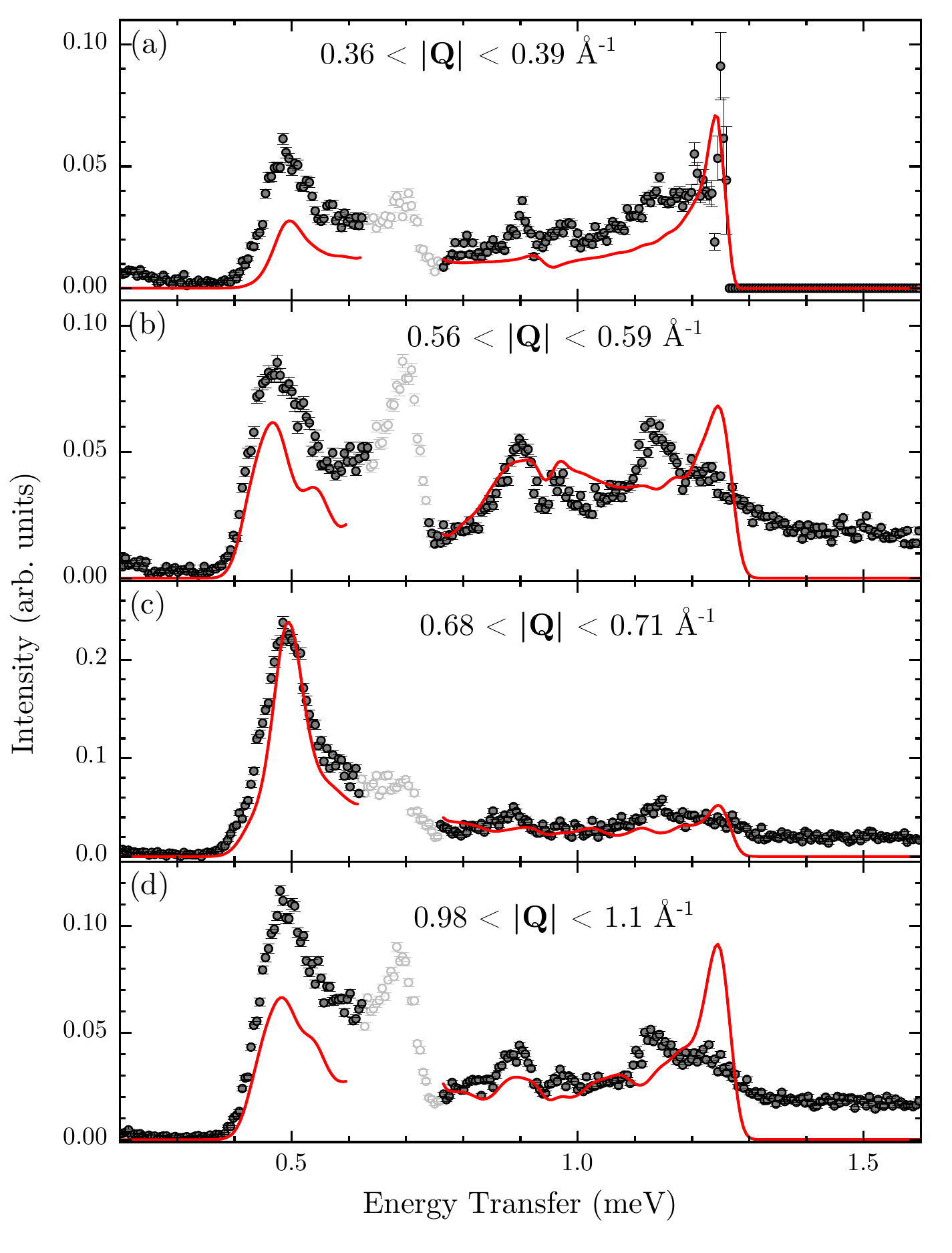}
    \caption{Representative line cuts of the inelastic neutron scattering data for [Ni(pym-D$_{4}$)(D$_{2}$O)$_{4}$]SO$_{4}\cdot$D$_{2}$O. The data are integrated over (a) $0.36\leq |\textbf{Q}|\leq 0.39$\,$\si{\angstrom}^{-1}$, (b) $0.56\leq |\textbf{Q}|\leq 0.59$\,$\si{\angstrom}^{-1}$, (c) $0.68\leq |\textbf{Q}|\leq 0.71$\,$\si{\angstrom}^{-1}$ and (d) $0.98\leq |\textbf{Q}|\leq 1.1$\,$\si{\angstrom}^{-1}$. The red lines are modelled using Eq.~\ref{eq: chiral_Hamitonian} and the parameters $J_{0}=6.81(1)$\,K, $J'_{1 \rm a}= -0.091(1)$\,K and $D=-3.02(1)$\,K, determined by fitting the full 2D dataset. Data points are shown as black circles, and the grey circles indicate the data masked during fitting.}
    \label{fig: Chiral_cuts}
\end{figure}

The dispersionless band at $0.70(1)$\,meV is not reproduced by the simulations, even when higher-order interactions are considered. Therefore, to aid in convergence when fitting $J_{0}$, $D$ and $J'_{1 \rm a}$ parameters, data between $0.6$\,meV and $0.75(1)$\,meV were masked. The values of $J_{0}$, $D$ and $J'_{1 \rm a}$ were optimised by using the Nelder-Mead simplex algorithm to fit the simulations to the $I(|\textbf{Q}|, \epsilon)$ 2D map. The resulting simulations, with the fitted parameters $J_{0}=6.81(3)$\,K ($0.586(9)$\,meV), $J'_{1 \rm a}= -0.09(1)$\,K ($0.0079(8)$\,meV) and $D=-3.02(1)$\,K ($0.2602(9)$\,meV), are shown in Fig.~\ref{fig: Chiral_let_maps}(c) and representative energy cuts with a comparison of the model and the data are shown in Fig.~\ref{fig: Chiral_cuts}. The value of $D$ obtained from the fit is in reasonable agreement with the value implied by the $0.20(8)$\,meV in-gap peak.  The SIA parameter used in SpinW simulations is renormalised by a factor $\left [ 1 - 1/(2S) \right ]$ to account for the non-linear contributions to the SIA which are omitted in LSWT~\cite{Wheeler_2009_renorm, Dahlbom_2023}. The results imply a ratio $D/J_{0}=0.443(5)$ and Eq.~\ref{eq: chiral_DJ_ratio} yields an estimated canting angle of $\theta\approx12.6^{\circ}$. These values are slightly smaller than the $D/J_{0}=0.65(4)$ and $\theta=17.8(9)^{\circ}$ estimated from the elastic neutron diffraction data. This may indicate that the LSWT model over-estimates the value $J_{0}$ or underestimates $D$. Indeed, LSWT is a semiclassical technique which is largely successful in predicting spin dynamics in large-$S$ classical systems, while in small-$S$ and low-dimensional quantum systems, it can fall short of capturing the full excitation spectra~\cite{Songvilay_2021,Manson_Pajerowski_2023,bai_2021,Dahlbom_2023}. The $S=1$ Ni(II) spin, low-dimensionality ($|J'_{1\rm a}/J_{0}|\approx 0.01$) and suppressed ordered moment size present here, does indicate a need for more robust spin-wave calculations taking into account additional quantum corrections. It was also recently argued that some details of $S = 1$ magnetic excitation spectra can only be accounted for by generalizing spin–wave modelling to include SU(3) degrees of freedom~\cite{bai_2021}. When this is done, additional features within the dispersive band, similar to that observed in the spectra presented here, were reproduced by the simulations~\cite{bai_2021}. This is a developing issue that requires further investigation.

Nevertheless, the LSWT model presented here succeeds in two keys aspects. First, it reproduces the main dispersive features, as demonstrated by comparing the simulated dispersion and the data [see Fig.~\ref{fig: Chiral_let_maps}(b)]. Second, it provides a reasonable estimate of the sizes of $J_{0}$, $D$ and the net interchain interactions which stabilize the observed chiral magnetic structure. Hence, while more complex and computationally expensive methods may provide additional insights, LSWT sufficiently captures the details required for this work.

\subsection{Monte-Carlo simulations}\label{sec: Chiral_MC}

To check the Hamiltonian parameters of [Ni(pym)(H$_{2}$O)$_{4}$]SO$_{4}\cdot$H$_{2}$O, determined through INS experiments, MC simulations of the low-temperature $M(H)$ data were performed. For each $H$ value, the spin configuration of a cluster of 16 ions arranged in four chains is determined by minimizing the energy of the system using the Hamiltonian in Eq.~\ref{eq: chiral_Hamitonian}, with periodic boundary conditions. The resulting simulated $M(H)$ and d$M$/d$H$, with the field applied parallel and perpendicular to the chain, are shown as blue dash-dot lines in Fig.~\ref{fig: Chiral_MH}. 

When the applied magnetic field is aligned parallel to the chain, the spin-flop transition is reproduced at $\mu_{0}H_{\rm SF} = 3.5(2)$,T [see Fig.~\ref{fig: Chiral_MH}(c)]. As the field increases, the simulated $M(H)$ successfully captures the characteristic concave rise observed in the experimental data. However, the high-field peak in d$M$/d$H$ is predicted to occur at $16.6(2)$,T, which is higher than the experimentally observed value of $13.91(5)$,T. Beyond this peak, both the data and the simulations show that $M(H)$ continues to increase with the applied field.

In systems with an alternating single-ion anisotropy (SIA) axis, there exist principal directions along which the SIA can lower the system's energy by allowing the spins to cant away from the field direction~\cite{Vaidya_pseudo, Vaidya_staggered}. As a result, true magnetic saturation is only achieved in the infinite field limit. In [Ni(pym)(H$_{2}$O)$_{4}$]SO$_{4}\cdot$H$_{2}$O, for high fields applied along the $c$-axis ($H \parallel c$), the chiral nature of the SIA axis is expected to give rise to a chiral ferromagnetic (FM) phase rather than a fully spin-polarized state. In the chiral FM phase, the $c$-axis components of the spin exhibit FM alignment while the $ab$-plane components display a four-fold chiral modulation along the chain, similar to the chiral AFM zero-field state. As the field strength approaches infinity, the canting angle $\theta$ tends toward zero, resulting in a continued increase in magnetization up to the highest applied fields.

The $H \perp c$ data were best reproduced by the simulation when the applied field is within the $ab$ plane and tilted by roughly $10^{\circ}$ from $[100]$ direction [see Fig.~\ref{fig: Chiral_MH}(b) and (d)], which is within the alignment error. The simulation $M(H)$ exhibits a concave rise in $M(H)$ with increasing field, mirroring the data. Moreover, the broad peak in d$M$/d$H$ at $8.9(5)$\,T is qualitatively reproduced by the simulation, albeit shifted to a higher field of $10.3(5)$\,T. At fields above $20$\,T, the simulation begins to overestimate $M(H)$, suggesting the presence of a small $g$-factor anisotropy. Accurately quantifying this anisotropy would require measurements directly sensitive to the $g$-factor, such as electron spin resonance performed at high excitation frequencies to overcome the effects of zero-field splitting.

While these simulations capture the overall characteristic of the measured $M(H)$ data well, there are systematic discrepancies between the positions of the salient features, such as the apparent saturation features in Fig.~\ref{fig: Chiral_MH}(c) and the decrease in d$M$/d$H$ in Fig.~\ref{fig: Chiral_MH}(d). These discrepancies suggest that, although the model qualitatively captures the magnetic properties of [Ni(pym)(H$_{2}$O)$_{4}$]SO$_{4}\cdot$H$_{2}$O, the $D/J_0$ value might be slightly underestimated by the LSWT model, consistent with earlier discussions.

\section{Summary and conclusions}

The magnetic properties of the chiral $S=1$ chain compound [Ni(pym)(H$_{2}$O)$_{4}$]SO$_{4}\cdot$H$_{2}$O have been characterized using a range of experimental techniques. In this system, Ni(II) ions are linked into chains via pym ligands, with the Ni(II) octahedra exhibiting a four-fold chiral modulation in their orientations along the chain. Adjacent chains are connected through complex hydrogen-bonded networks, which are expected to help mediate interchain exchange interactions. These interactions stabilize the three-dimensional long-range magnetic order below $T_{\rm N}=1.82(2)$\,K, as determined by $\mu^{+}$SR measurements. Powder neutron diffraction reveals a commensurate chiral AFM structure, characterized by AFM alignment of the spin components along the $c$-axis and a $90^\circ$clockwise rotation of the $ab$-plane components between neighboring spins. 

The Hamiltonian parameters, $J_{0} = 6.81(1)$\,K, $J'_{1 \rm a} = -0.091(1)$\,K, and $D = -3.02(1)$\,K were determined through LSWT analysis of INS data. While this model captures the main features of the spin-wave spectra, it underestimates the canting angle ($\theta \approx 12.6^{\circ}$ predicted vs. $\theta = 17.8(9)^{\circ}$ observed) and fails to reproduce dispersionless features observed in the spectra. We attribute these discrepancies to limitations of LSWT in capturing the physics of low-dimensional, anisotropic quantum systems and highlight the need for the development of a spin-wave model more suited to $S=1$ quantum spin chains.

Our results demonstrate that the chiral magnetic order in this material does not rely on typically required DM interactions, geometric frustration, or higher-order exchange mechanisms. Instead, the dominant factor appears to be easy-axis anisotropy, which mirrors the four-fold chiral pattern of the Ni octahedra. Unlike chiral orders induced by other mechanisms, the SIA-driven chiral magnetic order observed here is commensurate and rigidly coupled to the underlying crystal chirality. While this precludes the possibility of stabilising skyrmions using this method, it introduces a new design principle for realizing chiral magnetic materials with potentially robust and tunable properties.~\cite{bousquet_structural_2025}.

Chirality is a key factor in optical phenomena, underpinning effects such as magnetic circular dichroism and magneto-chiral dichroism. These effects are enhanced in chiral magnetic systems at the x-ray absorption edges~\cite{saito_magnetic_2008, sakamoto_observation_2021, Okamoto_2024}, and have been linked to spin-mediated optical activity~\cite{bordacs_chirality_2012}.  Future work could explore these magneto-optical effects experimentally in [Ni(pym)(H$_{2}$O)$_{4}$]SO$_{4}\cdot$H$_{2}$O.

Finally, linear $S=1$ easy-axis chains with comparable values of $|D|/J_{0}$ and $J'_{1\rm a}/J_{0}$ are expected to exhibit Ising AFM order, rather than entering the Haldane phase~\cite{Wierschem_2014}. In the case of [Ni(pym)(H$_{2}$O)$_{4}$]SO$_{4} \cdot$H$_{2}$O, the chirality of the crystal structure imposes a chiral magnetic order through the SIA. Notably, in coordination polymer systems, the $D/J_{0}$ ratio can be tuned through external parameters such as applied pressure or chemical modification via ligand substitution~\cite{Schlueter_2012, Cortijo_2013, Cortijo_2014, Liu_2016_correlations, Kubus_2018, Blackmore-correlations, Pajerowski_2022, Coak_2023, Geers_preesure_2023, Povarov_2024, Vaidya_pseudo}. This tunability suggests that the SIA-induced chiral magnetic order in [Ni(pym)(H$_{2}$O)$_{4}$]SO$_{4} \cdot$H$_{2}$O, and potentially in related $S=1$ chain systems, could be manipulated to explore the interplay between the Haldane phase and chirality.
\\
\begin{acknowledgments}
    We thank the late Jamie Manson for instigating this work, contribution to the design and growth of the samples and for his numerous invaluable contributions. We thank T. Orton and P. Ruddy for their technical assistance. We also thank R. Coldea, D. M. Pajerowsk and C. Stock for valuable discussions. SV thanks the UK Engineering and Physical Sciences Research Council (EPSRC) for supporting his studentship. This project has received funding from the European Research Council (ERC) under the European Union’s Horizon 2020 research and innovation programme (Grant Agreement No. 681260) and EPSRC (Grant No. EP/N024028/1). A portion of this work was performed at the National High Magnetic Field Laboratory, which is supported by National Science Foundation Cooperative Agreements Nos. DMR-1644779 and DMR-2128556, the US Department of Energy (DoE) and the State of Florida. JS acknowledges support from the DoE BES FWP “Science of 100 T". We grateful for the NSF grant CHE-1827313. Part of this work was carried out at the Swiss Muon Source, Paul Scherrer Institut and we are grateful for the provision of beamtime. AHM is grateful to STFC-ISIS and EPSRC for PhD studentship funding. The work at EWU was supported by the National Science Foundation through grant no. DMR-2104167. Data presented in this paper will be made available at~\cite{Data}. The data collected on the WISH and LET instruments can be found in Ref.~\cite{WISH_data} and Ref.~\cite{Let_data} repectively. For the purpose of open access, the author has applied a Creative Commons Attribution (CC-BY) licence to any Author Accepted Manuscript version arising from this submission.
\end{acknowledgments}

\bibliography{main.bib}

\begin{thebibliography}{63}%
\makeatletter
\providecommand \@ifxundefined [1]{%
 \@ifx{#1\undefined}
}%
\providecommand \@ifnum [1]{%
 \ifnum #1\expandafter \@firstoftwo
 \else \expandafter \@secondoftwo
 \fi
}%
\providecommand \@ifx [1]{%
 \ifx #1\expandafter \@firstoftwo
 \else \expandafter \@secondoftwo
 \fi
}%
\providecommand \natexlab [1]{#1}%
\providecommand \enquote  [1]{``#1''}%
\providecommand \bibnamefont  [1]{#1}%
\providecommand \bibfnamefont [1]{#1}%
\providecommand \citenamefont [1]{#1}%
\providecommand \href@noop [0]{\@secondoftwo}%
\providecommand \href [0]{\begingroup \@sanitize@url \@href}%
\providecommand \@href[1]{\@@startlink{#1}\@@href}%
\providecommand \@@href[1]{\endgroup#1\@@endlink}%
\providecommand \@sanitize@url [0]{\catcode `\\12\catcode `\$12\catcode
  `\&12\catcode `\#12\catcode `\^12\catcode `\_12\catcode `\%12\relax}%
\providecommand \@@startlink[1]{}%
\providecommand \@@endlink[0]{}%
\providecommand \url  [0]{\begingroup\@sanitize@url \@url }%
\providecommand \@url [1]{\endgroup\@href {#1}{\urlprefix }}%
\providecommand \urlprefix  [0]{URL }%
\providecommand \Eprint [0]{\href }%
\providecommand \doibase [0]{https://doi.org/}%
\providecommand \selectlanguage [0]{\@gobble}%
\providecommand \bibinfo  [0]{\@secondoftwo}%
\providecommand \bibfield  [0]{\@secondoftwo}%
\providecommand \translation [1]{[#1]}%
\providecommand \BibitemOpen [0]{}%
\providecommand \bibitemStop [0]{}%
\providecommand \bibitemNoStop [0]{.\EOS\space}%
\providecommand \EOS [0]{\spacefactor3000\relax}%
\providecommand \BibitemShut  [1]{\csname bibitem#1\endcsname}%
\let\auto@bib@innerbib\@empty
\bibitem [{\citenamefont {Bousquet}\ \emph {et~al.}(2025)\citenamefont
  {Bousquet}, \citenamefont {Fava}, \citenamefont {Romestan}, \citenamefont
  {Gómez-Ortiz}, \citenamefont {McCabe},\ and\ \citenamefont
  {Romero}}]{bousquet_structural_2025}%
  \BibitemOpen
  \bibfield  {author} {\bibinfo {author} {\bibfnamefont {E.}~\bibnamefont
  {Bousquet}}, \bibinfo {author} {\bibfnamefont {M.}~\bibnamefont {Fava}},
  \bibinfo {author} {\bibfnamefont {Z.}~\bibnamefont {Romestan}}, \bibinfo
  {author} {\bibfnamefont {F.}~\bibnamefont {Gómez-Ortiz}}, \bibinfo {author}
  {\bibfnamefont {E.~E.}\ \bibnamefont {McCabe}},\ and\ \bibinfo {author}
  {\bibfnamefont {A.~H.}\ \bibnamefont {Romero}},\ }\bibfield  {title}
  {\bibinfo {title} {Structural chirality and related properties in periodic
  inorganic solids: review and perspectives},\ }\href
  {https://doi.org/10.1088/1361-648X/adb674} {\bibfield  {journal} {\bibinfo
  {journal} {Journal of Physics: Condensed Matter}\ }\textbf {\bibinfo {volume}
  {37}},\ \bibinfo {pages} {163004} (\bibinfo {year} {2025})},\ \bibinfo {note}
  {publisher: IOP Publishing}\BibitemShut {NoStop}%
\bibitem [{\citenamefont {Yan}\ and\ \citenamefont
  {Felser}(2017)}]{Binghai_2017}%
  \BibitemOpen
  \bibfield  {author} {\bibinfo {author} {\bibfnamefont {B.}~\bibnamefont
  {Yan}}\ and\ \bibinfo {author} {\bibfnamefont {C.}~\bibnamefont {Felser}},\
  }\bibfield  {title} {\bibinfo {title} {{Topological Materials: {W}eyl
  Semimetals}},\ }\href
  {https://doi.org/https://doi.org/10.1146/annurev-conmatphys-031016-025458}
  {\bibfield  {journal} {\bibinfo  {journal} {Annual Review of Condensed Matter
  Physics}\ }\textbf {\bibinfo {volume} {8}},\ \bibinfo {pages} {337} (\bibinfo
  {year} {2017})}\BibitemShut {NoStop}%
\bibitem [{\citenamefont {Saito}\ \emph {et~al.}(2008)\citenamefont {Saito},
  \citenamefont {Ishikawa}, \citenamefont {Taniguchi},\ and\ \citenamefont
  {Arima}}]{saito_magnetic_2008}%
  \BibitemOpen
  \bibfield  {author} {\bibinfo {author} {\bibfnamefont {M.}~\bibnamefont
  {Saito}}, \bibinfo {author} {\bibfnamefont {K.}~\bibnamefont {Ishikawa}},
  \bibinfo {author} {\bibfnamefont {K.}~\bibnamefont {Taniguchi}},\ and\
  \bibinfo {author} {\bibfnamefont {T.}~\bibnamefont {Arima}},\ }\bibfield
  {title} {\bibinfo {title} {Magnetic control of crystal chirality and the
  existence of a large magneto-optical dichroism effect in {CuB}$_2${O}$_4$},\
  }\href {https://doi.org/10.1103/PhysRevLett.101.117402} {\bibfield  {journal}
  {\bibinfo  {journal} {Physical Review Letters}\ }\textbf {\bibinfo {volume}
  {101}},\ \bibinfo {pages} {117402} (\bibinfo {year} {2008})}\BibitemShut
  {NoStop}%
\bibitem [{\citenamefont {Sessoli}\ \emph {et~al.}(2015)\citenamefont
  {Sessoli}, \citenamefont {Boulon}, \citenamefont {Caneschi}, \citenamefont
  {Mannini}, \citenamefont {Poggini}, \citenamefont {Wilhelm},\ and\
  \citenamefont {Rogalev}}]{sessoli_strong_2015}%
  \BibitemOpen
  \bibfield  {author} {\bibinfo {author} {\bibfnamefont {R.}~\bibnamefont
  {Sessoli}}, \bibinfo {author} {\bibfnamefont {M.-E.}\ \bibnamefont {Boulon}},
  \bibinfo {author} {\bibfnamefont {A.}~\bibnamefont {Caneschi}}, \bibinfo
  {author} {\bibfnamefont {M.}~\bibnamefont {Mannini}}, \bibinfo {author}
  {\bibfnamefont {L.}~\bibnamefont {Poggini}}, \bibinfo {author} {\bibfnamefont
  {F.}~\bibnamefont {Wilhelm}},\ and\ \bibinfo {author} {\bibfnamefont
  {A.}~\bibnamefont {Rogalev}},\ }\bibfield  {title} {\bibinfo {title} {Strong
  magneto-chiral dichroism in a paramagnetic molecular helix observed by hard
  {X}-rays},\ }\href {https://doi.org/10.1038/nphys3152} {\bibfield  {journal}
  {\bibinfo  {journal} {Nature Physics}\ }\textbf {\bibinfo {volume} {11}},\
  \bibinfo {pages} {69} (\bibinfo {year} {2015})}\BibitemShut {NoStop}%
\bibitem [{\citenamefont {Bordács}\ \emph {et~al.}(2012)\citenamefont
  {Bordács}, \citenamefont {Kézsmárki}, \citenamefont {Szaller},
  \citenamefont {Demkó}, \citenamefont {Kida}, \citenamefont {Murakawa},
  \citenamefont {Onose}, \citenamefont {Shimano}, \citenamefont {Rõõm},
  \citenamefont {Nagel}, \citenamefont {Miyahara}, \citenamefont {Furukawa},\
  and\ \citenamefont {Tokura}}]{bordacs_chirality_2012}%
  \BibitemOpen
  \bibfield  {author} {\bibinfo {author} {\bibfnamefont {S.}~\bibnamefont
  {Bordács}}, \bibinfo {author} {\bibfnamefont {I.}~\bibnamefont
  {Kézsmárki}}, \bibinfo {author} {\bibfnamefont {D.}~\bibnamefont
  {Szaller}}, \bibinfo {author} {\bibfnamefont {L.}~\bibnamefont {Demkó}},
  \bibinfo {author} {\bibfnamefont {N.}~\bibnamefont {Kida}}, \bibinfo {author}
  {\bibfnamefont {H.}~\bibnamefont {Murakawa}}, \bibinfo {author}
  {\bibfnamefont {Y.}~\bibnamefont {Onose}}, \bibinfo {author} {\bibfnamefont
  {R.}~\bibnamefont {Shimano}}, \bibinfo {author} {\bibfnamefont
  {T.}~\bibnamefont {Rõõm}}, \bibinfo {author} {\bibfnamefont
  {U.}~\bibnamefont {Nagel}}, \bibinfo {author} {\bibfnamefont
  {S.}~\bibnamefont {Miyahara}}, \bibinfo {author} {\bibfnamefont
  {N.}~\bibnamefont {Furukawa}},\ and\ \bibinfo {author} {\bibfnamefont
  {Y.}~\bibnamefont {Tokura}},\ }\bibfield  {title} {\bibinfo {title}
  {Chirality of matter shows up via spin excitations},\ }\href
  {https://www.nature.com/articles/nphys2387} {\bibfield  {journal} {\bibinfo
  {journal} {Nature Physics}\ }\textbf {\bibinfo {volume} {8}},\ \bibinfo
  {pages} {734} (\bibinfo {year} {2012})}\BibitemShut {NoStop}%
\bibitem [{\citenamefont {Nakagawa}\ \emph {et~al.}(2017)\citenamefont
  {Nakagawa}, \citenamefont {Abe}, \citenamefont {Toyoda}, \citenamefont
  {Kimura}, \citenamefont {Zaccaro}, \citenamefont {Gautier-Luneau},
  \citenamefont {Luneau}, \citenamefont {Kousaka}, \citenamefont {Sera},
  \citenamefont {Sera}, \citenamefont {Inoue}, \citenamefont {Akimitsu},
  \citenamefont {Tokunaga},\ and\ \citenamefont
  {Arima}}]{nakagawa_magneto-chiral_2017}%
  \BibitemOpen
  \bibfield  {author} {\bibinfo {author} {\bibfnamefont {N.}~\bibnamefont
  {Nakagawa}}, \bibinfo {author} {\bibfnamefont {N.}~\bibnamefont {Abe}},
  \bibinfo {author} {\bibfnamefont {S.}~\bibnamefont {Toyoda}}, \bibinfo
  {author} {\bibfnamefont {S.}~\bibnamefont {Kimura}}, \bibinfo {author}
  {\bibfnamefont {J.}~\bibnamefont {Zaccaro}}, \bibinfo {author} {\bibfnamefont
  {I.}~\bibnamefont {Gautier-Luneau}}, \bibinfo {author} {\bibfnamefont
  {D.}~\bibnamefont {Luneau}}, \bibinfo {author} {\bibfnamefont
  {Y.}~\bibnamefont {Kousaka}}, \bibinfo {author} {\bibfnamefont
  {A.}~\bibnamefont {Sera}}, \bibinfo {author} {\bibfnamefont {M.}~\bibnamefont
  {Sera}}, \bibinfo {author} {\bibfnamefont {K.}~\bibnamefont {Inoue}},
  \bibinfo {author} {\bibfnamefont {J.}~\bibnamefont {Akimitsu}}, \bibinfo
  {author} {\bibfnamefont {Y.}~\bibnamefont {Tokunaga}},\ and\ \bibinfo
  {author} {\bibfnamefont {T.}~\bibnamefont {Arima}},\ }\bibfield  {title}
  {\bibinfo {title} {Magneto-chiral dichroism of {CsCuCl} 3},\ }\href
  {https://doi.org/10.1103/PhysRevB.96.121102} {\bibfield  {journal} {\bibinfo
  {journal} {Physical Review B}\ }\textbf {\bibinfo {volume} {96}},\ \bibinfo
  {pages} {121102} (\bibinfo {year} {2017})}\BibitemShut {NoStop}%
\bibitem [{\citenamefont {Mühlbauer}\ \emph {et~al.}(2009)\citenamefont
  {Mühlbauer}, \citenamefont {Binz}, \citenamefont {Jonietz}, \citenamefont
  {Pfleiderer}, \citenamefont {Rosch}, \citenamefont {Neubauer}, \citenamefont
  {Georgii},\ and\ \citenamefont {Böni}}]{Muhlbauer_2009}%
  \BibitemOpen
  \bibfield  {author} {\bibinfo {author} {\bibfnamefont {S.}~\bibnamefont
  {Mühlbauer}}, \bibinfo {author} {\bibfnamefont {B.}~\bibnamefont {Binz}},
  \bibinfo {author} {\bibfnamefont {F.}~\bibnamefont {Jonietz}}, \bibinfo
  {author} {\bibfnamefont {C.}~\bibnamefont {Pfleiderer}}, \bibinfo {author}
  {\bibfnamefont {A.}~\bibnamefont {Rosch}}, \bibinfo {author} {\bibfnamefont
  {A.}~\bibnamefont {Neubauer}}, \bibinfo {author} {\bibfnamefont
  {R.}~\bibnamefont {Georgii}},\ and\ \bibinfo {author} {\bibfnamefont
  {P.}~\bibnamefont {Böni}},\ }\bibfield  {title} {\bibinfo {title} {Skyrmion
  lattice in a chiral magnet},\ }\href
  {https://doi.org/10.1126/science.1166767} {\bibfield  {journal} {\bibinfo
  {journal} {Science}\ }\textbf {\bibinfo {volume} {323}},\ \bibinfo {pages}
  {915} (\bibinfo {year} {2009})},\ \Eprint
  {https://arxiv.org/abs/https://www.science.org/doi/pdf/10.1126/science.1166767}
  {https://www.science.org/doi/pdf/10.1126/science.1166767} \BibitemShut
  {NoStop}%
\bibitem [{\citenamefont {Fert}\ \emph {et~al.}(2017)\citenamefont {Fert},
  \citenamefont {Reyren},\ and\ \citenamefont {Cros}}]{fert_2017}%
  \BibitemOpen
  \bibfield  {author} {\bibinfo {author} {\bibfnamefont {A.}~\bibnamefont
  {Fert}}, \bibinfo {author} {\bibfnamefont {N.}~\bibnamefont {Reyren}},\ and\
  \bibinfo {author} {\bibfnamefont {V.}~\bibnamefont {Cros}},\ }\bibfield
  {title} {\bibinfo {title} {Magnetic skyrmions: advances in physics and
  potential applications},\ }\href
  {https://www.nature.com/articles/natrevmats201731} {\bibfield  {journal}
  {\bibinfo  {journal} {Nature Reviews Materials}\ }\textbf {\bibinfo {volume}
  {2}},\ \bibinfo {pages} {1} (\bibinfo {year} {2017})}\BibitemShut {NoStop}%
\bibitem [{\citenamefont {Lancaster}(2019)}]{Lancaster_2019}%
  \BibitemOpen
  \bibfield  {author} {\bibinfo {author} {\bibfnamefont {T.}~\bibnamefont
  {Lancaster}},\ }\bibfield  {title} {\bibinfo {title} {Skyrmions in magnetic
  materials},\ }\href {https://doi.org/10.1080/00107514.2019.1699352}
  {\bibfield  {journal} {\bibinfo  {journal} {Contemporary Physics}\ }\textbf
  {\bibinfo {volume} {60}},\ \bibinfo {pages} {246} (\bibinfo {year}
  {2019})}\BibitemShut {NoStop}%
\bibitem [{\citenamefont {Mayoh}\ \emph {et~al.}(2022)\citenamefont {Mayoh},
  \citenamefont {Bouaziz}, \citenamefont {Hall}, \citenamefont {Staunton},
  \citenamefont {Lees},\ and\ \citenamefont {Balakrishnan}}]{Mayoh_2022_THE}%
  \BibitemOpen
  \bibfield  {author} {\bibinfo {author} {\bibfnamefont {D.~A.}\ \bibnamefont
  {Mayoh}}, \bibinfo {author} {\bibfnamefont {J.}~\bibnamefont {Bouaziz}},
  \bibinfo {author} {\bibfnamefont {A.~E.}\ \bibnamefont {Hall}}, \bibinfo
  {author} {\bibfnamefont {J.~B.}\ \bibnamefont {Staunton}}, \bibinfo {author}
  {\bibfnamefont {M.~R.}\ \bibnamefont {Lees}},\ and\ \bibinfo {author}
  {\bibfnamefont {G.}~\bibnamefont {Balakrishnan}},\ }\bibfield  {title}
  {\bibinfo {title} {Giant topological and planar hall effect in
  {Cr}$_{1/3}${NbS}$_{2}$},\ }\href
  {https://doi.org/10.1103/PhysRevResearch.4.013134} {\bibfield  {journal}
  {\bibinfo  {journal} {Physical Review Research}\ }\textbf {\bibinfo {volume}
  {4}},\ \bibinfo {pages} {013134} (\bibinfo {year} {2022})}\BibitemShut
  {NoStop}%
\bibitem [{\citenamefont {Rikken}(2011)}]{Geert_2011}%
  \BibitemOpen
  \bibfield  {author} {\bibinfo {author} {\bibfnamefont {G.~L. J.~A.}\
  \bibnamefont {Rikken}},\ }\bibfield  {title} {\bibinfo {title} {A new twist
  on spintronics},\ }\href {https://doi.org/10.1126/science.1201663} {\bibfield
   {journal} {\bibinfo  {journal} {Science}\ }\textbf {\bibinfo {volume}
  {331}},\ \bibinfo {pages} {864} (\bibinfo {year} {2011})}\BibitemShut
  {NoStop}%
\bibitem [{\citenamefont {Emori}\ \emph {et~al.}(2013)\citenamefont {Emori},
  \citenamefont {Bauer}, \citenamefont {Ahn}, \citenamefont {Martinez},\ and\
  \citenamefont {Beach}}]{Emori_2013}%
  \BibitemOpen
  \bibfield  {author} {\bibinfo {author} {\bibfnamefont {S.}~\bibnamefont
  {Emori}}, \bibinfo {author} {\bibfnamefont {U.}~\bibnamefont {Bauer}},
  \bibinfo {author} {\bibfnamefont {S.-M.}\ \bibnamefont {Ahn}}, \bibinfo
  {author} {\bibfnamefont {E.}~\bibnamefont {Martinez}},\ and\ \bibinfo
  {author} {\bibfnamefont {G.~S.}\ \bibnamefont {Beach}},\ }\bibfield  {title}
  {\bibinfo {title} {Current-driven dynamics of chiral ferromagnetic domain
  walls},\ }\href {https://www.nature.com/articles/nmat3675} {\bibfield
  {journal} {\bibinfo  {journal} {Nature Materials}\ }\textbf {\bibinfo
  {volume} {12}},\ \bibinfo {pages} {611} (\bibinfo {year} {2013})}\BibitemShut
  {NoStop}%
\bibitem [{\citenamefont {Wang}\ \emph
  {et~al.}(2022{\natexlab{a}})\citenamefont {Wang}, \citenamefont
  {Bheemarasetty}, \citenamefont {Duan}, \citenamefont {Zhou},\ and\
  \citenamefont {Xiao}}]{WANG_2022_skyrmions_rev}%
  \BibitemOpen
  \bibfield  {author} {\bibinfo {author} {\bibfnamefont {K.}~\bibnamefont
  {Wang}}, \bibinfo {author} {\bibfnamefont {V.}~\bibnamefont {Bheemarasetty}},
  \bibinfo {author} {\bibfnamefont {J.}~\bibnamefont {Duan}}, \bibinfo {author}
  {\bibfnamefont {S.}~\bibnamefont {Zhou}},\ and\ \bibinfo {author}
  {\bibfnamefont {G.}~\bibnamefont {Xiao}},\ }\bibfield  {title} {\bibinfo
  {title} {Fundamental physics and applications of skyrmions: a review},\
  }\href {https://doi.org/https://doi.org/10.1016/j.jmmm.2022.169905}
  {\bibfield  {journal} {\bibinfo  {journal} {Journal of Magnetism and Magnetic
  Materials}\ }\textbf {\bibinfo {volume} {563}},\ \bibinfo {pages} {169905}
  (\bibinfo {year} {2022}{\natexlab{a}})}\BibitemShut {NoStop}%
\bibitem [{\citenamefont {Wang}\ \emph
  {et~al.}(2022{\natexlab{b}})\citenamefont {Wang}, \citenamefont {Dai},
  \citenamefont {Chow},\ and\ \citenamefont {Chen}}]{WANG_2022_THE_rev}%
  \BibitemOpen
  \bibfield  {author} {\bibinfo {author} {\bibfnamefont {H.}~\bibnamefont
  {Wang}}, \bibinfo {author} {\bibfnamefont {Y.}~\bibnamefont {Dai}}, \bibinfo
  {author} {\bibfnamefont {G.-M.}\ \bibnamefont {Chow}},\ and\ \bibinfo
  {author} {\bibfnamefont {J.}~\bibnamefont {Chen}},\ }\bibfield  {title}
  {\bibinfo {title} {Topological hall transport: Materials, mechanisms and
  potential applications},\ }\href
  {https://doi.org/https://doi.org/10.1016/j.pmatsci.2022.100971} {\bibfield
  {journal} {\bibinfo  {journal} {Progress in Materials Science}\ }\textbf
  {\bibinfo {volume} {130}},\ \bibinfo {pages} {100971} (\bibinfo {year}
  {2022}{\natexlab{b}})}\BibitemShut {NoStop}%
\bibitem [{\citenamefont {Cheong}\ and\ \citenamefont
  {Xu}(2022)}]{cheong_magnetic_2022}%
  \BibitemOpen
  \bibfield  {author} {\bibinfo {author} {\bibfnamefont {S.-W.}\ \bibnamefont
  {Cheong}}\ and\ \bibinfo {author} {\bibfnamefont {X.}~\bibnamefont {Xu}},\
  }\bibfield  {title} {\bibinfo {title} {Magnetic chirality},\ }\href
  {https://doi.org/10.1038/s41535-022-00447-5} {\bibfield  {journal} {\bibinfo
  {journal} {npj Quantum Materials}\ }\textbf {\bibinfo {volume} {7}},\
  \bibinfo {pages} {40} (\bibinfo {year} {2022})}\BibitemShut {NoStop}%
\bibitem [{\citenamefont {Marty}\ \emph {et~al.}(2008)\citenamefont {Marty},
  \citenamefont {Simonet}, \citenamefont {Ressouche}, \citenamefont {Ballou},
  \citenamefont {Lejay},\ and\ \citenamefont {Bordet}}]{Marty_2008}%
  \BibitemOpen
  \bibfield  {author} {\bibinfo {author} {\bibfnamefont {K.}~\bibnamefont
  {Marty}}, \bibinfo {author} {\bibfnamefont {V.}~\bibnamefont {Simonet}},
  \bibinfo {author} {\bibfnamefont {E.}~\bibnamefont {Ressouche}}, \bibinfo
  {author} {\bibfnamefont {R.}~\bibnamefont {Ballou}}, \bibinfo {author}
  {\bibfnamefont {P.}~\bibnamefont {Lejay}},\ and\ \bibinfo {author}
  {\bibfnamefont {P.}~\bibnamefont {Bordet}},\ }\bibfield  {title} {\bibinfo
  {title} {Single domain magnetic helicity and triangular chirality in
  structurally enantiopure {Ba}$_{3}${NbFe}$_{3}${Si}$_{2}${O}$_{14}$},\ }\href
  {https://doi.org/10.1103/PhysRevLett.101.247201} {\bibfield  {journal}
  {\bibinfo  {journal} {Physical Review Letters}\ }\textbf {\bibinfo {volume}
  {101}},\ \bibinfo {pages} {247201} (\bibinfo {year} {2008})}\BibitemShut
  {NoStop}%
\bibitem [{\citenamefont {Johnson}\ \emph {et~al.}(2013)\citenamefont
  {Johnson}, \citenamefont {Cao}, \citenamefont {Chapon}, \citenamefont
  {Fabrizi}, \citenamefont {Perks}, \citenamefont {Manuel}, \citenamefont
  {Yang}, \citenamefont {Oh}, \citenamefont {Cheong},\ and\ \citenamefont
  {Radaelli}}]{Rog_2013}%
  \BibitemOpen
  \bibfield  {author} {\bibinfo {author} {\bibfnamefont {R.~D.}\ \bibnamefont
  {Johnson}}, \bibinfo {author} {\bibfnamefont {K.}~\bibnamefont {Cao}},
  \bibinfo {author} {\bibfnamefont {L.~C.}\ \bibnamefont {Chapon}}, \bibinfo
  {author} {\bibfnamefont {F.}~\bibnamefont {Fabrizi}}, \bibinfo {author}
  {\bibfnamefont {N.}~\bibnamefont {Perks}}, \bibinfo {author} {\bibfnamefont
  {P.}~\bibnamefont {Manuel}}, \bibinfo {author} {\bibfnamefont {J.~J.}\
  \bibnamefont {Yang}}, \bibinfo {author} {\bibfnamefont {Y.~S.}\ \bibnamefont
  {Oh}}, \bibinfo {author} {\bibfnamefont {S.-W.}\ \bibnamefont {Cheong}},\
  and\ \bibinfo {author} {\bibfnamefont {P.~G.}\ \bibnamefont {Radaelli}},\
  }\bibfield  {title} {\bibinfo {title} {{MnSb}$_{2}${O}$_{6}$: a polar magnet
  with a chiral crystal structure},\ }\href
  {https://doi.org/10.1103/PhysRevLett.111.017202} {\bibfield  {journal}
  {\bibinfo  {journal} {Physical Review Letters}\ }\textbf {\bibinfo {volume}
  {111}},\ \bibinfo {pages} {017202} (\bibinfo {year} {2013})}\BibitemShut
  {NoStop}%
\bibitem [{\citenamefont {Khanh}\ \emph {et~al.}(2020)\citenamefont {Khanh},
  \citenamefont {Nakajima}, \citenamefont {Yu}, \citenamefont {Gao},
  \citenamefont {Shibata}, \citenamefont {Hirschberger}, \citenamefont
  {Yamasaki}, \citenamefont {Sagayama}, \citenamefont {Nakao}, \citenamefont
  {Peng}, \citenamefont {Nakajima}, \citenamefont {Takagi}, \citenamefont
  {Arima}, \citenamefont {Tokura},\ and\ \citenamefont
  {Seki}}]{khanh_2020_centro}%
  \BibitemOpen
  \bibfield  {author} {\bibinfo {author} {\bibfnamefont {N.~D.}\ \bibnamefont
  {Khanh}}, \bibinfo {author} {\bibfnamefont {T.}~\bibnamefont {Nakajima}},
  \bibinfo {author} {\bibfnamefont {X.}~\bibnamefont {Yu}}, \bibinfo {author}
  {\bibfnamefont {S.}~\bibnamefont {Gao}}, \bibinfo {author} {\bibfnamefont
  {K.}~\bibnamefont {Shibata}}, \bibinfo {author} {\bibfnamefont
  {M.}~\bibnamefont {Hirschberger}}, \bibinfo {author} {\bibfnamefont
  {Y.}~\bibnamefont {Yamasaki}}, \bibinfo {author} {\bibfnamefont
  {H.}~\bibnamefont {Sagayama}}, \bibinfo {author} {\bibfnamefont
  {H.}~\bibnamefont {Nakao}}, \bibinfo {author} {\bibfnamefont
  {L.}~\bibnamefont {Peng}}, \bibinfo {author} {\bibfnamefont {K.}~\bibnamefont
  {Nakajima}}, \bibinfo {author} {\bibfnamefont {R.}~\bibnamefont {Takagi}},
  \bibinfo {author} {\bibfnamefont {T.-h.}\ \bibnamefont {Arima}}, \bibinfo
  {author} {\bibfnamefont {Y.}~\bibnamefont {Tokura}},\ and\ \bibinfo {author}
  {\bibfnamefont {S.}~\bibnamefont {Seki}},\ }\bibfield  {title} {\bibinfo
  {title} {Nanometric square skyrmion lattice in a centrosymmetric tetragonal
  magnet},\ }\href {https://www.nature.com/articles/s41565-020-0684-7}
  {\bibfield  {journal} {\bibinfo  {journal} {Nature Nanotechnology}\ }\textbf
  {\bibinfo {volume} {15}},\ \bibinfo {pages} {444} (\bibinfo {year}
  {2020})}\BibitemShut {NoStop}%
\bibitem [{\citenamefont {Huang}\ and\ \citenamefont
  {Affleck}(2004)}]{Huang_2004}%
  \BibitemOpen
  \bibfield  {author} {\bibinfo {author} {\bibfnamefont {H.}~\bibnamefont
  {Huang}}\ and\ \bibinfo {author} {\bibfnamefont {I.}~\bibnamefont
  {Affleck}},\ }\bibfield  {title} {\bibinfo {title} {Susceptibility and
  dzyaloshinskii-moriya interaction in the haldane-gap compound
  {Ni}({C}$_{2}${H}$_{8}${N}$_{2}$)$_{2}${NO}$_{2}$({ClO}$_{4}$)},\ }\href
  {https://doi.org/10.1103/PhysRevB.69.184414} {\bibfield  {journal} {\bibinfo
  {journal} {Phys. Rev. B}\ }\textbf {\bibinfo {volume} {69}},\ \bibinfo
  {pages} {184414} (\bibinfo {year} {2004})}\BibitemShut {NoStop}%
\bibitem [{\citenamefont {Vaidya}\ \emph {et~al.}(2025)\citenamefont {Vaidya},
  \citenamefont {Curley}, \citenamefont {Manuel}, \citenamefont {Stewart},
  \citenamefont {Le}, \citenamefont {Balz}, \citenamefont {Shiroka},
  \citenamefont {Blundell}, \citenamefont {Wheeler}, \citenamefont
  {Calderon-Lin}, \citenamefont {Manson}, \citenamefont {Manson}, \citenamefont
  {Singleton}, \citenamefont {Lancaster}, \citenamefont {Johnson},\ and\
  \citenamefont {Goddard}}]{Vaidya_staggered}%
  \BibitemOpen
  \bibfield  {author} {\bibinfo {author} {\bibfnamefont {S.}~\bibnamefont
  {Vaidya}}, \bibinfo {author} {\bibfnamefont {S.~P.~M.}\ \bibnamefont
  {Curley}}, \bibinfo {author} {\bibfnamefont {P.}~\bibnamefont {Manuel}},
  \bibinfo {author} {\bibfnamefont {J.~R.}\ \bibnamefont {Stewart}}, \bibinfo
  {author} {\bibfnamefont {M.~D.}\ \bibnamefont {Le}}, \bibinfo {author}
  {\bibfnamefont {C.}~\bibnamefont {Balz}}, \bibinfo {author} {\bibfnamefont
  {T.}~\bibnamefont {Shiroka}}, \bibinfo {author} {\bibfnamefont {S.~J.}\
  \bibnamefont {Blundell}}, \bibinfo {author} {\bibfnamefont {K.~A.}\
  \bibnamefont {Wheeler}}, \bibinfo {author} {\bibfnamefont {I.}~\bibnamefont
  {Calderon-Lin}}, \bibinfo {author} {\bibfnamefont {Z.~E.}\ \bibnamefont
  {Manson}}, \bibinfo {author} {\bibfnamefont {J.~L.}\ \bibnamefont {Manson}},
  \bibinfo {author} {\bibfnamefont {J.}~\bibnamefont {Singleton}}, \bibinfo
  {author} {\bibfnamefont {T.}~\bibnamefont {Lancaster}}, \bibinfo {author}
  {\bibfnamefont {R.~D.}\ \bibnamefont {Johnson}},\ and\ \bibinfo {author}
  {\bibfnamefont {P.~A.}\ \bibnamefont {Goddard}},\ }\bibfield  {title}
  {\bibinfo {title} {Magnetic properties of a staggered ${S}=1$ chain with an
  alternating single-ion anisotropy direction},\ }\href
  {https://doi.org/10.1103/PhysRevB.111.014421} {\bibfield  {journal} {\bibinfo
   {journal} {Phys. Rev. B}\ }\textbf {\bibinfo {volume} {111}},\ \bibinfo
  {pages} {014421} (\bibinfo {year} {2025})}\BibitemShut {NoStop}%
\bibitem [{\citenamefont {Feyerherm}\ \emph {et~al.}(2004)\citenamefont
  {Feyerherm}, \citenamefont {Loose}, \citenamefont {Ishida}, \citenamefont
  {Nogami}, \citenamefont {Kreitlow}, \citenamefont {Baabe}, \citenamefont
  {Litterst}, \citenamefont {S{\"u}llow}, \citenamefont {Klauss},\ and\
  \citenamefont {Doll}}]{Feyerherm_FeCl2_2004}%
  \BibitemOpen
  \bibfield  {author} {\bibinfo {author} {\bibfnamefont {R.}~\bibnamefont
  {Feyerherm}}, \bibinfo {author} {\bibfnamefont {A.}~\bibnamefont {Loose}},
  \bibinfo {author} {\bibfnamefont {T.}~\bibnamefont {Ishida}}, \bibinfo
  {author} {\bibfnamefont {T.}~\bibnamefont {Nogami}}, \bibinfo {author}
  {\bibfnamefont {J.}~\bibnamefont {Kreitlow}}, \bibinfo {author}
  {\bibfnamefont {D.}~\bibnamefont {Baabe}}, \bibinfo {author} {\bibfnamefont
  {F.~J.}\ \bibnamefont {Litterst}}, \bibinfo {author} {\bibfnamefont
  {S.}~\bibnamefont {S{\"u}llow}}, \bibinfo {author} {\bibfnamefont {H.-H.}\
  \bibnamefont {Klauss}},\ and\ \bibinfo {author} {\bibfnamefont
  {K.}~\bibnamefont {Doll}},\ }\bibfield  {title} {\bibinfo {title} {{Weak
  ferromagnetism with very large canting in a chiral lattice:
  $\mathrm{Fe}(\mathrm{pyrimidine})_{2}\mathrm{Cl}_{2}$}},\ }\href
  {https://doi.org/10.1103/PhysRevB.69.134427} {\bibfield  {journal} {\bibinfo
  {journal} {Phys. Rev. B}\ }\textbf {\bibinfo {volume} {69}},\ \bibinfo
  {pages} {134427} (\bibinfo {year} {2004})}\BibitemShut {NoStop}%
\bibitem [{\citenamefont {Pitcairn}\ \emph {et~al.}(2024)\citenamefont
  {Pitcairn}, \citenamefont {Ongkiko}, \citenamefont {Iliceto}, \citenamefont
  {Speakman}, \citenamefont {Calder}, \citenamefont {Cochran}, \citenamefont
  {Paddison}, \citenamefont {Liu}, \citenamefont {Argent}, \citenamefont
  {Morris},\ and\ \citenamefont {Cliffe}}]{XCl2L_Jem_2024}%
  \BibitemOpen
  \bibfield  {author} {\bibinfo {author} {\bibfnamefont {J.}~\bibnamefont
  {Pitcairn}}, \bibinfo {author} {\bibfnamefont {M.~A.}\ \bibnamefont
  {Ongkiko}}, \bibinfo {author} {\bibfnamefont {A.}~\bibnamefont {Iliceto}},
  \bibinfo {author} {\bibfnamefont {P.~J.}\ \bibnamefont {Speakman}}, \bibinfo
  {author} {\bibfnamefont {S.}~\bibnamefont {Calder}}, \bibinfo {author}
  {\bibfnamefont {M.~J.}\ \bibnamefont {Cochran}}, \bibinfo {author}
  {\bibfnamefont {J.~A.~M.}\ \bibnamefont {Paddison}}, \bibinfo {author}
  {\bibfnamefont {C.}~\bibnamefont {Liu}}, \bibinfo {author} {\bibfnamefont
  {S.~P.}\ \bibnamefont {Argent}}, \bibinfo {author} {\bibfnamefont {A.~J.}\
  \bibnamefont {Morris}},\ and\ \bibinfo {author} {\bibfnamefont {M.~J.}\
  \bibnamefont {Cliffe}},\ }\bibfield  {title} {\bibinfo {title} {Controlling
  noncollinear ferromagnetism in van der {W}aals metal–organic magnets},\
  }\href {https://doi.org/10.1021/jacs.4c04102} {\bibfield  {journal} {\bibinfo
   {journal} {Journal of the American Chemical Society}\ }\textbf {\bibinfo
  {volume} {146}},\ \bibinfo {pages} {19146} (\bibinfo {year}
  {2024})}\BibitemShut {NoStop}%
\bibitem [{\citenamefont {Povarov}\ \emph {et~al.}(2020)\citenamefont
  {Povarov}, \citenamefont {Facheris}, \citenamefont {Velja}, \citenamefont
  {Blosser}, \citenamefont {Yan}, \citenamefont {Gvasaliya},\ and\
  \citenamefont {Zheludev}}]{Povarov_2020}%
  \BibitemOpen
  \bibfield  {author} {\bibinfo {author} {\bibfnamefont {K.~Y.}\ \bibnamefont
  {Povarov}}, \bibinfo {author} {\bibfnamefont {L.}~\bibnamefont {Facheris}},
  \bibinfo {author} {\bibfnamefont {S.}~\bibnamefont {Velja}}, \bibinfo
  {author} {\bibfnamefont {D.}~\bibnamefont {Blosser}}, \bibinfo {author}
  {\bibfnamefont {Z.}~\bibnamefont {Yan}}, \bibinfo {author} {\bibfnamefont
  {S.}~\bibnamefont {Gvasaliya}},\ and\ \bibinfo {author} {\bibfnamefont
  {A.}~\bibnamefont {Zheludev}},\ }\bibfield  {title} {\bibinfo {title}
  {Magnetization plateaux cascade in the frustrated quantum antiferromagnet
  {Cs}$_{2}${CoBr}$_{4}$},\ }\href
  {https://doi.org/10.1103/PhysRevResearch.2.043384} {\bibfield  {journal}
  {\bibinfo  {journal} {Phys. Rev. Res.}\ }\textbf {\bibinfo {volume} {2}},\
  \bibinfo {pages} {043384} (\bibinfo {year} {2020})}\BibitemShut {NoStop}%
\bibitem [{\citenamefont {Facheris}\ \emph {et~al.}(2024)\citenamefont
  {Facheris}, \citenamefont {Nabi}, \citenamefont {Povarov}, \citenamefont
  {Yan}, \citenamefont {Moshe}, \citenamefont {Nagel}, \citenamefont {R\~o\
  om}, \citenamefont {Podlesnyak}, \citenamefont {Ressouche}, \citenamefont
  {Beauvois}, \citenamefont {Stewart}, \citenamefont {Manuel}, \citenamefont
  {Khalyavin}, \citenamefont {Orlandi},\ and\ \citenamefont
  {Zheludev}}]{Facheris_2024}%
  \BibitemOpen
  \bibfield  {author} {\bibinfo {author} {\bibfnamefont {L.}~\bibnamefont
  {Facheris}}, \bibinfo {author} {\bibfnamefont {S.~D.}\ \bibnamefont {Nabi}},
  \bibinfo {author} {\bibfnamefont {K.~Y.}\ \bibnamefont {Povarov}}, \bibinfo
  {author} {\bibfnamefont {Z.}~\bibnamefont {Yan}}, \bibinfo {author}
  {\bibfnamefont {A.~G.}\ \bibnamefont {Moshe}}, \bibinfo {author}
  {\bibfnamefont {U.}~\bibnamefont {Nagel}}, \bibinfo {author} {\bibfnamefont
  {T.}~\bibnamefont {R\~o\ om}}, \bibinfo {author} {\bibfnamefont
  {A.}~\bibnamefont {Podlesnyak}}, \bibinfo {author} {\bibfnamefont
  {E.}~\bibnamefont {Ressouche}}, \bibinfo {author} {\bibfnamefont
  {K.}~\bibnamefont {Beauvois}}, \bibinfo {author} {\bibfnamefont {J.~R.}\
  \bibnamefont {Stewart}}, \bibinfo {author} {\bibfnamefont {P.}~\bibnamefont
  {Manuel}}, \bibinfo {author} {\bibfnamefont {D.}~\bibnamefont {Khalyavin}},
  \bibinfo {author} {\bibfnamefont {F.}~\bibnamefont {Orlandi}},\ and\ \bibinfo
  {author} {\bibfnamefont {A.}~\bibnamefont {Zheludev}},\ }\bibfield  {title}
  {\bibinfo {title} {Magnetic field induced phases and spin hamiltonian in
  {Cs$_{2}$CoBr$_{4}$}},\ }\href {https://doi.org/10.1103/PhysRevB.109.104433}
  {\bibfield  {journal} {\bibinfo  {journal} {Phys. Rev. B}\ }\textbf {\bibinfo
  {volume} {109}},\ \bibinfo {pages} {104433} (\bibinfo {year}
  {2024})}\BibitemShut {NoStop}%
\bibitem [{\citenamefont {Vaidya}\ \emph {et~al.}(2024)\citenamefont {Vaidya},
  \citenamefont {Hern\'andez-Meli\'an}, \citenamefont {Tidey}, \citenamefont
  {Curley}, \citenamefont {Sharma}, \citenamefont {Manuel}, \citenamefont
  {Wang}, \citenamefont {Hannaford}, \citenamefont {Blundell}, \citenamefont
  {Manson}, \citenamefont {Manson}, \citenamefont {Singleton}, \citenamefont
  {Lancaster}, \citenamefont {Johnson},\ and\ \citenamefont
  {Goddard}}]{Vaidya_pseudo}%
  \BibitemOpen
  \bibfield  {author} {\bibinfo {author} {\bibfnamefont {S.}~\bibnamefont
  {Vaidya}}, \bibinfo {author} {\bibfnamefont {A.}~\bibnamefont
  {Hern\'andez-Meli\'an}}, \bibinfo {author} {\bibfnamefont {J.~P.}\
  \bibnamefont {Tidey}}, \bibinfo {author} {\bibfnamefont {S.~P.~M.}\
  \bibnamefont {Curley}}, \bibinfo {author} {\bibfnamefont {S.}~\bibnamefont
  {Sharma}}, \bibinfo {author} {\bibfnamefont {P.}~\bibnamefont {Manuel}},
  \bibinfo {author} {\bibfnamefont {C.}~\bibnamefont {Wang}}, \bibinfo {author}
  {\bibfnamefont {G.~L.}\ \bibnamefont {Hannaford}}, \bibinfo {author}
  {\bibfnamefont {S.~J.}\ \bibnamefont {Blundell}}, \bibinfo {author}
  {\bibfnamefont {Z.~E.}\ \bibnamefont {Manson}}, \bibinfo {author}
  {\bibfnamefont {J.~L.}\ \bibnamefont {Manson}}, \bibinfo {author}
  {\bibfnamefont {J.}~\bibnamefont {Singleton}}, \bibinfo {author}
  {\bibfnamefont {T.}~\bibnamefont {Lancaster}}, \bibinfo {author}
  {\bibfnamefont {R.~D.}\ \bibnamefont {Johnson}},\ and\ \bibinfo {author}
  {\bibfnamefont {P.~A.}\ \bibnamefont {Goddard}},\ }\bibfield  {title}
  {\bibinfo {title} {Pseudo-easy-axis anisotropy in antiferromagnetic ${S}=1$
  diamond-lattice systems},\ }\href
  {https://doi.org/10.1103/PhysRevB.110.174438} {\bibfield  {journal} {\bibinfo
   {journal} {Phys. Rev. B}\ }\textbf {\bibinfo {volume} {110}},\ \bibinfo
  {pages} {174438} (\bibinfo {year} {2024})}\BibitemShut {NoStop}%
\bibitem [{\citenamefont {Dender}\ \emph {et~al.}(1997)\citenamefont {Dender},
  \citenamefont {Hammar}, \citenamefont {Reich}, \citenamefont {Broholm},\ and\
  \citenamefont {Aeppli}}]{Dender_Cu_benzo_1997}%
  \BibitemOpen
  \bibfield  {author} {\bibinfo {author} {\bibfnamefont {D.~C.}\ \bibnamefont
  {Dender}}, \bibinfo {author} {\bibfnamefont {P.~R.}\ \bibnamefont {Hammar}},
  \bibinfo {author} {\bibfnamefont {D.~H.}\ \bibnamefont {Reich}}, \bibinfo
  {author} {\bibfnamefont {C.}~\bibnamefont {Broholm}},\ and\ \bibinfo {author}
  {\bibfnamefont {G.}~\bibnamefont {Aeppli}},\ }\bibfield  {title} {\bibinfo
  {title} {Direct observation of field-induced incommensurate fluctuations in a
  one-dimensional
  $\mathit{S}\phantom{\rule{0ex}{0ex}}=\phantom{\rule{0ex}{0ex}}1/2$
  antiferromagnet},\ }\href {https://doi.org/10.1103/PhysRevLett.79.1750}
  {\bibfield  {journal} {\bibinfo  {journal} {Phys. Rev. Lett.}\ }\textbf
  {\bibinfo {volume} {79}},\ \bibinfo {pages} {1750} (\bibinfo {year}
  {1997})}\BibitemShut {NoStop}%
\bibitem [{\citenamefont {Feyerherm}\ \emph {et~al.}(2000)\citenamefont
  {Feyerherm}, \citenamefont {Abens}, \citenamefont {G{\"u}nther},
  \citenamefont {Ishida}, \citenamefont {Mei{\ss}ner}, \citenamefont {Meschke},
  \citenamefont {Nogami},\ and\ \citenamefont {Steiner}}]{Feyerherm_stag_2000}%
  \BibitemOpen
  \bibfield  {author} {\bibinfo {author} {\bibfnamefont {R.}~\bibnamefont
  {Feyerherm}}, \bibinfo {author} {\bibfnamefont {S.}~\bibnamefont {Abens}},
  \bibinfo {author} {\bibfnamefont {D.}~\bibnamefont {G{\"u}nther}}, \bibinfo
  {author} {\bibfnamefont {T.}~\bibnamefont {Ishida}}, \bibinfo {author}
  {\bibfnamefont {M.}~\bibnamefont {Mei{\ss}ner}}, \bibinfo {author}
  {\bibfnamefont {M.}~\bibnamefont {Meschke}}, \bibinfo {author} {\bibfnamefont
  {T.}~\bibnamefont {Nogami}},\ and\ \bibinfo {author} {\bibfnamefont
  {M.}~\bibnamefont {Steiner}},\ }\bibfield  {title} {\bibinfo {title}
  {{M}agnetic-field induced gap and staggered susceptibility in the ${S}= 1/2$
  chain {[PMCu(NO$_{3}$)$_{2}$(H$_{2}$O)$_{2}$]$_{n}$ (PM= pyrimidine)}},\
  }\href {https://iopscience.iop.org/article/10.1088/0953-8984/12/39/312}
  {\bibfield  {journal} {\bibinfo  {journal} {Journal of Physics: Condensed
  Matter}\ }\textbf {\bibinfo {volume} {12}},\ \bibinfo {pages} {8495}
  (\bibinfo {year} {2000})}\BibitemShut {NoStop}%
\bibitem [{\citenamefont {Zvyagin}\ \emph {et~al.}(2004)\citenamefont
  {Zvyagin}, \citenamefont {Kolezhuk}, \citenamefont {Krzystek},\ and\
  \citenamefont {Feyerherm}}]{Zvyagin_SG_2004}%
  \BibitemOpen
  \bibfield  {author} {\bibinfo {author} {\bibfnamefont {S.~A.}\ \bibnamefont
  {Zvyagin}}, \bibinfo {author} {\bibfnamefont {A.~K.}\ \bibnamefont
  {Kolezhuk}}, \bibinfo {author} {\bibfnamefont {J.}~\bibnamefont {Krzystek}},\
  and\ \bibinfo {author} {\bibfnamefont {R.}~\bibnamefont {Feyerherm}},\
  }\bibfield  {title} {\bibinfo {title} {Excitation hierarchy of the quantum
  sine-gordon spin chain in a strong magnetic field},\ }\href
  {https://doi.org/10.1103/PhysRevLett.93.027201} {\bibfield  {journal}
  {\bibinfo  {journal} {Phys. Rev. Lett.}\ }\textbf {\bibinfo {volume} {93}},\
  \bibinfo {pages} {027201} (\bibinfo {year} {2004})}\BibitemShut {NoStop}%
\bibitem [{\citenamefont {Huddart}\ \emph {et~al.}(2021)\citenamefont
  {Huddart}, \citenamefont {Gomil\v{s}ek}, \citenamefont {Hicken},
  \citenamefont {Pratt}, \citenamefont {Blundell}, \citenamefont {Goddard},
  \citenamefont {Kaech}, \citenamefont {Manson},\ and\ \citenamefont
  {Lancaster}}]{Huddart_spin_transport_2021}%
  \BibitemOpen
  \bibfield  {author} {\bibinfo {author} {\bibfnamefont {B.~M.}\ \bibnamefont
  {Huddart}}, \bibinfo {author} {\bibfnamefont {M.}~\bibnamefont
  {Gomil\v{s}ek}}, \bibinfo {author} {\bibfnamefont {T.~J.}\ \bibnamefont
  {Hicken}}, \bibinfo {author} {\bibfnamefont {F.~L.}\ \bibnamefont {Pratt}},
  \bibinfo {author} {\bibfnamefont {S.~J.}\ \bibnamefont {Blundell}}, \bibinfo
  {author} {\bibfnamefont {P.~A.}\ \bibnamefont {Goddard}}, \bibinfo {author}
  {\bibfnamefont {S.~J.}\ \bibnamefont {Kaech}}, \bibinfo {author}
  {\bibfnamefont {J.~L.}\ \bibnamefont {Manson}},\ and\ \bibinfo {author}
  {\bibfnamefont {T.}~\bibnamefont {Lancaster}},\ }\bibfield  {title} {\bibinfo
  {title} {{Magnetic order and ballistic spin transport in a sine-Gordon spin
  chain}},\ }\href {https://doi.org/10.1103/PhysRevB.103.L060405} {\bibfield
  {journal} {\bibinfo  {journal} {Phys. Rev. B}\ }\textbf {\bibinfo {volume}
  {103}},\ \bibinfo {pages} {L060405} (\bibinfo {year} {2021})}\BibitemShut
  {NoStop}%
\bibitem [{\citenamefont {Liu}\ \emph {et~al.}(2019)\citenamefont {Liu},
  \citenamefont {Kittaka}, \citenamefont {Johnson}, \citenamefont {Lancaster},
  \citenamefont {Singleton}, \citenamefont {Sakakibara}, \citenamefont
  {Kohama}, \citenamefont {van Tol}, \citenamefont {Ardavan}, \citenamefont
  {Williams}, \citenamefont {Blundell}, \citenamefont {Manson}, \citenamefont
  {Manson},\ and\ \citenamefont {Goddard}}]{Liu_chiral_2019}%
  \BibitemOpen
  \bibfield  {author} {\bibinfo {author} {\bibfnamefont {J.}~\bibnamefont
  {Liu}}, \bibinfo {author} {\bibfnamefont {S.}~\bibnamefont {Kittaka}},
  \bibinfo {author} {\bibfnamefont {R.~D.}\ \bibnamefont {Johnson}}, \bibinfo
  {author} {\bibfnamefont {T.}~\bibnamefont {Lancaster}}, \bibinfo {author}
  {\bibfnamefont {J.}~\bibnamefont {Singleton}}, \bibinfo {author}
  {\bibfnamefont {T.}~\bibnamefont {Sakakibara}}, \bibinfo {author}
  {\bibfnamefont {Y.}~\bibnamefont {Kohama}}, \bibinfo {author} {\bibfnamefont
  {J.}~\bibnamefont {van Tol}}, \bibinfo {author} {\bibfnamefont
  {A.}~\bibnamefont {Ardavan}}, \bibinfo {author} {\bibfnamefont {B.~H.}\
  \bibnamefont {Williams}}, \bibinfo {author} {\bibfnamefont {S.~J.}\
  \bibnamefont {Blundell}}, \bibinfo {author} {\bibfnamefont {Z.~E.}\
  \bibnamefont {Manson}}, \bibinfo {author} {\bibfnamefont {J.~L.}\
  \bibnamefont {Manson}},\ and\ \bibinfo {author} {\bibfnamefont {P.~A.}\
  \bibnamefont {Goddard}},\ }\bibfield  {title} {\bibinfo {title}
  {Unconventional field-induced spin gap in an ${S}=1/2$ chiral staggered
  chain},\ }\href {https://doi.org/10.1103/PhysRevLett.122.057207} {\bibfield
  {journal} {\bibinfo  {journal} {Physical Review Letters}\ }\textbf {\bibinfo
  {volume} {122}},\ \bibinfo {pages} {057207} (\bibinfo {year}
  {2019})}\BibitemShut {NoStop}%
\bibitem [{sup()}]{supplementary}%
  \BibitemOpen
  \href@noop {} {}\bibinfo {note} {See Supplemental Material at [URL will be
  inserted by publisher].}\BibitemShut {Stop}%
\bibitem [{\citenamefont {Boča}(2004)}]{Boca_2004}%
  \BibitemOpen
  \bibfield  {author} {\bibinfo {author} {\bibfnamefont {R.}~\bibnamefont
  {Boča}},\ }\bibfield  {title} {\bibinfo {title} {Zero-field splitting in
  metal complexes},\ }\href
  {https://doi.org/https://doi.org/10.1016/j.ccr.2004.03.001} {\bibfield
  {journal} {\bibinfo  {journal} {Coordination Chemistry Reviews}\ }\textbf
  {\bibinfo {volume} {248}},\ \bibinfo {pages} {757} (\bibinfo {year}
  {2004})}\BibitemShut {NoStop}%
\bibitem [{\citenamefont {Manson}\ \emph {et~al.}(2020)\citenamefont {Manson},
  \citenamefont {Manson}, \citenamefont {Sargent}, \citenamefont {Villa},
  \citenamefont {Etten}, \citenamefont {Blackmore}, \citenamefont {Curley},
  \citenamefont {Williams}, \citenamefont {Brambleby}, \citenamefont {Goddard},
  \citenamefont {Ozarowski}, \citenamefont {Wilson}, \citenamefont {Huddart},
  \citenamefont {Lancaster}, \citenamefont {Johnson}, \citenamefont {Blundell},
  \citenamefont {Bendix}, \citenamefont {Wheeler}, \citenamefont {Lapidus},
  \citenamefont {Xiao}, \citenamefont {Birnbaum},\ and\ \citenamefont
  {Singleton}}]{Manson_2020}%
  \BibitemOpen
  \bibfield  {author} {\bibinfo {author} {\bibfnamefont {J.~L.}\ \bibnamefont
  {Manson}}, \bibinfo {author} {\bibfnamefont {Z.~E.}\ \bibnamefont {Manson}},
  \bibinfo {author} {\bibfnamefont {A.}~\bibnamefont {Sargent}}, \bibinfo
  {author} {\bibfnamefont {D.~Y.}\ \bibnamefont {Villa}}, \bibinfo {author}
  {\bibfnamefont {N.~L.}\ \bibnamefont {Etten}}, \bibinfo {author}
  {\bibfnamefont {W.~J.}\ \bibnamefont {Blackmore}}, \bibinfo {author}
  {\bibfnamefont {S.~P.}\ \bibnamefont {Curley}}, \bibinfo {author}
  {\bibfnamefont {R.~C.}\ \bibnamefont {Williams}}, \bibinfo {author}
  {\bibfnamefont {J.}~\bibnamefont {Brambleby}}, \bibinfo {author}
  {\bibfnamefont {P.~A.}\ \bibnamefont {Goddard}}, \bibinfo {author}
  {\bibfnamefont {A.}~\bibnamefont {Ozarowski}}, \bibinfo {author}
  {\bibfnamefont {M.~N.}\ \bibnamefont {Wilson}}, \bibinfo {author}
  {\bibfnamefont {B.~M.}\ \bibnamefont {Huddart}}, \bibinfo {author}
  {\bibfnamefont {T.}~\bibnamefont {Lancaster}}, \bibinfo {author}
  {\bibfnamefont {R.~D.}\ \bibnamefont {Johnson}}, \bibinfo {author}
  {\bibfnamefont {S.~J.}\ \bibnamefont {Blundell}}, \bibinfo {author}
  {\bibfnamefont {J.}~\bibnamefont {Bendix}}, \bibinfo {author} {\bibfnamefont
  {K.~A.}\ \bibnamefont {Wheeler}}, \bibinfo {author} {\bibfnamefont {S.~H.}\
  \bibnamefont {Lapidus}}, \bibinfo {author} {\bibfnamefont {F.}~\bibnamefont
  {Xiao}}, \bibinfo {author} {\bibfnamefont {S.}~\bibnamefont {Birnbaum}},\
  and\ \bibinfo {author} {\bibfnamefont {J.}~\bibnamefont {Singleton}},\
  }\bibfield  {title} {\bibinfo {title} {{Enhancing easy-plane anisotropy in
  bespoke Ni(II) quantum magnets}},\ }\href
  {https://doi.org/https://doi.org/10.1016/j.poly.2020.114379} {\bibfield
  {journal} {\bibinfo  {journal} {Polyhedron}\ }\textbf {\bibinfo {volume}
  {180}},\ \bibinfo {pages} {114379} (\bibinfo {year} {2020})}\BibitemShut
  {NoStop}%
\bibitem [{\citenamefont {Steele}\ \emph {et~al.}(2011)\citenamefont {Steele},
  \citenamefont {Lancaster}, \citenamefont {Blundell}, \citenamefont {Baker},
  \citenamefont {Pratt}, \citenamefont {Baines}, \citenamefont {Conner},
  \citenamefont {Southerland}, \citenamefont {Manson},\ and\ \citenamefont
  {Schlueter}}]{Schueter_2011}%
  \BibitemOpen
  \bibfield  {author} {\bibinfo {author} {\bibfnamefont {A.~J.}\ \bibnamefont
  {Steele}}, \bibinfo {author} {\bibfnamefont {T.}~\bibnamefont {Lancaster}},
  \bibinfo {author} {\bibfnamefont {S.~J.}\ \bibnamefont {Blundell}}, \bibinfo
  {author} {\bibfnamefont {P.~J.}\ \bibnamefont {Baker}}, \bibinfo {author}
  {\bibfnamefont {F.~L.}\ \bibnamefont {Pratt}}, \bibinfo {author}
  {\bibfnamefont {C.}~\bibnamefont {Baines}}, \bibinfo {author} {\bibfnamefont
  {M.~M.}\ \bibnamefont {Conner}}, \bibinfo {author} {\bibfnamefont {H.~I.}\
  \bibnamefont {Southerland}}, \bibinfo {author} {\bibfnamefont {J.~L.}\
  \bibnamefont {Manson}},\ and\ \bibinfo {author} {\bibfnamefont {J.~A.}\
  \bibnamefont {Schlueter}},\ }\bibfield  {title} {\bibinfo {title} {Magnetic
  order in quasi-two-dimensional molecular magnets investigated with muon-spin
  relaxation},\ }\href {https://doi.org/10.1103/PhysRevB.84.064412} {\bibfield
  {journal} {\bibinfo  {journal} {Phys. Rev. B}\ }\textbf {\bibinfo {volume}
  {84}},\ \bibinfo {pages} {064412} (\bibinfo {year} {2011})}\BibitemShut
  {NoStop}%
\bibitem [{\citenamefont {Blundell}\ \emph {et~al.}(2022)\citenamefont
  {Blundell}, \citenamefont {Renzi}, \citenamefont {Lancaster},\ and\
  \citenamefont {Pratt}}]{Blundell2022_muon}%
  \BibitemOpen
  \bibinfo {editor} {\bibfnamefont {S.~J.}\ \bibnamefont {Blundell}}, \bibinfo
  {editor} {\bibfnamefont {R.~D.}\ \bibnamefont {Renzi}}, \bibinfo {editor}
  {\bibfnamefont {T.}~\bibnamefont {Lancaster}},\ and\ \bibinfo {editor}
  {\bibfnamefont {F.~L.}\ \bibnamefont {Pratt}},\ eds.,\ \href@noop {} {\emph
  {\bibinfo {title} {Muon Spectroscopy - An Introduction}}}\ (\bibinfo
  {publisher} {Oxford University Press},\ \bibinfo {address} {Oxford},\
  \bibinfo {year} {2022})\BibitemShut {NoStop}%
\bibitem [{Note1()}]{Note1}%
  \BibitemOpen
  \bibinfo {note} {The exponent $\alpha $ describes the low-temperature
  behavior of the order parameter, while $\beta $ characterizes its behavior
  near $T_{\protect \rm N}$. For AFMs, $\alpha $ typically lies between 2 and
  3. However, in our fitting procedure, $\alpha $ tended to have large
  non-physical values. We therefore constrained $\alpha $ to a fixed value of
  3.}\BibitemShut {Stop}%
\bibitem [{\citenamefont {Chapon}\ \emph {et~al.}(2011)\citenamefont {Chapon},
  \citenamefont {Manuel}, \citenamefont {Radaelli}, \citenamefont {Benson},
  \citenamefont {Perrott}, \citenamefont {Ansell}, \citenamefont {Rhodes},
  \citenamefont {Raspino}, \citenamefont {Duxbury}, \citenamefont {Spill} \emph
  {et~al.}}]{WISH}%
  \BibitemOpen
  \bibfield  {author} {\bibinfo {author} {\bibfnamefont {L.~C.}\ \bibnamefont
  {Chapon}}, \bibinfo {author} {\bibfnamefont {P.}~\bibnamefont {Manuel}},
  \bibinfo {author} {\bibfnamefont {P.~G.}\ \bibnamefont {Radaelli}}, \bibinfo
  {author} {\bibfnamefont {C.}~\bibnamefont {Benson}}, \bibinfo {author}
  {\bibfnamefont {L.}~\bibnamefont {Perrott}}, \bibinfo {author} {\bibfnamefont
  {S.}~\bibnamefont {Ansell}}, \bibinfo {author} {\bibfnamefont {N.~J.}\
  \bibnamefont {Rhodes}}, \bibinfo {author} {\bibfnamefont {D.}~\bibnamefont
  {Raspino}}, \bibinfo {author} {\bibfnamefont {D.}~\bibnamefont {Duxbury}},
  \bibinfo {author} {\bibfnamefont {E.}~\bibnamefont {Spill}}, \emph {et~al.},\
  }\bibfield  {title} {\bibinfo {title} {{WISH: The new powder and single
  crystal magnetic diffractometer on the second target station}},\ }\href
  {https://doi.org/10.1080/10448632.2011.569650} {\bibfield  {journal}
  {\bibinfo  {journal} {Neutron News}\ }\textbf {\bibinfo {volume} {22}},\
  \bibinfo {pages} {22} (\bibinfo {year} {2011})}\BibitemShut {NoStop}%
\bibitem [{\citenamefont {Vaidya}\ \emph {et~al.}(2022)\citenamefont {Vaidya},
  \citenamefont {Manuel}, \citenamefont {Coak}, \citenamefont {Johnson},
  \citenamefont {Manson},\ and\ \citenamefont {Goddard}}]{WISH_data}%
  \BibitemOpen
  \bibfield  {author} {\bibinfo {author} {\bibfnamefont {S.}~\bibnamefont
  {Vaidya}}, \bibinfo {author} {\bibfnamefont {P.}~\bibnamefont {Manuel}},
  \bibinfo {author} {\bibfnamefont {M.}~\bibnamefont {Coak}}, \bibinfo {author}
  {\bibfnamefont {R.}~\bibnamefont {Johnson}}, \bibinfo {author} {\bibfnamefont
  {J.}~\bibnamefont {Manson}},\ and\ \bibinfo {author} {\bibfnamefont
  {P.}~\bibnamefont {Goddard}},\ }\href@noop {} {} (\bibinfo {year} {2022}),\
  \bibinfo {note} {{M}agnetic ground states of staggered and chiral
  integer-spin chains, STFC ISIS Neutron and Muon Source,
  \url{https://doi.org/10.5286/ISIS.E.RB2210219}}\BibitemShut {NoStop}%
\bibitem [{\citenamefont {Rodr{\'\i}guez-Carvajal}(1993)}]{Fullprof}%
  \BibitemOpen
  \bibfield  {author} {\bibinfo {author} {\bibfnamefont {J.}~\bibnamefont
  {Rodr{\'\i}guez-Carvajal}},\ }\bibfield  {title} {\bibinfo {title} {Recent
  advances in magnetic structure determination by neutron powder diffraction},\
  }\href {https://doi.org/10.1016/0921-4526(93)90108-I} {\bibfield  {journal}
  {\bibinfo  {journal} {Physica B: Condensed Matter}\ }\textbf {\bibinfo
  {volume} {192}},\ \bibinfo {pages} {55} (\bibinfo {year} {1993})}\BibitemShut
  {NoStop}%
\bibitem [{Iso()}]{Isodistort}%
  \BibitemOpen
  \href@noop {} {}\bibinfo {note} {H. T. Stokes, D. M. Hatch, and B. J.
  Campbell, ISODISTORT, ISOTROPY Software Suite,
  \url{iso.byu.edu},}\BibitemShut {NoStop}%
\bibitem [{\citenamefont {Campbell}\ \emph {et~al.}(2006)\citenamefont
  {Campbell}, \citenamefont {Stokes}, \citenamefont {Tanner},\ and\
  \citenamefont {Hatch}}]{Isodistort_2006}%
  \BibitemOpen
  \bibfield  {author} {\bibinfo {author} {\bibfnamefont {B.~J.}\ \bibnamefont
  {Campbell}}, \bibinfo {author} {\bibfnamefont {H.~T.}\ \bibnamefont
  {Stokes}}, \bibinfo {author} {\bibfnamefont {D.~E.}\ \bibnamefont {Tanner}},\
  and\ \bibinfo {author} {\bibfnamefont {D.~M.}\ \bibnamefont {Hatch}},\
  }\bibfield  {title} {\bibinfo {title} {{ISODISPLACE: a web-based tool for
  exploring structural distortions}},\ }\href
  {https://doi.org/10.1107/S0021889806014075} {\bibfield  {journal} {\bibinfo
  {journal} {Journal of Applied Crystallography}\ }\textbf {\bibinfo {volume}
  {39}},\ \bibinfo {pages} {607} (\bibinfo {year} {2006})}\BibitemShut
  {NoStop}%
\bibitem [{\citenamefont {Bewley}\ \emph {et~al.}(2011)\citenamefont {Bewley},
  \citenamefont {Taylor},\ and\ \citenamefont {Bennington.}}]{LET}%
  \BibitemOpen
  \bibfield  {author} {\bibinfo {author} {\bibfnamefont {R.}~\bibnamefont
  {Bewley}}, \bibinfo {author} {\bibfnamefont {J.}~\bibnamefont {Taylor}},\
  and\ \bibinfo {author} {\bibfnamefont {S.}~\bibnamefont {Bennington.}},\
  }\bibfield  {title} {\bibinfo {title} {{LET, a cold neutron multi-disk
  chopper spectrometer at ISIS}},\ }\href
  {https://doi.org/https://doi.org/10.1016/j.nima.2011.01.173} {\bibfield
  {journal} {\bibinfo  {journal} {Nuclear Instruments and Methods in Physics
  Research Section A: Accelerators, Spectrometers, Detectors and Associated
  Equipment}\ }\textbf {\bibinfo {volume} {637}},\ \bibinfo {pages} {128}
  (\bibinfo {year} {2011})}\BibitemShut {NoStop}%
\bibitem [{\citenamefont {Vaidya}\ \emph {et~al.}(2023)\citenamefont {Vaidya},
  \citenamefont {Steward}, \citenamefont {Johnson}, \citenamefont {Manson},\
  and\ \citenamefont {Goddard}}]{Let_data}%
  \BibitemOpen
  \bibfield  {author} {\bibinfo {author} {\bibfnamefont {S.}~\bibnamefont
  {Vaidya}}, \bibinfo {author} {\bibfnamefont {R.}~\bibnamefont {Steward}},
  \bibinfo {author} {\bibfnamefont {R.}~\bibnamefont {Johnson}}, \bibinfo
  {author} {\bibfnamefont {J.}~\bibnamefont {Manson}},\ and\ \bibinfo {author}
  {\bibfnamefont {P.}~\bibnamefont {Goddard}},\ }\href@noop {} {} (\bibinfo
  {year} {2023}),\ \bibinfo {note} {{M}easuring magnetic anisotropy and
  magnetic excitations in staggered and chiral spin-1 chains, STFC ISIS Neutron
  and Muon Source, \url{https://doi.org/10.5286/ISIS.E.RB2310305}}\BibitemShut
  {NoStop}%
\bibitem [{\citenamefont {Toth}\ and\ \citenamefont {Lake}(2015)}]{SpinW}%
  \BibitemOpen
  \bibfield  {author} {\bibinfo {author} {\bibfnamefont {S.}~\bibnamefont
  {Toth}}\ and\ \bibinfo {author} {\bibfnamefont {B.}~\bibnamefont {Lake}},\
  }\bibfield  {title} {\bibinfo {title} {{Linear spin wave theory for single-Q
  incommensurate magnetic structures}},\ }\href
  {https://iopscience.iop.org/article/10.1088/0953-8984/27/16/166002}
  {\bibfield  {journal} {\bibinfo  {journal} {Journal of Physics: Condensed
  Matter}\ }\textbf {\bibinfo {volume} {27}},\ \bibinfo {pages} {166002}
  (\bibinfo {year} {2015})}\BibitemShut {NoStop}%
\bibitem [{\citenamefont {Wheeler}\ \emph {et~al.}(2009)\citenamefont
  {Wheeler}, \citenamefont {Coldea}, \citenamefont
  {Wawrzy\ifmmode~\acute{n}\else \'{n}\fi{}ska}, \citenamefont {S\"orgel},
  \citenamefont {Jansen}, \citenamefont {Koza}, \citenamefont {Taylor},
  \citenamefont {Adroguer},\ and\ \citenamefont
  {Shannon}}]{Wheeler_2009_renorm}%
  \BibitemOpen
  \bibfield  {author} {\bibinfo {author} {\bibfnamefont {E.~M.}\ \bibnamefont
  {Wheeler}}, \bibinfo {author} {\bibfnamefont {R.}~\bibnamefont {Coldea}},
  \bibinfo {author} {\bibfnamefont {E.}~\bibnamefont
  {Wawrzy\ifmmode~\acute{n}\else \'{n}\fi{}ska}}, \bibinfo {author}
  {\bibfnamefont {T.}~\bibnamefont {S\"orgel}}, \bibinfo {author}
  {\bibfnamefont {M.}~\bibnamefont {Jansen}}, \bibinfo {author} {\bibfnamefont
  {M.~M.}\ \bibnamefont {Koza}}, \bibinfo {author} {\bibfnamefont
  {J.}~\bibnamefont {Taylor}}, \bibinfo {author} {\bibfnamefont
  {P.}~\bibnamefont {Adroguer}},\ and\ \bibinfo {author} {\bibfnamefont
  {N.}~\bibnamefont {Shannon}},\ }\bibfield  {title} {\bibinfo {title} {{Spin
  dynamics of the frustrated easy-axis triangular antiferromagnet
  $2H{\text{-AgNiO}}_{2}$ explored by inelastic neutron scattering}},\ }\href
  {https://doi.org/10.1103/PhysRevB.79.104421} {\bibfield  {journal} {\bibinfo
  {journal} {Phys. Rev. B}\ }\textbf {\bibinfo {volume} {79}},\ \bibinfo
  {pages} {104421} (\bibinfo {year} {2009})}\BibitemShut {NoStop}%
\bibitem [{\citenamefont {Dahlbom}\ \emph {et~al.}(2023)\citenamefont
  {Dahlbom}, \citenamefont {Zhang}, \citenamefont {Laraib}, \citenamefont
  {M.Pajerowski}, \citenamefont {Barros},\ and\ \citenamefont
  {Batista}}]{Dahlbom_2023}%
  \BibitemOpen
  \bibfield  {author} {\bibinfo {author} {\bibfnamefont {D.}~\bibnamefont
  {Dahlbom}}, \bibinfo {author} {\bibfnamefont {H.}~\bibnamefont {Zhang}},
  \bibinfo {author} {\bibfnamefont {Z.}~\bibnamefont {Laraib}}, \bibinfo
  {author} {\bibfnamefont {D.}~\bibnamefont {M.Pajerowski}}, \bibinfo {author}
  {\bibfnamefont {K.}~\bibnamefont {Barros}},\ and\ \bibinfo {author}
  {\bibfnamefont {C.}~\bibnamefont {Batista}},\ }\bibfield  {title} {\bibinfo
  {title} {Renormalized classical theory of quantum magnets},\ }\href
  {https://arxiv.org/abs/2304.03874} {\bibfield  {journal} {\bibinfo  {journal}
  {arXiv preprint arXiv:2304.03874}\ } (\bibinfo {year} {2023})}\BibitemShut
  {NoStop}%
\bibitem [{\citenamefont {Songvilay}\ \emph {et~al.}(2021)\citenamefont
  {Songvilay}, \citenamefont {Petit}, \citenamefont {Damay}, \citenamefont
  {Roux}, \citenamefont {Qureshi}, \citenamefont {Walker}, \citenamefont
  {Rodriguez-Rivera}, \citenamefont {Gao}, \citenamefont {Cheong},\ and\
  \citenamefont {Stock}}]{Songvilay_2021}%
  \BibitemOpen
  \bibfield  {author} {\bibinfo {author} {\bibfnamefont {M.}~\bibnamefont
  {Songvilay}}, \bibinfo {author} {\bibfnamefont {S.}~\bibnamefont {Petit}},
  \bibinfo {author} {\bibfnamefont {F.}~\bibnamefont {Damay}}, \bibinfo
  {author} {\bibfnamefont {G.}~\bibnamefont {Roux}}, \bibinfo {author}
  {\bibfnamefont {N.}~\bibnamefont {Qureshi}}, \bibinfo {author} {\bibfnamefont
  {H.~C.}\ \bibnamefont {Walker}}, \bibinfo {author} {\bibfnamefont {J.~A.}\
  \bibnamefont {Rodriguez-Rivera}}, \bibinfo {author} {\bibfnamefont
  {B.}~\bibnamefont {Gao}}, \bibinfo {author} {\bibfnamefont {S.-W.}\
  \bibnamefont {Cheong}},\ and\ \bibinfo {author} {\bibfnamefont
  {C.}~\bibnamefont {Stock}},\ }\bibfield  {title} {\bibinfo {title} {From one-
  to two-magnon excitations in the ${S}=3/2$ magnet
  ${\beta}$\ensuremath{-}{CaCr}$_{2}${O}$_{4}$},\ }\href
  {https://doi.org/10.1103/PhysRevLett.126.017201} {\bibfield  {journal}
  {\bibinfo  {journal} {Physical Review Letters}\ }\textbf {\bibinfo {volume}
  {126}},\ \bibinfo {pages} {017201} (\bibinfo {year} {2021})}\BibitemShut
  {NoStop}%
\bibitem [{\citenamefont {Manson}\ \emph {et~al.}(2023)\citenamefont {Manson},
  \citenamefont {Pajerowski}, \citenamefont {Donovan}, \citenamefont {Twamley},
  \citenamefont {Goddard}, \citenamefont {Johnson}, \citenamefont {Bendix},
  \citenamefont {Singleton}, \citenamefont {Lancaster}, \citenamefont
  {Blundell}, \citenamefont {Herbrych}, \citenamefont {Baker}, \citenamefont
  {Steele}, \citenamefont {Pratt}, \citenamefont {Franke-Chaudet},
  \citenamefont {McDonald}, \citenamefont {Plonczak},\ and\ \citenamefont
  {Manuel}}]{Manson_Pajerowski_2023}%
  \BibitemOpen
  \bibfield  {author} {\bibinfo {author} {\bibfnamefont {J.~L.}\ \bibnamefont
  {Manson}}, \bibinfo {author} {\bibfnamefont {D.~M.}\ \bibnamefont
  {Pajerowski}}, \bibinfo {author} {\bibfnamefont {J.~M.}\ \bibnamefont
  {Donovan}}, \bibinfo {author} {\bibfnamefont {B.}~\bibnamefont {Twamley}},
  \bibinfo {author} {\bibfnamefont {P.~A.}\ \bibnamefont {Goddard}}, \bibinfo
  {author} {\bibfnamefont {R.}~\bibnamefont {Johnson}}, \bibinfo {author}
  {\bibfnamefont {J.}~\bibnamefont {Bendix}}, \bibinfo {author} {\bibfnamefont
  {J.}~\bibnamefont {Singleton}}, \bibinfo {author} {\bibfnamefont
  {T.}~\bibnamefont {Lancaster}}, \bibinfo {author} {\bibfnamefont {S.~J.}\
  \bibnamefont {Blundell}}, \bibinfo {author} {\bibfnamefont {J.}~\bibnamefont
  {Herbrych}}, \bibinfo {author} {\bibfnamefont {P.~J.}\ \bibnamefont {Baker}},
  \bibinfo {author} {\bibfnamefont {A.~J.}\ \bibnamefont {Steele}}, \bibinfo
  {author} {\bibfnamefont {F.~L.}\ \bibnamefont {Pratt}}, \bibinfo {author}
  {\bibfnamefont {I.}~\bibnamefont {Franke-Chaudet}}, \bibinfo {author}
  {\bibfnamefont {R.~D.}\ \bibnamefont {McDonald}}, \bibinfo {author}
  {\bibfnamefont {A.}~\bibnamefont {Plonczak}},\ and\ \bibinfo {author}
  {\bibfnamefont {P.}~\bibnamefont {Manuel}},\ }\bibfield  {title} {\bibinfo
  {title} {Spatially anisotropic ${S}=1$ square-lattice antiferromagnet with
  single-ion anisotropy realized in a
  {Ni(II)}pyrazine-$n,{n}^{\ensuremath{'}}$-dioxide coordination polymer},\
  }\href {https://doi.org/10.1103/PhysRevB.108.094425} {\bibfield  {journal}
  {\bibinfo  {journal} {Physical Review B}\ }\textbf {\bibinfo {volume}
  {108}},\ \bibinfo {pages} {094425} (\bibinfo {year} {2023})}\BibitemShut
  {NoStop}%
\bibitem [{\citenamefont {Bai}\ \emph {et~al.}(2021)\citenamefont {Bai},
  \citenamefont {Zhang}, \citenamefont {Dun}, \citenamefont {Zhang},
  \citenamefont {Huang}, \citenamefont {Zhou}, \citenamefont {Stone},
  \citenamefont {Kolesnikov}, \citenamefont {Ye}, \citenamefont {Batista},\
  and\ \citenamefont {Mourigal}}]{bai_2021}%
  \BibitemOpen
  \bibfield  {author} {\bibinfo {author} {\bibfnamefont {X.}~\bibnamefont
  {Bai}}, \bibinfo {author} {\bibfnamefont {S.-S.}\ \bibnamefont {Zhang}},
  \bibinfo {author} {\bibfnamefont {Z.}~\bibnamefont {Dun}}, \bibinfo {author}
  {\bibfnamefont {H.}~\bibnamefont {Zhang}}, \bibinfo {author} {\bibfnamefont
  {Q.}~\bibnamefont {Huang}}, \bibinfo {author} {\bibfnamefont
  {H.}~\bibnamefont {Zhou}}, \bibinfo {author} {\bibfnamefont {M.~B.}\
  \bibnamefont {Stone}}, \bibinfo {author} {\bibfnamefont {A.~I.}\ \bibnamefont
  {Kolesnikov}}, \bibinfo {author} {\bibfnamefont {F.}~\bibnamefont {Ye}},
  \bibinfo {author} {\bibfnamefont {C.~D.}\ \bibnamefont {Batista}},\ and\
  \bibinfo {author} {\bibfnamefont {M.}~\bibnamefont {Mourigal}},\ }\bibfield
  {title} {\bibinfo {title} {Hybridized quadrupolar excitations in the
  spin-anisotropic frustrated magnet {FeI}$_2$},\ }\href
  {https://www.nature.com/articles/s41567-020-01110-1} {\bibfield  {journal}
  {\bibinfo  {journal} {Nature Physics}\ }\textbf {\bibinfo {volume} {17}},\
  \bibinfo {pages} {467} (\bibinfo {year} {2021})}\BibitemShut {NoStop}%
\bibitem [{\citenamefont {Sakamoto}\ \emph {et~al.}(2021)\citenamefont
  {Sakamoto}, \citenamefont {Higo}, \citenamefont {Shiga}, \citenamefont
  {Amemiya}, \citenamefont {Nakatsuji},\ and\ \citenamefont
  {Miwa}}]{sakamoto_observation_2021}%
  \BibitemOpen
  \bibfield  {author} {\bibinfo {author} {\bibfnamefont {S.}~\bibnamefont
  {Sakamoto}}, \bibinfo {author} {\bibfnamefont {T.}~\bibnamefont {Higo}},
  \bibinfo {author} {\bibfnamefont {M.}~\bibnamefont {Shiga}}, \bibinfo
  {author} {\bibfnamefont {K.}~\bibnamefont {Amemiya}}, \bibinfo {author}
  {\bibfnamefont {S.}~\bibnamefont {Nakatsuji}},\ and\ \bibinfo {author}
  {\bibfnamefont {S.}~\bibnamefont {Miwa}},\ }\bibfield  {title} {\bibinfo
  {title} {Observation of spontaneous {X}-ray magnetic circular dichroism in a
  chiral antiferromagnet},\ }\href
  {https://doi.org/10.1103/PhysRevB.104.134431} {\bibfield  {journal} {\bibinfo
   {journal} {Physical Review B}\ }\textbf {\bibinfo {volume} {104}},\ \bibinfo
  {pages} {134431} (\bibinfo {year} {2021})}\BibitemShut {NoStop}%
\bibitem [{\citenamefont {Okamoto}\ \emph {et~al.}(2024)\citenamefont
  {Okamoto}, \citenamefont {Wang}, \citenamefont {Chu}, \citenamefont {Shiu},
  \citenamefont {Singh}, \citenamefont {Huang}, \citenamefont {Mou},
  \citenamefont {Teh}, \citenamefont {Jeng}, \citenamefont {Du}, \citenamefont
  {Xu}, \citenamefont {Cheong}, \citenamefont {Du}, \citenamefont {Chen},
  \citenamefont {Fujimori},\ and\ \citenamefont {Huang}}]{Okamoto_2024}%
  \BibitemOpen
  \bibfield  {author} {\bibinfo {author} {\bibfnamefont {J.}~\bibnamefont
  {Okamoto}}, \bibinfo {author} {\bibfnamefont {R.-P.}\ \bibnamefont {Wang}},
  \bibinfo {author} {\bibfnamefont {Y.-Y.}\ \bibnamefont {Chu}}, \bibinfo
  {author} {\bibfnamefont {H.-W.}\ \bibnamefont {Shiu}}, \bibinfo {author}
  {\bibfnamefont {A.}~\bibnamefont {Singh}}, \bibinfo {author} {\bibfnamefont
  {H.-Y.}\ \bibnamefont {Huang}}, \bibinfo {author} {\bibfnamefont {C.-Y.}\
  \bibnamefont {Mou}}, \bibinfo {author} {\bibfnamefont {S.}~\bibnamefont
  {Teh}}, \bibinfo {author} {\bibfnamefont {H.-T.}\ \bibnamefont {Jeng}},
  \bibinfo {author} {\bibfnamefont {K.}~\bibnamefont {Du}}, \bibinfo {author}
  {\bibfnamefont {X.}~\bibnamefont {Xu}}, \bibinfo {author} {\bibfnamefont
  {S.-W.}\ \bibnamefont {Cheong}}, \bibinfo {author} {\bibfnamefont {C.-H.}\
  \bibnamefont {Du}}, \bibinfo {author} {\bibfnamefont {C.-T.}\ \bibnamefont
  {Chen}}, \bibinfo {author} {\bibfnamefont {A.}~\bibnamefont {Fujimori}},\
  and\ \bibinfo {author} {\bibfnamefont {D.-J.}\ \bibnamefont {Huang}},\
  }\bibfield  {title} {\bibinfo {title} {Giant {X}-ray circular dichroism in a
  time-reversal invariant antiferromagnet},\ }\href
  {https://doi.org/https://doi.org/10.1002/adma.202309172} {\bibfield
  {journal} {\bibinfo  {journal} {Advanced Materials}\ }\textbf {\bibinfo
  {volume} {36}},\ \bibinfo {pages} {2309172} (\bibinfo {year}
  {2024})}\BibitemShut {NoStop}%
\bibitem [{\citenamefont {Wierschem}\ and\ \citenamefont
  {Sengupta}(2014)}]{Wierschem_2014}%
  \BibitemOpen
  \bibfield  {author} {\bibinfo {author} {\bibfnamefont {K.}~\bibnamefont
  {Wierschem}}\ and\ \bibinfo {author} {\bibfnamefont {P.}~\bibnamefont
  {Sengupta}},\ }\bibfield  {title} {\bibinfo {title} {{Quenching the Haldane
  Gap in Spin-1 Heisenberg Antiferromagnets}},\ }\href
  {https://doi.org/10.1103/PhysRevLett.112.247203} {\bibfield  {journal}
  {\bibinfo  {journal} {Phys. Rev. Lett.}\ }\textbf {\bibinfo {volume} {112}},\
  \bibinfo {pages} {247203} (\bibinfo {year} {2014})}\BibitemShut {NoStop}%
\bibitem [{\citenamefont {Schlueter}\ \emph {et~al.}(2012)\citenamefont
  {Schlueter}, \citenamefont {Park}, \citenamefont {Halder}, \citenamefont
  {Armand}, \citenamefont {Dunmars}, \citenamefont {Chapman}, \citenamefont
  {Manson}, \citenamefont {Singleton}, \citenamefont {McDonald}, \citenamefont
  {Plonczak}, \citenamefont {Kang}, \citenamefont {Lee}, \citenamefont
  {Whangbo}, \citenamefont {Lancaster}, \citenamefont {Steele}, \citenamefont
  {Franke}, \citenamefont {Wright}, \citenamefont {Blundell}, \citenamefont
  {Pratt}, \citenamefont {deGeorge}, \citenamefont {Turnbull},\ and\
  \citenamefont {Landee}}]{Schlueter_2012}%
  \BibitemOpen
  \bibfield  {author} {\bibinfo {author} {\bibfnamefont {J.~A.}\ \bibnamefont
  {Schlueter}}, \bibinfo {author} {\bibfnamefont {H.}~\bibnamefont {Park}},
  \bibinfo {author} {\bibfnamefont {G.~J.}\ \bibnamefont {Halder}}, \bibinfo
  {author} {\bibfnamefont {W.~R.}\ \bibnamefont {Armand}}, \bibinfo {author}
  {\bibfnamefont {C.}~\bibnamefont {Dunmars}}, \bibinfo {author} {\bibfnamefont
  {K.~W.}\ \bibnamefont {Chapman}}, \bibinfo {author} {\bibfnamefont {J.~L.}\
  \bibnamefont {Manson}}, \bibinfo {author} {\bibfnamefont {J.}~\bibnamefont
  {Singleton}}, \bibinfo {author} {\bibfnamefont {R.}~\bibnamefont {McDonald}},
  \bibinfo {author} {\bibfnamefont {A.}~\bibnamefont {Plonczak}}, \bibinfo
  {author} {\bibfnamefont {J.}~\bibnamefont {Kang}}, \bibinfo {author}
  {\bibfnamefont {C.}~\bibnamefont {Lee}}, \bibinfo {author} {\bibfnamefont
  {M.-H.}\ \bibnamefont {Whangbo}}, \bibinfo {author} {\bibfnamefont
  {T.}~\bibnamefont {Lancaster}}, \bibinfo {author} {\bibfnamefont {A.~J.}\
  \bibnamefont {Steele}}, \bibinfo {author} {\bibfnamefont {I.}~\bibnamefont
  {Franke}}, \bibinfo {author} {\bibfnamefont {J.~D.}\ \bibnamefont {Wright}},
  \bibinfo {author} {\bibfnamefont {S.~J.}\ \bibnamefont {Blundell}}, \bibinfo
  {author} {\bibfnamefont {F.~L.}\ \bibnamefont {Pratt}}, \bibinfo {author}
  {\bibfnamefont {J.}~\bibnamefont {deGeorge}}, \bibinfo {author}
  {\bibfnamefont {M.~M.}\ \bibnamefont {Turnbull}},\ and\ \bibinfo {author}
  {\bibfnamefont {C.~P.}\ \bibnamefont {Landee}},\ }\bibfield  {title}
  {\bibinfo {title} {Importance of halogen$\cdot\cdot\cdot$halogen contacts for
  the structural and magnetic properties of
  {CuX}$_{2}$(pyrazine-${N}$,${N'}$-dioxide)({H}$_{2}${O})$_{2}$ (${X}$= {Cl}
  and {Br})},\ }\href {https://pubs.acs.org/doi/full/10.1021/ic201924q}
  {\bibfield  {journal} {\bibinfo  {journal} {Inorganic Chemistry}\ }\textbf
  {\bibinfo {volume} {51}},\ \bibinfo {pages} {2121} (\bibinfo {year}
  {2012})}\BibitemShut {NoStop}%
\bibitem [{\citenamefont {Cortijo}\ \emph {et~al.}(2013)\citenamefont
  {Cortijo}, \citenamefont {Herrero}, \citenamefont {Jimenez-Aparicio},\ and\
  \citenamefont {Matesanz}}]{Cortijo_2013}%
  \BibitemOpen
  \bibfield  {author} {\bibinfo {author} {\bibfnamefont {M.}~\bibnamefont
  {Cortijo}}, \bibinfo {author} {\bibfnamefont {S.}~\bibnamefont {Herrero}},
  \bibinfo {author} {\bibfnamefont {R.}~\bibnamefont {Jimenez-Aparicio}},\ and\
  \bibinfo {author} {\bibfnamefont {E.}~\bibnamefont {Matesanz}},\ }\bibfield
  {title} {\bibinfo {title} {Modulation of the magnetic properties of
  two-dimensional compounds [{Ni}${X}_{2}${(N--N)}] by tailoring their crystal
  structure},\ }\href {https://doi.org/10.1021/ic400632c} {\bibfield  {journal}
  {\bibinfo  {journal} {Inorganic Chemistry}\ }\textbf {\bibinfo {volume}
  {52}},\ \bibinfo {pages} {7087} (\bibinfo {year} {2013})}\BibitemShut
  {NoStop}%
\bibitem [{\citenamefont {Cortijo}\ \emph {et~al.}(2014)\citenamefont
  {Cortijo}, \citenamefont {Herrero}, \citenamefont {Jimenez-Aparicio},
  \citenamefont {Perles}, \citenamefont {Priego},\ and\ \citenamefont
  {Torroba}}]{Cortijo_2014}%
  \BibitemOpen
  \bibfield  {author} {\bibinfo {author} {\bibfnamefont {M.}~\bibnamefont
  {Cortijo}}, \bibinfo {author} {\bibfnamefont {S.}~\bibnamefont {Herrero}},
  \bibinfo {author} {\bibfnamefont {R.}~\bibnamefont {Jimenez-Aparicio}},
  \bibinfo {author} {\bibfnamefont {J.}~\bibnamefont {Perles}}, \bibinfo
  {author} {\bibfnamefont {J.~L.}\ \bibnamefont {Priego}},\ and\ \bibinfo
  {author} {\bibfnamefont {J.}~\bibnamefont {Torroba}},\ }\bibfield  {title}
  {\bibinfo {title} {Tuning of adsorption and magnetic properties in a series
  of self-templated isostructural {Ni(II)} metal-organic frameworks},\ }\href
  {https://doi.org/10.1021/cg401590w} {\bibfield  {journal} {\bibinfo
  {journal} {Crystal Growth \& Design}\ }\textbf {\bibinfo {volume} {14}},\
  \bibinfo {pages} {716} (\bibinfo {year} {2014})}\BibitemShut {NoStop}%
\bibitem [{\citenamefont {Liu}\ \emph {et~al.}(2016)\citenamefont {Liu},
  \citenamefont {Goddard}, \citenamefont {Singleton}, \citenamefont
  {Brambleby}, \citenamefont {Foronda}, \citenamefont {Möller}, \citenamefont
  {Kohama}, \citenamefont {Ghannadzadeh}, \citenamefont {Ardavan},
  \citenamefont {Blundell}, \citenamefont {Lancaster}, \citenamefont {Xiao},
  \citenamefont {Williams}, \citenamefont {Pratt}, \citenamefont {Baker},
  \citenamefont {Wierschem}, \citenamefont {Lapidus}, \citenamefont {Stone},
  \citenamefont {Stephens}, \citenamefont {Bendix}, \citenamefont {Woods},
  \citenamefont {Carreiro}, \citenamefont {Tran}, \citenamefont {Villa},\ and\
  \citenamefont {Manson}}]{Liu_2016_correlations}%
  \BibitemOpen
  \bibfield  {author} {\bibinfo {author} {\bibfnamefont {J.}~\bibnamefont
  {Liu}}, \bibinfo {author} {\bibfnamefont {P.~A.}\ \bibnamefont {Goddard}},
  \bibinfo {author} {\bibfnamefont {J.}~\bibnamefont {Singleton}}, \bibinfo
  {author} {\bibfnamefont {J.}~\bibnamefont {Brambleby}}, \bibinfo {author}
  {\bibfnamefont {F.}~\bibnamefont {Foronda}}, \bibinfo {author} {\bibfnamefont
  {J.~S.}\ \bibnamefont {Möller}}, \bibinfo {author} {\bibfnamefont
  {Y.}~\bibnamefont {Kohama}}, \bibinfo {author} {\bibfnamefont
  {S.}~\bibnamefont {Ghannadzadeh}}, \bibinfo {author} {\bibfnamefont
  {A.}~\bibnamefont {Ardavan}}, \bibinfo {author} {\bibfnamefont {S.~J.}\
  \bibnamefont {Blundell}}, \bibinfo {author} {\bibfnamefont {T.}~\bibnamefont
  {Lancaster}}, \bibinfo {author} {\bibfnamefont {F.}~\bibnamefont {Xiao}},
  \bibinfo {author} {\bibfnamefont {R.~C.}\ \bibnamefont {Williams}}, \bibinfo
  {author} {\bibfnamefont {F.~L.}\ \bibnamefont {Pratt}}, \bibinfo {author}
  {\bibfnamefont {P.~J.}\ \bibnamefont {Baker}}, \bibinfo {author}
  {\bibfnamefont {K.}~\bibnamefont {Wierschem}}, \bibinfo {author}
  {\bibfnamefont {S.~H.}\ \bibnamefont {Lapidus}}, \bibinfo {author}
  {\bibfnamefont {K.~H.}\ \bibnamefont {Stone}}, \bibinfo {author}
  {\bibfnamefont {P.~W.}\ \bibnamefont {Stephens}}, \bibinfo {author}
  {\bibfnamefont {J.}~\bibnamefont {Bendix}}, \bibinfo {author} {\bibfnamefont
  {T.~J.}\ \bibnamefont {Woods}}, \bibinfo {author} {\bibfnamefont {K.~E.}\
  \bibnamefont {Carreiro}}, \bibinfo {author} {\bibfnamefont {H.~E.}\
  \bibnamefont {Tran}}, \bibinfo {author} {\bibfnamefont {C.~J.}\ \bibnamefont
  {Villa}},\ and\ \bibinfo {author} {\bibfnamefont {J.~L.}\ \bibnamefont
  {Manson}},\ }\bibfield  {title} {\bibinfo {title} {Antiferromagnetism in a
  family of ${S}=1$ square lattice coordination polymers {NiX}$_2$(pyz)$_2$ {(X
  = Cl, Br, I, NCS; pyz = Pyrazine)}},\ }\href
  {https://doi.org/10.1021/acs.inorgchem.5b02991} {\bibfield  {journal}
  {\bibinfo  {journal} {Inorganic Chemistry}\ }\textbf {\bibinfo {volume}
  {55}},\ \bibinfo {pages} {3515} (\bibinfo {year} {2016})}\BibitemShut
  {NoStop}%
\bibitem [{\citenamefont {Kubus}\ \emph {et~al.}(2018)\citenamefont {Kubus},
  \citenamefont {Lanza}, \citenamefont {Scatena}, \citenamefont {Dos~Santos},
  \citenamefont {Wehinger}, \citenamefont {Casati}, \citenamefont {Fiolka},
  \citenamefont {Keller}, \citenamefont {Macchi}, \citenamefont {Rüegg},\ and\
  \citenamefont {Krämer}}]{Kubus_2018}%
  \BibitemOpen
  \bibfield  {author} {\bibinfo {author} {\bibfnamefont {M.}~\bibnamefont
  {Kubus}}, \bibinfo {author} {\bibfnamefont {A.}~\bibnamefont {Lanza}},
  \bibinfo {author} {\bibfnamefont {R.}~\bibnamefont {Scatena}}, \bibinfo
  {author} {\bibfnamefont {L.~H.~R.}\ \bibnamefont {Dos~Santos}}, \bibinfo
  {author} {\bibfnamefont {B.}~\bibnamefont {Wehinger}}, \bibinfo {author}
  {\bibfnamefont {N.}~\bibnamefont {Casati}}, \bibinfo {author} {\bibfnamefont
  {C.}~\bibnamefont {Fiolka}}, \bibinfo {author} {\bibfnamefont
  {L.}~\bibnamefont {Keller}}, \bibinfo {author} {\bibfnamefont
  {P.}~\bibnamefont {Macchi}}, \bibinfo {author} {\bibfnamefont
  {C.}~\bibnamefont {Rüegg}},\ and\ \bibinfo {author} {\bibfnamefont {K.~W.}\
  \bibnamefont {Krämer}},\ }\bibfield  {title} {\bibinfo {title} {Quasi-2d
  {H}eisenberg antiferromagnets [{Cu}${X}$(pyz)$_{2}$]({BF}$_{4}$) with ${X}$=
  {Cl} and {Br}},\ }\href {https://doi.org/10.1021/acs.inorgchem.7b03150}
  {\bibfield  {journal} {\bibinfo  {journal} {Inorganic Chemistry}\ }\textbf
  {\bibinfo {volume} {57}},\ \bibinfo {pages} {4934} (\bibinfo {year}
  {2018})}\BibitemShut {NoStop}%
\bibitem [{\citenamefont {Blackmore}\ \emph {et~al.}(2022)\citenamefont
  {Blackmore}, \citenamefont {Curley}, \citenamefont {Williams}, \citenamefont
  {Vaidya}, \citenamefont {Singleton}, \citenamefont {Birnbaum}, \citenamefont
  {Ozarowski}, \citenamefont {Schlueter}, \citenamefont {Chen}, \citenamefont
  {Gillon}, \citenamefont {Goukassov}, \citenamefont {Kibalin}, \citenamefont
  {Villa}, \citenamefont {Villa}, \citenamefont {Manson},\ and\ \citenamefont
  {Goddard}}]{Blackmore-correlations}%
  \BibitemOpen
  \bibfield  {author} {\bibinfo {author} {\bibfnamefont {W.~J.~A.}\
  \bibnamefont {Blackmore}}, \bibinfo {author} {\bibfnamefont {S.~P.~M.}\
  \bibnamefont {Curley}}, \bibinfo {author} {\bibfnamefont {R.~C.}\
  \bibnamefont {Williams}}, \bibinfo {author} {\bibfnamefont {S.}~\bibnamefont
  {Vaidya}}, \bibinfo {author} {\bibfnamefont {J.}~\bibnamefont {Singleton}},
  \bibinfo {author} {\bibfnamefont {S.}~\bibnamefont {Birnbaum}}, \bibinfo
  {author} {\bibfnamefont {A.}~\bibnamefont {Ozarowski}}, \bibinfo {author}
  {\bibfnamefont {J.~A.}\ \bibnamefont {Schlueter}}, \bibinfo {author}
  {\bibfnamefont {Y.-S.}\ \bibnamefont {Chen}}, \bibinfo {author}
  {\bibfnamefont {B.}~\bibnamefont {Gillon}}, \bibinfo {author} {\bibfnamefont
  {A.}~\bibnamefont {Goukassov}}, \bibinfo {author} {\bibfnamefont
  {I.}~\bibnamefont {Kibalin}}, \bibinfo {author} {\bibfnamefont {D.~Y.}\
  \bibnamefont {Villa}}, \bibinfo {author} {\bibfnamefont {J.~A.}\ \bibnamefont
  {Villa}}, \bibinfo {author} {\bibfnamefont {J.~L.}\ \bibnamefont {Manson}},\
  and\ \bibinfo {author} {\bibfnamefont {P.~A.}\ \bibnamefont {Goddard}},\
  }\bibfield  {title} {\bibinfo {title} {Magneto-structural correlations in
  {Ni}$^{2+}$–{Halide···Halide–Ni}$^{2+}$ chains},\ }\href
  {https://doi.org/10.1021/acs.inorgchem.1c02483} {\bibfield  {journal}
  {\bibinfo  {journal} {Inorganic Chemistry}\ }\textbf {\bibinfo {volume}
  {61}},\ \bibinfo {pages} {141} (\bibinfo {year} {2022})}\BibitemShut
  {NoStop}%
\bibitem [{\citenamefont {Pajerowski}\ \emph {et~al.}(2022)\citenamefont
  {Pajerowski}, \citenamefont {Podlesnyak}, \citenamefont {Herbrych},\ and\
  \citenamefont {Manson}}]{Pajerowski_2022}%
  \BibitemOpen
  \bibfield  {author} {\bibinfo {author} {\bibfnamefont {D.~M.}\ \bibnamefont
  {Pajerowski}}, \bibinfo {author} {\bibfnamefont {A.~P.}\ \bibnamefont
  {Podlesnyak}}, \bibinfo {author} {\bibfnamefont {J.}~\bibnamefont
  {Herbrych}},\ and\ \bibinfo {author} {\bibfnamefont {J.}~\bibnamefont
  {Manson}},\ }\bibfield  {title} {\bibinfo {title} {High-pressure inelastic
  neutron scattering study of the anisotropic ${S}=1$ spin chain
  {Ni}({HF}$_{2}$)(3\ensuremath{-}{Clpyradine})$_{4}$]{BF}$_{4}$},\ }\href
  {https://doi.org/10.1103/PhysRevB.105.134420} {\bibfield  {journal} {\bibinfo
   {journal} {Physical Review B}\ }\textbf {\bibinfo {volume} {105}},\ \bibinfo
  {pages} {134420} (\bibinfo {year} {2022})}\BibitemShut {NoStop}%
\bibitem [{\citenamefont {Coak}\ \emph {et~al.}(2023)\citenamefont {Coak},
  \citenamefont {Curley}, \citenamefont {Hawkhead}, \citenamefont {Tidey},
  \citenamefont {Graf}, \citenamefont {Clark}, \citenamefont {Sengupta},
  \citenamefont {Manson}, \citenamefont {Lancaster}, \citenamefont {Goddard},\
  and\ \citenamefont {Manson}}]{Coak_2023}%
  \BibitemOpen
  \bibfield  {author} {\bibinfo {author} {\bibfnamefont {M.~J.}\ \bibnamefont
  {Coak}}, \bibinfo {author} {\bibfnamefont {S.~P.~M.}\ \bibnamefont {Curley}},
  \bibinfo {author} {\bibfnamefont {Z.}~\bibnamefont {Hawkhead}}, \bibinfo
  {author} {\bibfnamefont {J.~P.}\ \bibnamefont {Tidey}}, \bibinfo {author}
  {\bibfnamefont {D.}~\bibnamefont {Graf}}, \bibinfo {author} {\bibfnamefont
  {S.~J.}\ \bibnamefont {Clark}}, \bibinfo {author} {\bibfnamefont
  {P.}~\bibnamefont {Sengupta}}, \bibinfo {author} {\bibfnamefont {Z.~E.}\
  \bibnamefont {Manson}}, \bibinfo {author} {\bibfnamefont {T.}~\bibnamefont
  {Lancaster}}, \bibinfo {author} {\bibfnamefont {P.~A.}\ \bibnamefont
  {Goddard}},\ and\ \bibinfo {author} {\bibfnamefont {J.~L.}\ \bibnamefont
  {Manson}},\ }\bibfield  {title} {\bibinfo {title} {Asymmetric phase diagram
  and dimensional crossover in a system of spin-$\frac{1}{2}$ dimers under
  applied hydrostatic pressure},\ }\href
  {https://doi.org/10.1103/PhysRevB.108.224431} {\bibfield  {journal} {\bibinfo
   {journal} {Physical Review B}\ }\textbf {\bibinfo {volume} {108}},\ \bibinfo
  {pages} {224431} (\bibinfo {year} {2023})}\BibitemShut {NoStop}%
\bibitem [{\citenamefont {Geers}\ \emph {et~al.}(2023)\citenamefont {Geers},
  \citenamefont {Jarvis}, \citenamefont {Liu}, \citenamefont {Saxena},
  \citenamefont {Pitcairn}, \citenamefont {Myatt}, \citenamefont {Hallweger},
  \citenamefont {Kronawitter}, \citenamefont {Kieslich}, \citenamefont {Ling},
  \citenamefont {Cairns}, \citenamefont {Daisenberger}, \citenamefont {Fabelo},
  \citenamefont {Ca\~nadillas Delgado},\ and\ \citenamefont
  {Cliffe}}]{Geers_preesure_2023}%
  \BibitemOpen
  \bibfield  {author} {\bibinfo {author} {\bibfnamefont {M.}~\bibnamefont
  {Geers}}, \bibinfo {author} {\bibfnamefont {D.~M.}\ \bibnamefont {Jarvis}},
  \bibinfo {author} {\bibfnamefont {C.}~\bibnamefont {Liu}}, \bibinfo {author}
  {\bibfnamefont {S.~S.}\ \bibnamefont {Saxena}}, \bibinfo {author}
  {\bibfnamefont {J.}~\bibnamefont {Pitcairn}}, \bibinfo {author}
  {\bibfnamefont {E.}~\bibnamefont {Myatt}}, \bibinfo {author} {\bibfnamefont
  {S.~A.}\ \bibnamefont {Hallweger}}, \bibinfo {author} {\bibfnamefont {S.~M.}\
  \bibnamefont {Kronawitter}}, \bibinfo {author} {\bibfnamefont
  {G.}~\bibnamefont {Kieslich}}, \bibinfo {author} {\bibfnamefont
  {S.}~\bibnamefont {Ling}}, \bibinfo {author} {\bibfnamefont {A.~B.}\
  \bibnamefont {Cairns}}, \bibinfo {author} {\bibfnamefont {D.}~\bibnamefont
  {Daisenberger}}, \bibinfo {author} {\bibfnamefont {O.}~\bibnamefont
  {Fabelo}}, \bibinfo {author} {\bibfnamefont {L.}~\bibnamefont {Ca\~nadillas
  Delgado}},\ and\ \bibinfo {author} {\bibfnamefont {M.~J.}\ \bibnamefont
  {Cliffe}},\ }\bibfield  {title} {\bibinfo {title} {High-pressure behavior of
  the magnetic van der waals molecular framework {Ni(NCS)}$_{2}$},\ }\href
  {https://doi.org/10.1103/PhysRevB.108.144439} {\bibfield  {journal} {\bibinfo
   {journal} {Physical Review B}\ }\textbf {\bibinfo {volume} {108}},\ \bibinfo
  {pages} {144439} (\bibinfo {year} {2023})}\BibitemShut {NoStop}%
\bibitem [{\citenamefont {Povarov}\ \emph {et~al.}(2024)\citenamefont
  {Povarov}, \citenamefont {Graf}, \citenamefont {Hauspurg}, \citenamefont
  {Zherlitsyn}, \citenamefont {Wosnitza}, \citenamefont {Sakurai},
  \citenamefont {Ohta}, \citenamefont {Kimura}, \citenamefont {Nojiri},
  \citenamefont {Garlea}, \citenamefont {Zheludev}, \citenamefont
  {Paduan-Filho}, \citenamefont {Nicklas},\ and\ \citenamefont
  {Zvyagin}}]{Povarov_2024}%
  \BibitemOpen
  \bibfield  {author} {\bibinfo {author} {\bibfnamefont {K.~Y.}\ \bibnamefont
  {Povarov}}, \bibinfo {author} {\bibfnamefont {D.~E.}\ \bibnamefont {Graf}},
  \bibinfo {author} {\bibfnamefont {A.}~\bibnamefont {Hauspurg}}, \bibinfo
  {author} {\bibfnamefont {S.}~\bibnamefont {Zherlitsyn}}, \bibinfo {author}
  {\bibfnamefont {J.}~\bibnamefont {Wosnitza}}, \bibinfo {author}
  {\bibfnamefont {T.}~\bibnamefont {Sakurai}}, \bibinfo {author} {\bibfnamefont
  {H.}~\bibnamefont {Ohta}}, \bibinfo {author} {\bibfnamefont {S.}~\bibnamefont
  {Kimura}}, \bibinfo {author} {\bibfnamefont {H.}~\bibnamefont {Nojiri}},
  \bibinfo {author} {\bibfnamefont {V.~O.}\ \bibnamefont {Garlea}}, \bibinfo
  {author} {\bibfnamefont {A.}~\bibnamefont {Zheludev}}, \bibinfo {author}
  {\bibfnamefont {A.}~\bibnamefont {Paduan-Filho}}, \bibinfo {author}
  {\bibfnamefont {M.}~\bibnamefont {Nicklas}},\ and\ \bibinfo {author}
  {\bibfnamefont {S.~A.}\ \bibnamefont {Zvyagin}},\ }\bibfield  {title}
  {\bibinfo {title} {Pressure-tuned quantum criticality in the large-{D}
  antiferromagnetic {DTN}},\ }\href
  {https://doi.org/10.1038/s41467-024-46527-x} {\bibfield  {journal} {\bibinfo
  {journal} {Nature Communications}\ }\textbf {\bibinfo {volume} {15}},\
  \bibinfo {pages} {2295} (\bibinfo {year} {2024})}\BibitemShut {NoStop}%
\bibitem [{Dat()}]{Data}%
  \BibitemOpen
  \href@noop {} {}\bibinfo {note} {S. Vaidya \textit{et al}., Data for
  "Single-ion anisotropy driven chiral magnetism in a Spin-1 antiferromagnetic
  chain” (2025), [URL to be inserted].}\BibitemShut {Stop}%
\end{thebibliography}%


\begin{thebibliography}{12}%
\makeatletter
\providecommand \@ifxundefined [1]{%
 \@ifx{#1\undefined}
}%
\providecommand \@ifnum [1]{%
 \ifnum #1\expandafter \@firstoftwo
 \else \expandafter \@secondoftwo
 \fi
}%
\providecommand \@ifx [1]{%
 \ifx #1\expandafter \@firstoftwo
 \else \expandafter \@secondoftwo
 \fi
}%
\providecommand \natexlab [1]{#1}%
\providecommand \enquote  [1]{``#1''}%
\providecommand \bibnamefont  [1]{#1}%
\providecommand \bibfnamefont [1]{#1}%
\providecommand \citenamefont [1]{#1}%
\providecommand \href@noop [0]{\@secondoftwo}%
\providecommand \href [0]{\begingroup \@sanitize@url \@href}%
\providecommand \@href[1]{\@@startlink{#1}\@@href}%
\providecommand \@@href[1]{\endgroup#1\@@endlink}%
\providecommand \@sanitize@url [0]{\catcode `\\12\catcode `\$12\catcode `\&12\catcode `\#12\catcode `\^12\catcode `\_12\catcode `\%12\relax}%
\providecommand \@@startlink[1]{}%
\providecommand \@@endlink[0]{}%
\providecommand \url  [0]{\begingroup\@sanitize@url \@url }%
\providecommand \@url [1]{\endgroup\@href {#1}{\urlprefix }}%
\providecommand \urlprefix  [0]{URL }%
\providecommand \Eprint [0]{\href }%
\providecommand \doibase [0]{https://doi.org/}%
\providecommand \selectlanguage [0]{\@gobble}%
\providecommand \bibinfo  [0]{\@secondoftwo}%
\providecommand \bibfield  [0]{\@secondoftwo}%
\providecommand \translation [1]{[#1]}%
\providecommand \BibitemOpen [0]{}%
\providecommand \bibitemStop [0]{}%
\providecommand \bibitemNoStop [0]{.\EOS\space}%
\providecommand \EOS [0]{\spacefactor3000\relax}%
\providecommand \BibitemShut  [1]{\csname bibitem#1\endcsname}%
\let\auto@bib@innerbib\@empty
\bibitem [{\citenamefont {Sheldrick}(2015{\natexlab{a}})}]{SheldrickT}%
  \BibitemOpen
  \bibfield  {author} {\bibinfo {author} {\bibfnamefont {G.~M.}\ \bibnamefont {Sheldrick}},\ }\bibfield  {title} {\bibinfo {title} {{{\it SHELXT} {--} Integrated space-group and crystal-structure determination}},\ }\href {https://doi.org/10.1107/S2053273314026370} {\bibfield  {journal} {\bibinfo  {journal} {Acta Crystallographica Section A}\ }\textbf {\bibinfo {volume} {71}},\ \bibinfo {pages} {3} (\bibinfo {year} {2015}{\natexlab{a}})}\BibitemShut {NoStop}%
\bibitem [{\citenamefont {Sheldrick}(2015{\natexlab{b}})}]{SheldrickL}%
  \BibitemOpen
  \bibfield  {author} {\bibinfo {author} {\bibfnamefont {G.~M.}\ \bibnamefont {Sheldrick}},\ }\bibfield  {title} {\bibinfo {title} {{Crystal structure refinement with {\it SHELXL}}},\ }\href {https://doi.org/10.1107/S2053229614024218} {\bibfield  {journal} {\bibinfo  {journal} {Acta Crystallographica Section C}\ }\textbf {\bibinfo {volume} {71}},\ \bibinfo {pages} {3} (\bibinfo {year} {2015}{\natexlab{b}})}\BibitemShut {NoStop}%
\bibitem [{\citenamefont {Dolomanov}\ \emph {et~al.}(2009)\citenamefont {Dolomanov}, \citenamefont {Bourhis}, \citenamefont {Gildea}, \citenamefont {Howard},\ and\ \citenamefont {Puschmann}}]{Dolomanov_olex}%
  \BibitemOpen
  \bibfield  {author} {\bibinfo {author} {\bibfnamefont {O.~V.}\ \bibnamefont {Dolomanov}}, \bibinfo {author} {\bibfnamefont {L.~J.}\ \bibnamefont {Bourhis}}, \bibinfo {author} {\bibfnamefont {R.~J.}\ \bibnamefont {Gildea}}, \bibinfo {author} {\bibfnamefont {J.~A.~K.}\ \bibnamefont {Howard}},\ and\ \bibinfo {author} {\bibfnamefont {H.}~\bibnamefont {Puschmann}},\ }\bibfield  {title} {\bibinfo {title} {{{\it OLEX2}: a complete structure solution, refinement and analysis program}},\ }\href {https://doi.org/10.1107/S0021889808042726} {\bibfield  {journal} {\bibinfo  {journal} {Journal of Applied Crystallography}\ }\textbf {\bibinfo {volume} {42}},\ \bibinfo {pages} {339} (\bibinfo {year} {2009})}\BibitemShut {NoStop}%
\bibitem [{\citenamefont {Bhattacharjee}\ \emph {et~al.}(1981)\citenamefont {Bhattacharjee}, \citenamefont {Chakravarty}, \citenamefont {Richardson},\ and\ \citenamefont {Scalapino}}]{Bhattacharjee_1981}%
  \BibitemOpen
  \bibfield  {author} {\bibinfo {author} {\bibfnamefont {J.}~\bibnamefont {Bhattacharjee}}, \bibinfo {author} {\bibfnamefont {S.}~\bibnamefont {Chakravarty}}, \bibinfo {author} {\bibfnamefont {J.~L.}\ \bibnamefont {Richardson}},\ and\ \bibinfo {author} {\bibfnamefont {D.~J.}\ \bibnamefont {Scalapino}},\ }\bibfield  {title} {\bibinfo {title} {Some properties of a one-dimensional ising chain with an inverse-square interaction},\ }\href {https://doi.org/10.1103/PhysRevB.24.3862} {\bibfield  {journal} {\bibinfo  {journal} {Phys. Rev. B}\ }\textbf {\bibinfo {volume} {24}},\ \bibinfo {pages} {3862} (\bibinfo {year} {1981})}\BibitemShut {NoStop}%
\bibitem [{\citenamefont {Rodr{\'\i}guez-Carvajal}(1993)}]{Fullprof}%
  \BibitemOpen
  \bibfield  {author} {\bibinfo {author} {\bibfnamefont {J.}~\bibnamefont {Rodr{\'\i}guez-Carvajal}},\ }\bibfield  {title} {\bibinfo {title} {Recent advances in magnetic structure determination by neutron powder diffraction},\ }\href {https://doi.org/10.1016/0921-4526(93)90108-I} {\bibfield  {journal} {\bibinfo  {journal} {Physica B: Condensed Matter}\ }\textbf {\bibinfo {volume} {192}},\ \bibinfo {pages} {55} (\bibinfo {year} {1993})}\BibitemShut {NoStop}%
\bibitem [{\citenamefont {Chapon}\ \emph {et~al.}(2011)\citenamefont {Chapon}, \citenamefont {Manuel}, \citenamefont {Radaelli}, \citenamefont {Benson}, \citenamefont {Perrott}, \citenamefont {Ansell}, \citenamefont {Rhodes}, \citenamefont {Raspino}, \citenamefont {Duxbury}, \citenamefont {Spill} \emph {et~al.}}]{Wish}%
  \BibitemOpen
  \bibfield  {author} {\bibinfo {author} {\bibfnamefont {L.~C.}\ \bibnamefont {Chapon}}, \bibinfo {author} {\bibfnamefont {P.}~\bibnamefont {Manuel}}, \bibinfo {author} {\bibfnamefont {P.~G.}\ \bibnamefont {Radaelli}}, \bibinfo {author} {\bibfnamefont {C.}~\bibnamefont {Benson}}, \bibinfo {author} {\bibfnamefont {L.}~\bibnamefont {Perrott}}, \bibinfo {author} {\bibfnamefont {S.}~\bibnamefont {Ansell}}, \bibinfo {author} {\bibfnamefont {N.~J.}\ \bibnamefont {Rhodes}}, \bibinfo {author} {\bibfnamefont {D.}~\bibnamefont {Raspino}}, \bibinfo {author} {\bibfnamefont {D.}~\bibnamefont {Duxbury}}, \bibinfo {author} {\bibfnamefont {E.}~\bibnamefont {Spill}}, \emph {et~al.},\ }\bibfield  {title} {\bibinfo {title} {{WISH: The new powder and single crystal magnetic diffractometer on the second target station}},\ }\href {https://doi.org/10.1080/10448632.2011.569650} {\bibfield  {journal} {\bibinfo  {journal} {Neutron News}\ }\textbf {\bibinfo {volume} {22}},\ \bibinfo {pages} {22} (\bibinfo {year} {2011})}\BibitemShut
  {NoStop}%
\bibitem [{\citenamefont {Bewley}\ \emph {et~al.}(2011)\citenamefont {Bewley}, \citenamefont {Taylor},\ and\ \citenamefont {Bennington.}}]{LET}%
  \BibitemOpen
  \bibfield  {author} {\bibinfo {author} {\bibfnamefont {R.}~\bibnamefont {Bewley}}, \bibinfo {author} {\bibfnamefont {J.}~\bibnamefont {Taylor}},\ and\ \bibinfo {author} {\bibfnamefont {S.}~\bibnamefont {Bennington.}},\ }\bibfield  {title} {\bibinfo {title} {{LET, a cold neutron multi-disk chopper spectrometer at ISIS}},\ }\href {https://doi.org/https://doi.org/10.1016/j.nima.2011.01.173} {\bibfield  {journal} {\bibinfo  {journal} {Nuclear Instruments and Methods in Physics Research Section A: Accelerators, Spectrometers, Detectors and Associated Equipment}\ }\textbf {\bibinfo {volume} {637}},\ \bibinfo {pages} {128} (\bibinfo {year} {2011})}\BibitemShut {NoStop}%
\bibitem [{\citenamefont {Toth}\ and\ \citenamefont {Lake}(2015)}]{SpinW}%
  \BibitemOpen
  \bibfield  {author} {\bibinfo {author} {\bibfnamefont {S.}~\bibnamefont {Toth}}\ and\ \bibinfo {author} {\bibfnamefont {B.}~\bibnamefont {Lake}},\ }\bibfield  {title} {\bibinfo {title} {{Linear spin wave theory for single-Q incommensurate magnetic structures}},\ }\href {https://iopscience.iop.org/article/10.1088/0953-8984/27/16/166002} {\bibfield  {journal} {\bibinfo  {journal} {Journal of Physics: Condensed Matter}\ }\textbf {\bibinfo {volume} {27}},\ \bibinfo {pages} {166002} (\bibinfo {year} {2015})}\BibitemShut {NoStop}%
\bibitem [{\citenamefont {Pianet}\ \emph {et~al.}(2017)\citenamefont {Pianet}, \citenamefont {Urdampilleta}, \citenamefont {Colin}, \citenamefont {Cl\'erac},\ and\ \citenamefont {Coulon}}]{Pianet_2017_DW}%
  \BibitemOpen
  \bibfield  {author} {\bibinfo {author} {\bibfnamefont {V.}~\bibnamefont {Pianet}}, \bibinfo {author} {\bibfnamefont {M.}~\bibnamefont {Urdampilleta}}, \bibinfo {author} {\bibfnamefont {T.}~\bibnamefont {Colin}}, \bibinfo {author} {\bibfnamefont {R.}~\bibnamefont {Cl\'erac}},\ and\ \bibinfo {author} {\bibfnamefont {C.}~\bibnamefont {Coulon}},\ }\bibfield  {title} {\bibinfo {title} {Domain walls in single-chain magnets},\ }\href {https://doi.org/10.1103/PhysRevB.96.214429} {\bibfield  {journal} {\bibinfo  {journal} {Phys. Rev. B}\ }\textbf {\bibinfo {volume} {96}},\ \bibinfo {pages} {214429} (\bibinfo {year} {2017})}\BibitemShut {NoStop}%
\bibitem [{\citenamefont {Vaidya}\ \emph {et~al.}(2025)\citenamefont {Vaidya}, \citenamefont {Curley}, \citenamefont {Manuel}, \citenamefont {Stewart}, \citenamefont {Le}, \citenamefont {Balz}, \citenamefont {Shiroka}, \citenamefont {Blundell}, \citenamefont {Wheeler}, \citenamefont {Calderon-Lin}, \citenamefont {Manson}, \citenamefont {Manson}, \citenamefont {Singleton}, \citenamefont {Lancaster}, \citenamefont {Johnson},\ and\ \citenamefont {Goddard}}]{Vaidya_staggered}%
  \BibitemOpen
  \bibfield  {author} {\bibinfo {author} {\bibfnamefont {S.}~\bibnamefont {Vaidya}}, \bibinfo {author} {\bibfnamefont {S.~P.~M.}\ \bibnamefont {Curley}}, \bibinfo {author} {\bibfnamefont {P.}~\bibnamefont {Manuel}}, \bibinfo {author} {\bibfnamefont {J.~R.}\ \bibnamefont {Stewart}}, \bibinfo {author} {\bibfnamefont {M.~D.}\ \bibnamefont {Le}}, \bibinfo {author} {\bibfnamefont {C.}~\bibnamefont {Balz}}, \bibinfo {author} {\bibfnamefont {T.}~\bibnamefont {Shiroka}}, \bibinfo {author} {\bibfnamefont {S.~J.}\ \bibnamefont {Blundell}}, \bibinfo {author} {\bibfnamefont {K.~A.}\ \bibnamefont {Wheeler}}, \bibinfo {author} {\bibfnamefont {I.}~\bibnamefont {Calderon-Lin}}, \bibinfo {author} {\bibfnamefont {Z.~E.}\ \bibnamefont {Manson}}, \bibinfo {author} {\bibfnamefont {J.~L.}\ \bibnamefont {Manson}}, \bibinfo {author} {\bibfnamefont {J.}~\bibnamefont {Singleton}}, \bibinfo {author} {\bibfnamefont {T.}~\bibnamefont {Lancaster}}, \bibinfo {author} {\bibfnamefont {R.~D.}\ \bibnamefont {Johnson}},\ and\ \bibinfo {author}
  {\bibfnamefont {P.~A.}\ \bibnamefont {Goddard}},\ }\bibfield  {title} {\bibinfo {title} {Magnetic properties of a staggered ${S}=1$ chain with an alternating single-ion anisotropy direction},\ }\href {https://doi.org/10.1103/PhysRevB.111.014421} {\bibfield  {journal} {\bibinfo  {journal} {Phys. Rev. B}\ }\textbf {\bibinfo {volume} {111}},\ \bibinfo {pages} {014421} (\bibinfo {year} {2025})}\BibitemShut {NoStop}%
\bibitem [{\citenamefont {Bezanson}\ \emph {et~al.}(2017)\citenamefont {Bezanson}, \citenamefont {Edelman}, \citenamefont {Karpinski},\ and\ \citenamefont {Shah}}]{bezanson2017julia}%
  \BibitemOpen
  \bibfield  {author} {\bibinfo {author} {\bibfnamefont {J.}~\bibnamefont {Bezanson}}, \bibinfo {author} {\bibfnamefont {A.}~\bibnamefont {Edelman}}, \bibinfo {author} {\bibfnamefont {S.}~\bibnamefont {Karpinski}},\ and\ \bibinfo {author} {\bibfnamefont {V.~B.}\ \bibnamefont {Shah}},\ }\bibfield  {title} {\bibinfo {title} {Julia: A fresh approach to numerical computing},\ }\href {https://doi.org/10.1137/141000671} {\bibfield  {journal} {\bibinfo  {journal} {SIAM review}\ }\textbf {\bibinfo {volume} {59}},\ \bibinfo {pages} {65} (\bibinfo {year} {2017})}\BibitemShut {NoStop}%
\bibitem [{\citenamefont {Hannay}\ and\ \citenamefont {Nye}(2004)}]{Hannay_2004_fibo}%
  \BibitemOpen
  \bibfield  {author} {\bibinfo {author} {\bibfnamefont {J.}~\bibnamefont {Hannay}}\ and\ \bibinfo {author} {\bibfnamefont {J.}~\bibnamefont {Nye}},\ }\bibfield  {title} {\bibinfo {title} {Fibonacci numerical integration on a sphere},\ }\href {https://iopscience.iop.org/article/10.1088/0305-4470/37/48/005} {\bibfield  {journal} {\bibinfo  {journal} {Journal of Physics A: Mathematical and General}\ }\textbf {\bibinfo {volume} {37}},\ \bibinfo {pages} {11591} (\bibinfo {year} {2004})}\BibitemShut {NoStop}%
\end{thebibliography}%

\end{document}


\title{Supplementary Material accompanying Single-ion anisotropy driven chiral magnetism in a Spin-1 antiferromagnetic chain}

\author{S. Vaidya}
\email{s.vaidya@warwick.ac.uk}
\affiliation{Department of Physics, University of Warwick, Gibbet Hill Road, Coventry, CV4 7AL, UK}
\author{S. P. M. Curley}
\affiliation{Department of Physics, University of Warwick, Gibbet Hill Road, Coventry, CV4 7AL, UK}
\author{P. Manuel}
\author{J. Ross Stewart}
\author{M. Duc Le}
\affiliation{ISIS Pulsed Neutron Source, STFC Rutherford Appleton Laboratory, Didcot, Oxfordshire OX11 0QX, United Kingdom}
\author{A. Hern\'{a}ndez-Meli\'{a}n}
\affiliation{Department of Physics, Durham University, Durham DH1 3LE, United Kingdom}
\author{T. J. Hicken}
\affiliation{Department of Physics, Durham University, Durham DH1 3LE, United Kingdom}
\affiliation{Center for Neutron and Muon Sciences, Paul Scherrer Institut, Forschungsstrasse 111, 5232 Villigen PSI, Switzerland}
\author{C. Wang}
\author{H. Luetkens}
\author{J. Krieger}
\affiliation{Center for Neutron and Muon Sciences, Paul Scherrer Institut, Forschungsstrasse 111, 5232 Villigen PSI, Switzerland}
\author{S. J. Blundell}
\affiliation{Department of Physics, Clarendon Laboratory, University of Oxford, Parks Road, Oxford, OX1 3PU, United Kingdom}
\author{T. Lancaster}
\affiliation{Department of Physics, Durham University, Durham DH1 3LE, United Kingdom}
\author{K. A. Wheeler}
\affiliation{Department of Chemistry, Whitworth University, Spokane, Washington 99251, USA}
\author{D. Y. Villa}
\affiliation{Department of Chemistry and Biochemistry, Eastern Washington University, Cheney, Washington 99004, USA}
\affiliation{National High Magnetic Field Laboratory (NHMFL), Los Alamos National Laboratory, Los Alamos, NM, USA}
\author{Z. E. Manson}
 \author{J. A. Villa}
\author{J.~L.~Manson}\thanks{Deceased 7 June 2023.}
\affiliation{Department of Chemistry and Biochemistry, Eastern Washington University, Cheney, Washington 99004, USA}
\author{J. Singleton}
\affiliation{National High Magnetic Field Laboratory (NHMFL), Los Alamos National Laboratory, Los Alamos, NM, USA}
\author{R. D. Johnson}
\affiliation{Department of Physics and Astronomy, University College London, Gower Street, London WC1E 6BT, United Kingdom}
\affiliation{London Centre for Nanotechnology, University College London, London WC1H 0AH, United Kingdom}
\author{P. A. Goddard}
\email{p.goddard@warwick.ac.uk}
\affiliation{Department of Physics, University of Warwick, Gibbet Hill Road, Coventry, CV4 7AL, UK}

\maketitle
\tableofcontents
\section{Further Experimental Details and Results}
\subsection{Synthesis}

All chemical reagents were obtained from commercial sources and used as received. All reactions were performed in 25 ml glass beakers. An aqueous solution of NiSO$_{4}$ ($0.306$\,g, $1.1435$\,mmol) was slowly added to an aqueous solution of pyrimidine ($01079$\,g, $1.035$\,mmol) to afford a clear green solution. Slow evaporation of the solution at room temperature yields dark blue and smaller light blue crystallites.

\subsection{Single-crystal x-ray diffraction}

Single-crystal x-ray diffraction measurements were conducted on [Ni(pym)(H$_{2}$O)$_{4}$]SO$_{4}\cdot$H$_{2}$O crystals measuring $0.30 \cross 0.28 \cross 0.25$\,mm$^{3}$. A Bruker Venture D8 diffractometer equipped with a Bruker PHOTON II detector and Mo K$_{\alpha}$ ($0.71073$\,$\si{\angstrom}$) generated by a microfocus sealed x-ray tube was used to collect data at $T=100$\,K. The structure was solved using ShelXT~\cite{SheldrickT} and refined using ShelXL~\cite{SheldrickL} as implemented through Olex2~\cite{Dolomanov_olex}. C--H hydrogen atoms were restrained to calculated distances, while O--H hydrogen atoms were located in the Fourier difference maps. All non-H atomic displacements were treated anisotropically while those of hydrogen atoms were treated isotropically and riding on the donor atom. Table~\ref{tab: X-ray} provides full details of the structural refinement. The structure has been deposited and may be found at Deposition Number CCDC XXXX.

\begin{table}
\caption{Single Crystal x-ray data and refinement details for [Ni(pym)(H$_{2}$O)$_{4}$]SO$_{4}\cdot$H$_{2}$O.}
\begin{tabular}{p{6cm}p{5cm}}
\hline\hline
\multicolumn{2}{c}{Crystal data}\\
\colrule
Emp. formula                &NiC$_{4}$H$_{12}$N$_{2}$O$_{4}$S$\cdot$H$_{2}$O \\
$T$\,(K)                    & $100$ \\
Crystal system              & Tetragonal \\
Space group                 & $P4_{1}$  \\
$a,c$\,$(\si{\angstrom})$     & $7.9395(4)$, $18.4594(10)$\\
$V\,(\si{\angstrom}^{3})$ (as used)	&$1163.60(13)$\\
$Z$                         &4\\
Crystal size (mm$^{3}$) & $0.30 \cross 0.28 \cross 0.25$\\
\colrule
\multicolumn{2}{c}{Data collection}\\
\colrule
Instrument & Bruker Venture D8 diffractometer \\
Radiation & Mo K$_{\alpha}$ ($0.71073$\,$\si{\angstrom}$)\\
No. of measured reflections & 14733\\
No. of independent reflections  & 2356 \\
No. of measured reflections [$I\geq2\sigma(I)$]& 2311\\
$R_{\text{int}}$                &$0.063$  \\
$h$-index range & $-9\leq h \leq 9$\\
$k$-index range & $-9\leq k \leq 9$\\
$l$-index range & $-23\leq l \leq 23$\\
Absorption Correction & Multi-Scan method (SADABS) \\
\colrule
\multicolumn{2}{c}{Refinement}\\
\colrule
\multirow{3}{*}{R indexes [$F^{2}\geq2\sigma(F^{2})$]}&R\,=\,$0.022$ \\                                                                            &$w$R\,=\,$0.0548$ \\
                                                      &$S$\,=\,$1.06$\\
Data/param./restr. &$2356$/$195$/$11$\\
$\Delta\rho_{\text{max}}$, $\Delta\rho_{\text{max}}$\,(e$\si{\angstrom}^{-3}$)&$0.40$, $-0.66$\\
\hline\hline
\label{tab: X-ray}
\end{tabular}
\end{table}

\subsection{SQUID Magnetometry}

\begin{figure}
    \includegraphics[width= 0.9\linewidth]{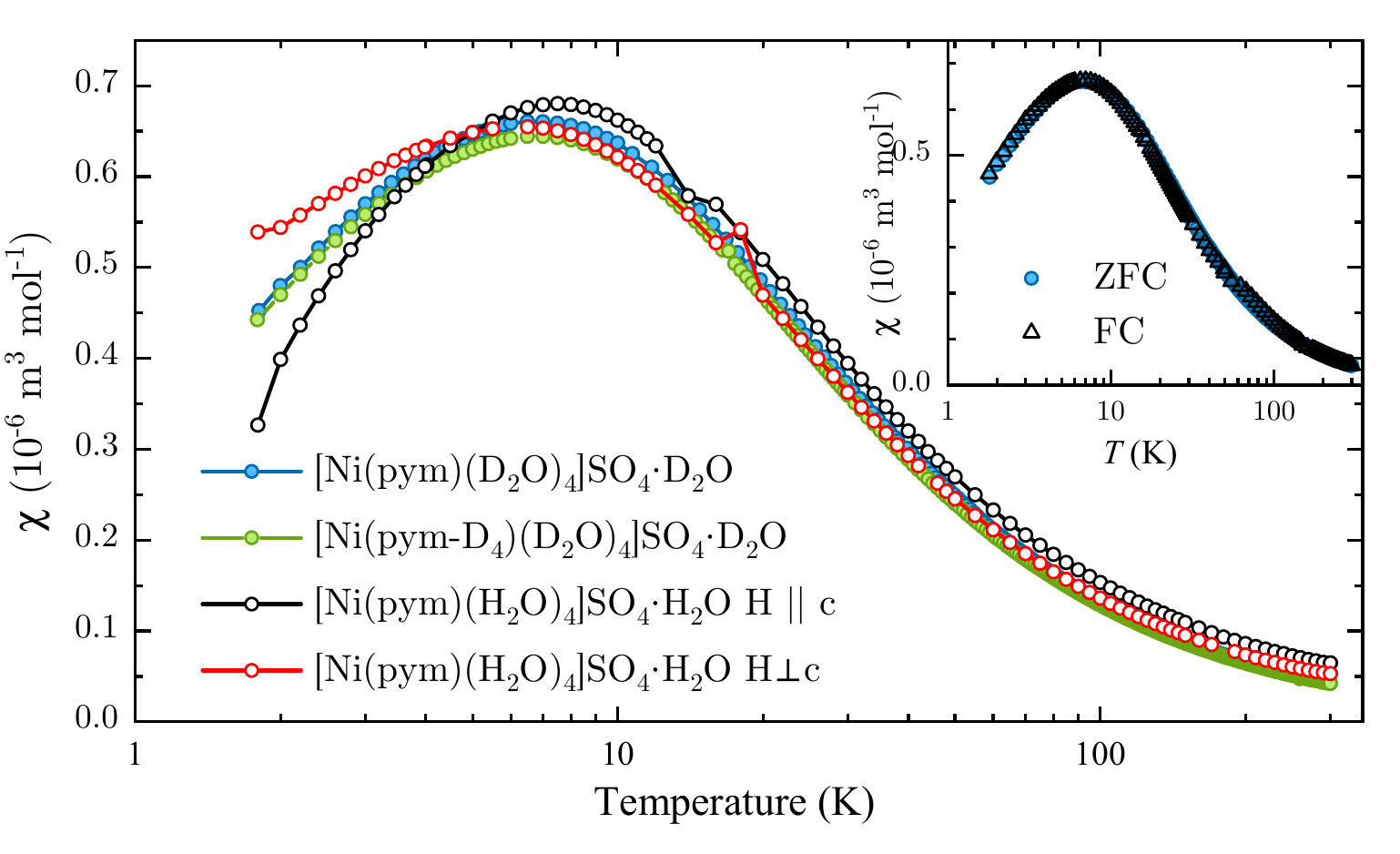}
    \caption{Comparison of the zero-field-cooled (ZFC) magnetic suceptibilites ($\chi(T)$) of the fully deuterated [Ni(pym-D$_4$)(D$_{2}$O)$_{4}$]SO$_{4}\cdot$D$_{2}$O, partially deuterated [Ni(pym)(D$_{2}$O)$_{4}$]SO$_{4}\cdot$D$_{2}$O and hydrogenated single crystal [Ni(pym)(H$_{2}$O)$_{4}$]SO$_{4}\cdot$H$_{2}$O samples. Inset: ZFC and field-cooled $\chi(T)$ curves of [Ni(pym)(D$_{2}$O)$_{4}$]SO$_{4}\cdot$D$_{2}$O lie on top of each other.}
    \label{fig: chiral_supp_chi}
\end{figure}

\begin{table}[h!]
    \centering
    \begin{tabular}{p{2.5cm}|c|c|>{\centering\arraybackslash}p{2.5cm}|>{\centering\arraybackslash}p{2.5cm}}
    \hline\hline
    \multirow{2}{*}{Parameter} & \multirow{2}{*}{[Ni(pym)(D$_{2}$O)$_{4}$]SO$_{4}\cdot$D$_{2}$O} & \multirow{2}{*}{[Ni(pym-D$_4$)(D$_{2}$O)$_{4}$]SO$_{4}\cdot$D$_{2}$O} & \multicolumn{2}{c}{[Ni(pym)(H$_{2}$O)$_{4}$]SO$_{4}\cdot$H$_{2}$O} \\
     & & & $H \parallel c$ & $H \parallel ab$ \\
    \hline
    $g$ & $2.18(1)$ & $2.15(2)$ & $2.16(1)$ & $2.07(1)$\\
    $\theta_{\rm c}$ (K) & $-9.19(6)$ & $-9.2(2)$ & $-7.2(8)$& $-6(1)$ \\
    $\chi_{0}$ (m$^{3}$mol$^{-1}$) & $-5.73(4) \times 10^{-9}$ & $-4.5(1) \times 10^{-9}$ & $-15.5(5) \times 10^{-9}$ & $-8.0(5)\times 10^{-9}$ \\
    \hline\hline
    \label{tab: chiral_supp_CW_params}
    \end{tabular}
    \caption{Relevant magnetic parameters for different Ni complexes under various field orientations.}
\end{table}

Temperature-dependent magnetic susceptibility measurements, $\chi(T)$, were carried out using a Quantum Design MPMS XL SQUID magnetometer. For the non-deuterated [Ni(pym)(H$_{2}$O)$_{4}$]SO$_{4}\cdot$H$_{2}$O samples, measurements were performed on multiple co-aligned crystallites under an applied magnetic field of $\mu_{0}H = 0.1$\,T, oriented both parallel and perpendicular to the crystallographic $c$-axis. For the partially-deuterated ([Ni(pym)(D$_{2}$O)$_{4}$]SO$_{4}\cdot$D$_{2}$O) and fully-deuterated ([Ni(pym-D$_4$)(D$_{2}$O)$_{4}$]SO$_{4}\cdot$D$_{2}$O) compounds, data was collected using powder samples. These were suspended in Vaseline to prevent grain movement during measurement and enclosed in gelatin capsules with low magnetic background, which were then placed inside a plastic drinking straw for the measurement. 

Samples were zero-field cooled (ZFC) to $T = 1.8$\,K, and ZFC measurements were performed during warming to $T = 300$\,K. Subsequently, field-cooled (FC) measurements were carried out during cooling to $T = 1.8$\,K. The ZFC magnetic susceptibility, $\chi(T)$, for all samples is shown in Fig.~\ref{fig: chiral_supp_chi}. The inset to Fig.~\ref{fig: chiral_supp_chi}  shows the ZFC and FC $\chi(T)$ curves for [Ni(pym)(D$_{2}$O)$_{4}$]SO$_{4}\cdot$D$_{2}$O, which lie on top of each other. 

As discussed in the main text, $\chi(T)$ for all samples exhibit Curie-Weiss (CW) behavior above $50$\,K. Fitting $\chi(T)^{-1}$ to CW law, $\chi(T) = N_{\text{A}}\mu_{0}g^{2}\mu_{\text{B}}^{2}S(S+1)/3k_{\text{B}}(T - \theta_{\rm c}) + \chi_{0}$, yields the parameters shown in Table~\ref{tab: chiral_supp_CW_params}. The CW temperature, $\theta_{\rm c}$, is negative for all samples and in close agreement with each other. The $g$-factors are typical of Ni(II) ions in octahedral environment and are also in close agreement. Below $50$\,K, $\chi(T)$ for all samples exhibit a broad maximum at $T_{\chi \rm max}\sim6.5$\,K, indicative of short-range antiferromagnetic (AFM) correlations along the chain. In the single crystal measurements, the $\chi(T)$ for $H \parallel c$ shows a rapid drop to zero for $T<T_{\chi \rm max}$, while the $H \parallel ab$ data show hints of a plateau. This suggests that the four-fold SIA results in a spin-gap when the applied field is parallel to the $c$-axis~\cite{Bhattacharjee_1981}. 

\subsection{Pulsed-field Magnetometry} 
Isothermal pulsed-field magnetization measurements were performed at the National High Magnetic Field Laboratory in Los Alamos, USA. Fields of up to $65$\,T with a typical rise time of $\approx10$\,ms were used. Multiple single-crystallite samples were coaligned and packed into a PCTFE ampoule (inner diameter $1.0$\,mm) and sealed with vacuum grease to prevent sample movement. The ampoule can be moved in and out of a $1500$-turn, $1.5$\,mm bore, $1.5$\,mm-long compensated-coil susceptometer constructed from $50$-gauge high-purity copper wire. When the sample is in the coil, the voltage induced in the coil is proportional to the rate of change in $M$ over time (d$M$/d$t$). The signal is integrated and the background data, measured with an empty coil under the same conditions, is subtracted to obtain $M(H)$. The magnetic field value is measured using a coaxial $10$-turn coil and calibrated using observations of de Haas-van Alphen oscillations arising from the copper coils of the susceptometer. A $^{3}$He cryostat provides temperature control and is used to attain temperatures down to $500$\,mK. The $M(H)$ measured in pulsed-fields were normalised using $T=1.8$\,K and $4$\,K $M(H)$ data measured in the SQUID magnetometer.

\subsection{Muon-spin relaxation}

Zero-field muon-spin relaxation measurements were performed on powder samples of [Ni(pym)(H$_{2}$O)$_{4}$]SO$_{4}\cdot$H$_{2}$O using the FLAME instrument at the Swiss Muon Source, Paul Scherrer Institut, Switzerland, on warming from $20$\,mK in steps of $0.2$\,K.

\subsection{Powder x-ray diffraction}

\begin{figure}
\centering
    \includegraphics[width= 0.8\linewidth]{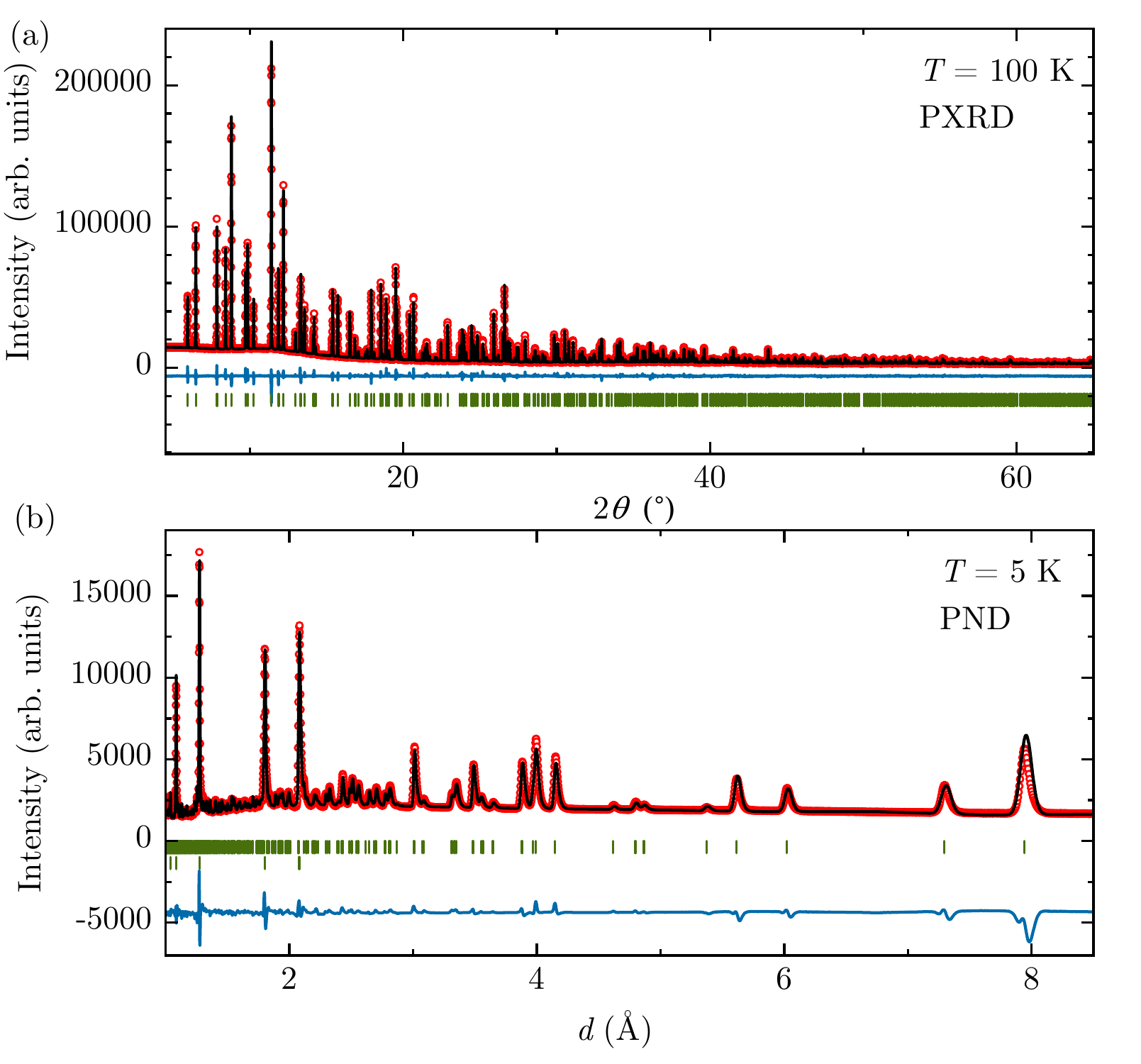}
    \caption{\label{fig: chiral_PXRD_PND}
     (a) Rietveld refinement of the $P4_{1}$ nuclear structure of [Ni(pym)(D$_{2}$O)$_{4}$]SO$_{4}\cdot$D$_{2}$O using powder x-ray diffraction (PXRD) data  collected at $T=100$\,K. (b) Powder neutron diffraction (PND) data collected at $T=5$\,K. In both panels, the experimental data are shown as red circles, the calculated fit as solid black lines, vertical tick marks indicate Bragg peak positions, and blue lines represent the difference curves. The refinement of the PND data includes the $P4_{1}$ nuclear structure and a Le Bail fit of the copper $Fm\bar{3}m$ phase.
}
\end{figure}

PXRD measurements on the partially deuterated, [Ni(pym)(D$_{2}$O)$_{4}$]SO$_{4}\cdot$D$_{2}$O, samples were performed using synchrotron radiation $\lambda = 0.82398$\,$\si{\angstrom}$ at the I11 beamline of the Diamond Light Source, Didcot, UK. Samples were ground into fine powder and used to fill quartz capillary tubes with an inner diameter of $0.5$\,mm. Data was collected using a MYTHEN detector at $T = 100$\,K. Rietveld analysis of the data was performed using the FULLPROF software~\cite{Fullprof}. A description of the refinement and structural parameters are given in Table~\ref{tab: powder_crystal} and the data is presented in Fig.~\ref{fig: chiral_PXRD_PND}(a). 

\subsection{Elastic neutron diffraction}

Elastic neutron diffraction measurements were carried out on the WISH instrument~\cite{Wish} at the ISIS Neutron and Muon Source, UK, using $\sim 1$\,g of partially deuterated powder sample of [Ni(pym)(D$_{2}$O)$_{4}$]SO$_{4}\cdot$D$_{2}$O. Data was collected over a temperature range of $0.28$-$5$\,K, with long counting times of $\approx5$\,hours at $0.28$\,K and $5$\,K and 30\,mins collection time was used for intermediate temperatures. Rietveld refinement of the nuclear structure is performed using the FULLPROF software~\cite{Fullprof} using the $T=5$\,K data collected across four detector banks with average fixed angles of $2\theta = 58.33$, $90.00$, $121.66$ and $152.83$. A description of the refinement and structural parameters and comparison to the PXRD parameters is given in Table~\ref{tab: powder_crystal}. The $5$\,K data and refinement are presented in Fig.~\ref{fig: chiral_PXRD_PND}.

\begin{table}
\caption{\label{tab: powder_crystal} Details of the crystal structure refinement of [Ni(pym)(D$_{2}$O)$_{4}$]SO$_{4}\cdot$D$_{2}$O using powder x-ray diffraction (PXRD) on the I11 instrument at $100$\,K and powder neutron diffraction (PND) measurement on WISH instrument at $5$\,K.}
\centering
\begin{tabular}{p{5cm} p{3.5cm} p{3.5cm}}
\hline
\hline
 & PXRD & PND\\
\colrule
Emp. formula                    & \multicolumn{2}{c}{NiC$_{4}$H$_{4}$D$_{8}$N$_{2}$O$_{4}$S$\cdot$D$_{2}$O} \\
Crystal system, Space group     & \multicolumn{2}{c}{Tetragonal, $P4_{1}$} \\ 
\colrule
Instrument                      & I11 & WISH\\
$T$\,(K)                        & $100$ & $5$\\
$a,c$\,$(\si{\angstrom})$     & $7.9427(1)$, $18.4552(1)$ & $7.9302(2)$, $18.4541(7)$\\
$V$\,$(\si{\angstrom}^{3})$     & $1164.281(1)$ & $1160.53(5)$  \\
$Z$                             & $4$ & $4$            \\
$R_{\text{WP}}$\,($\%$)         & $4.23$   & $4.04$         \\
$R_{\text{Bragg}}$\,($\%$)      & $7.578$  & $13.08$       \\
\hline
\hline
\end{tabular}
\end{table}

\subsection{Inelastic neutron scattering}

\begin{figure}
\centering
    \includegraphics[width= 0.5\linewidth]{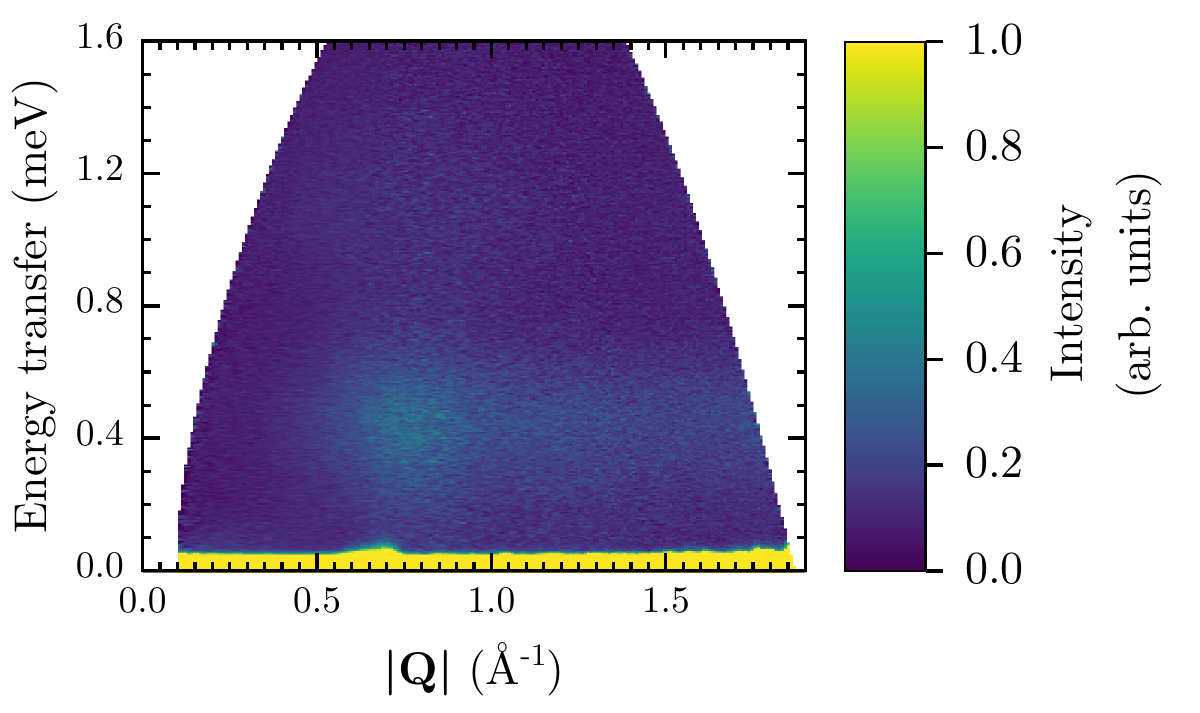}
    \caption{\label{fig: chiral_4K_LET} Time of flight inelastic neutron scattering (INS) spectra of [Ni(pym-D$_{4}$)(D$_{2}$O)$_{4}$]SO$_{4}\cdot$D$_{2}$O taken at $4.3$\,K and with neutron incident energy $\epsilon_{i} = 2.04$\,meV.}
\end{figure}

Inelastic neutron scattering measurements were performed on the LET instrument~\cite{LET} at the ISIS Neutron and Muon Source, UK, using $\sim 1$\,g of fully deuterated powder sample, [Ni(pym-D$_{4}$)(D$_{2}$O)$_{4}$]SO$_{4}\cdot$D$_{2}$O. Data was collected at INS measurements were carried out at $0.25$\,K, $4.3$\,K and $50$\,K using incident energy $\epsilon_{i}=2.04$\,meV, $3.36$\,meV, $6.55$\,meV and $18$\,meV. Figure~\ref{fig: chiral_4K_LET} shows the $4.3$\,K data collected with $\epsilon_{i}=2.04$\,meV.

\section{Calculations}
\subsection{Anisotropy tensor}

We use a site-dependent local anisotropy tensor, $K_{i}$ in the global $xyz$ laboratory frame to account for the chiral octahedral orientation. For LSWT simulation in SpinW~\cite{SpinW}, this laboratory frame is chosen such that the $x,y$ and $z$ axes are respectively parallel to the crystallographic $a,b$ and $c$ axes. The anisotropy energy of Ni(II) ions in the frame of the local octahedra is 
\begin{equation}
    \mathcal{H}_{\text{SIA}} = \mathbf{S}^{T}K^{\text{loc}}\mathbf{S} =
        \begin{pmatrix}
        S_{x}&S_{y}&S_{z}
    \end{pmatrix} \cdot
    \text{diag}\left[0,0,D\right] \cdot
    \begin{pmatrix}
        S_{x}\\S_{y}\\S_{z}
    \end{pmatrix}.
\end{equation}
The local $K^{\text{loc}}$ tensors are transformed to the global frame using Euler rotations
\begin{equation}
    K_{i} = Q_{i}^{T}K^{\text{loc}}Q_{i}.
\end{equation}
Here the transformation matrix is
\begin{equation}
    Q_{i} = R_{x}(\alpha)\cdot R_{z}(\gamma_{i}),
\end{equation}
with the $R_{z}(\gamma)$, $R_{x}(\alpha)$ rotation matrices,
\begin{equation}
    R_{x}(\alpha) = \begin{bmatrix}
        1         & 0              & 0\\
        0         & \cos(\alpha)  & \sin(\alpha)\\
        0         &-\sin(\alpha)  & \cos(\alpha)
    \end{bmatrix}, 
    R_{z}(\gamma_{i}) = \begin{bmatrix}
        \cos(\gamma_{i})&\sin(\gamma_{i})& 0\\
        -\sin(\gamma_{i})& \cos(\gamma_{i})& 0\\
        0           & 0           & 1
    \end{bmatrix}
\end{equation}
The angle $\alpha=49.26(8)^{\circ}$ is the tilting angle of the Ni octahedra away from the $c$-axis and $\gamma_{i} = 0^{\circ}, 90^{\circ}, 180^{\circ}$ and $270^{\circ}$ accounts for the four-fold rotation of the Ni octahedra about the $c$-axis.

\subsection{$D/J_{0}$ Estimate from canting angle}

\label{sec: chiral_MF}
\begin{figure}[h]
    \centering
    \includegraphics[width= \linewidth]{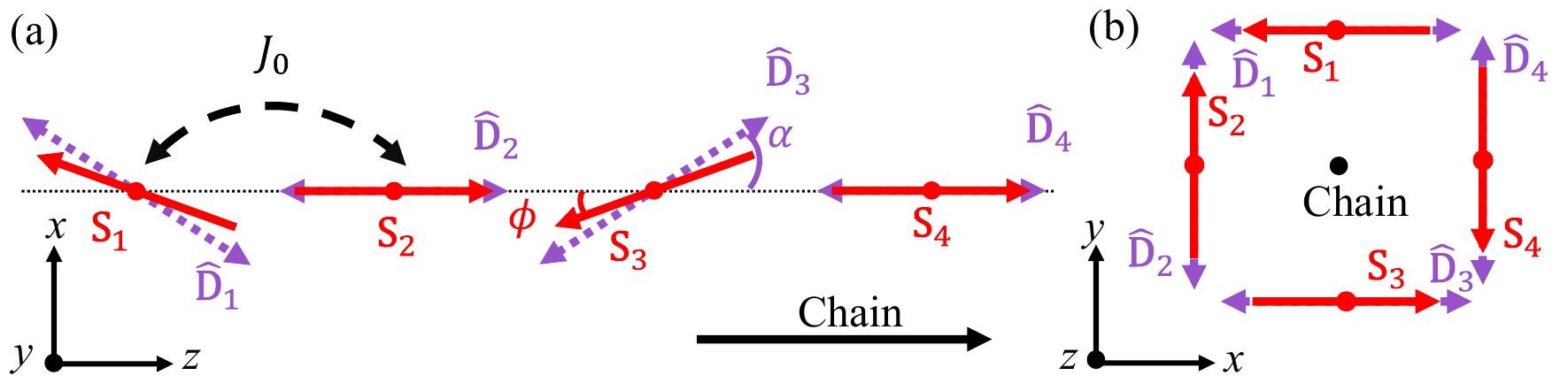}
    \caption{Schematic representation of the Ni chiral chain [Ni(pym)(H$_{2}$O)$_{4}$]SO$_{4}\cdot$H$_{2}$O viewed down the (a) $y$-axis and (b) $z$-axis (chain axis). The red arrows represent the magnetic moment direction with canting angle $\phi$ and the purple arrows show the local single-ion-anisotropy direction at angle $\alpha$ from the chain direction.}
    \label{fig: chiral_min_Hamitonian}
\end{figure}

Similar to Refs.~\cite{Pianet_2017_DW, Vaidya_staggered}, we employ a minimal mean-field Hamiltonian for Ni chiral chain, containing only $J_{0}$ and $D$ terms
\begin{equation}
\begin{split}
    \mathcal{H}=&J_{0}\sum^{N}_{\left \langle i,j \right \rangle}\mathbf{S}_{i}\cdot\mathbf{S}_{j} +D \sum^{N/4}_{n}\left[\sum^{4}_{i}\left (\hat{\mathbf{D}}_{i}\cdot\mathbf{S}_{i} \right )^{2} \right]. 
\end{split}
    \label{eq: chiral_min_Hamitonian}
\end{equation}
to estimate the SIA-induced canting angle ($\theta$) for [Ni(pym)(H$_{2}$O)$_{4}$]SO$_{4}\cdot$H$_{2}$O. Here, $\mathbf{S}_{i}$ is the spin vectors of each ion and $\left \langle i,j \right \rangle$ denotes the sum over unique newerest neighbour exchange bonds. The second term involves two summations: the inner sum accounts for the SIA for the four spin sites along the chain with unit vectors $\hat{\mathbf{D}_{1}} = [-\sin(\alpha), 0, \cos(\alpha)]$, $\hat{\mathbf{D}_{2}} = [0, -\sin(\alpha), \cos(\alpha)]$, $\hat{\mathbf{D}_{3}} = [\sin(\alpha), 0, \cos(\alpha)]$ and $\hat{\mathbf{D}_{4}} = [0, \sin(\alpha), \cos(\alpha)]$. The outer sum is over the whole system with $N$ spins. A schematic representation of Eq.~\ref{eq: chiral_min_Hamitonian} and the ground state structure is shown in Fig.~\ref{fig: chiral_min_Hamitonian}. The ground state energy per spin is
\begin{equation}
    \epsilon = -J\cos^{2}(\theta) + D\cos^2(\theta-\alpha).
\end{equation}
Minimising this energy with respect to $\theta$ yields
\begin{equation}
    \frac{D}{J_{0}} = \frac{\sin{2\theta}}{2\sin(\theta-\alpha)\cos(\theta-\alpha)}.
    \label{eq: chiral_DJ_ratio}
\end{equation}

\subsection{Monte-Carlo simulations}

Monte-Carlo simulations of $M(H)$ were performed using a routine written in Julia~\cite{bezanson2017julia}. The simulation utilises a cluster of 8 ions arranged in the configuration shown in Fig.~6(a) in the main text, with periodic boundary conditions. The magnetic moment of each ion is represented as a classical unit vector and the simulations aim to minimise the ground-state energy of the Hamiltonian in Eq.~1 at each value of $\mu_{0}H$ using simulated annealing loops. The powder average $M(H)$ is calculated by averaging the magnetisation curves for all 100 different field orientations, which are evenly spaced on a unit sphere using the golden ratio method~\cite{Hannay_2004_fibo}.

\bibliography{supp.bib}